\documentclass[final,3p,fleqn,12pt]{elsarticle}
\usepackage{amsthm}
\usepackage{mathtools}
\usepackage{tabularx}

\makeatletter
\def\ps@pprintTitle{%
 \let\@oddhead\@empty
 \let\@evenhead\@empty
 \def\@oddfoot{}%
 \let\@evenfoot\@oddfoot}
\makeatother

\usepackage{amssymb}
\usepackage{multicol}
\usepackage{amsmath}
\usepackage{upgreek}

\PassOptionsToPackage{numbers,sort&compress}{natbib}
\usepackage{natbib}

\usepackage{pifont}
\usepackage{xcolor}
\definecolor{darkgreen}{rgb}{0, 0.7, 0}

\newcommand{\cmarkgreen}{\textcolor{darkgreen}{\raisebox{0.11ex}{\ding{51}}}}
\newcommand{\xmarkred}{\textcolor{red}{\raisebox{0.11ex}{\ding{55}}}}
\usepackage{amsmath}
\usepackage{setspace}
\usepackage{amsfonts}
\usepackage{amsthm}
\usepackage{bbm}
\usepackage{mathtools}
\usepackage{subfig}
\usepackage{lineno}
\usepackage{algorithm}
\usepackage{algpseudocode}
\usepackage{hyperref}
\usepackage{float}
\usepackage{array,multirow}
\usepackage{color}
\usepackage{tikz}
\usepackage{graphicx}
\usepackage[flushleft]{threeparttable}
\usepackage{blindtext}
\usepackage[colorinlistoftodos,prependcaption,textsize=tiny]{todonotes}
\usepackage{regexpatch}
\usepackage[export]{adjustbox}
\usepackage{rotating}
\usepackage{amssymb,mleftright}
\usepackage{bm}
\usepackage{caption}
\usepackage{subfig}
\usepackage{booktabs}
\usepackage{lscape}

\definecolor{lightgray}{gray}{0.80}

\usetikzlibrary{decorations.pathreplacing}
\usepackage[listings,theorems,skins,breakable]{tcolorbox}
\tcbset{skin=enhanced}
\newtcolorbox{lbracebox}[1][Word]{%
   frame hidden,enlarge left by=2cm,width=\linewidth-2cm,%
  overlay unbroken = {\draw [decorate,decoration={brace,amplitude=10pt},]%
                     (frame.south west)-- (frame.north west)
                    node [black,midway,left,xshift=-.6cm] {#1};},%
}

\usepackage{scalerel,amssymb}

\makeatletter
\xpatchcmd{\@todo}{\setkeys{todonotes}{#1}}{\setkeys{todonotes}{inline,#1}}{}{}
\makeatother

\theoremstyle{plain}
\newtheorem{theorem}{Theorem}[section]
\newtheorem{proposition}[theorem]{Proposition}

\newtheorem{remark}[theorem]{Remark}

\theoremstyle{definition}

\newcommand{\mat}[1]{\mathbf{#1}}

\modulolinenumbers[5]

\usepackage[colorinlistoftodos]{todonotes}
\usepackage{mathrsfs}

\allowdisplaybreaks

\begin{document}

\begin{frontmatter}



\title{An entropy-variable macro-element HDG formulation with entropy stability and kinetic energy preservation for compressible flows}

\author[TUD,TUD2]{Vahid Badrkhani}
\ead{badrkhani@fdy.tu-darmstadt.de}
\corref{cor1}

\author[TUD]{Marco F.P. ten Eikelder}
\ead{marco.eikelder@tu-darmstadt.de}

\author[TUD2]{Christian Hasse}
\ead{hasse@stfs.tu-darmstadt.de}

\author[TUD]{Dominik Schillinger}
\ead{dominik.schillinger@tu-darmstadt.de }

\cortext[cor1]{Corresponding author}

\address[TUD]{Institute for Mechanics, Computational Mechanics Group, Technical University of Darmstadt, Germany}
\address[TUD2]{Institute for Simulation of Reactive Thermo-Fluid Systems, Technical University of Darmstadt, Germany}

\begin{abstract}
This paper presents an entropy-variable formulation for macro-element hybridized discontinuous Galerkin (HDG) discretizations of the compressible Navier–Stokes equations. The formulation extends the previously developed macro-element HDG framework by introducing entropy variables together with entropy-conservative, entropy-stable, and kinetic-energy-preserving numerical fluxes while retaining its computational efficiency. We derive the discrete entropy evolution equation and establish conditions for entropy conservation and entropy stability for arbitrary polynomial degrees and general macro-element discretizations. In contrast to the standard conservative-variable formulation, the entropy-variable formulation provides intrinsic stabilization within the $C^0$-continuous macro-elements, eliminating the need for additional SUPG stabilization. The proposed method is assessed via several benchmark problems including the inviscid isentropic vortex and the Taylor–Green vortex in both inviscid and turbulent regimes. Our numerical results demonstrate optimal convergence and substantially improved robustness in under-resolved flow regimes while preserving the order-of-magnitude computational advantage of the macro-element HDG framework over conventional HDG methods.
\end{abstract}

\begin{highlights}
\item We present an entropy-variable macro-element HDG formulation for 3D compressible flows.
\item Entropy-conservative, entropy-stable, and kinetic-energy-preserving fluxes are constructed within the macro-element HDG framework.
\item Entropy conservation and entropy stability are shown for arbitrary polynomial degree and general macro-element discretizations.
\item Entropy variables improve robustness in under-resolved and turbulent compressible-flow simulations.
\item Additional SUPG stabilization is unnecessary inside $C^0$-continuous macro-elements.
\item Benchmark results confirm optimal accuracy and up to one order-of-magnitude speedup over standard HDG methods for turbulent compressible flows.

\end{highlights}

\end{frontmatter}



\section{Introduction \label{Sec1}}

Discontinuous Galerkin (DG) methods \cite{arnold2002unified} are well-regarded for their solid mathematical foundation and effectiveness in solving conservation laws. These methods offer flexibility by supporting arbitrary polynomial orders on unstructured meshes and demonstrate favorable stability properties when applied to convective operators \cite{bassi1997high,cockburn2018discontinuous,hesthaven2007nodal,peraire2008compact}.
To enhance computational performance, Cockburn and co-workers introduced the hybridized discontinuous Galerkin (HDG) method in 2009 \cite{cockburn2009unified}. By introducing numerical traces on element boundaries, HDG methods substantially reduce the number of globally coupled degrees of freedom, enabling localized element-wise solves and improved scalability. These properties have made HDG methods attractive for large-scale simulations on modern parallel architectures \cite{HDGLES,peraire2010hybridizable}.

Despite these advantages, two major challenges remain for large-scale compressible flow simulations. First, standard HDG methods still incur considerable memory consumption and computational cost due to the storage of numerical traces and auxiliary variables and the solution of large coupled systems \cite{pazner2017stage,kronbichler2018performance}. Second, similar to other high-order DG discretizations, they become increasingly susceptible to numerical instabilities in under-resolved and turbulent flow regimes \cite{moura2017eddy,winters2018comparative}.

The present work addresses these two challenges by combining two complementary developments. The first is the macro-element HDG framework introduced in our earlier work \cite{badrkhani2023matrix,badrkhani2025matrix}. By embedding $C^0$-continuous Galerkin discretizations within discontinuous macro-elements, this framework substantially reduces both global and local degrees of freedom while preserving the key advantages of standard HDG. It naturally supports matrix-free implementations and highly parallel local solves, resulting in approximately an order-of-magnitude computational speedup.

The second, and primary contribution of the present work, is the development of an entropy-variable formulation for macro-element HDG discretizations. Entropy-variable formulations have proven highly effective in improving robustness of high-order DG methods for under-resolved and turbulent compressible flows by enforcing a discrete entropy inequality consistent with the second law of thermodynamics \cite{chandrashekar2013discontinuous,hiltebrand2014entropy,lv2014discontinuous,chen2017entropy,gassner2018br1,williams2018entropy,keim2025entropy}. A particularly successful strategy employs flux differencing together with entropy-conservative and entropy-stable numerical fluxes following the pioneering work of Tadmor \cite{osher1988convergence,tadmor1987numerical,tadmor2003entropy}. While these concepts have been extensively developed for DG and HDG methods, an entropy-variable formulation for the macro-element variant of the HDG method has, to the best of our knowledge, not yet been developed.

To address this gap, we develop an entropy-variable formulation for macro-element HDG discretizations. The proposed formulation combines the computational efficiency of the macro-element HDG framework with the robustness of entropy-variable methods by introducing entropy-conservative, entropy-stable, and kinetic-energy-preserving numerical fluxes. We derive the corresponding discrete entropy evolution equation and establish conditions for entropy conservation and entropy stability for arbitrary polynomial degrees and general macro-element discretizations. In contrast to the corresponding conservative-variable formulation, the entropy-variable formulation provides intrinsic stabilization within the $C^0$-continuous macro-elements, eliminating the need for additional SUPG stabilization. Numerical examples demonstrate improved robustness and reliability in subsonic, transonic, and turbulent flow regimes while retaining the order-of-magnitude computational advantage of the macro-element HDG framework over conventional HDG methods.

The remainder of this paper is organized as follows. Section \ref{Sec2} introduces the compressible Navier--Stokes equations together with entropy pairs and symmetrization. Sections \ref{Sec3} and \ref{Sec4} present the conservative-variable and entropy-variable macro-element HDG formulations, respectively. Section \ref{Sec5} introduces the entropy-conservative, entropy-stable, and kinetic-energy-preserving numerical fluxes together with the corresponding entropy analysis. Section \ref{Sec6} discusses the parallel solution strategy, including the inexact Newton method, static condensation, matrix-free implementation, and preconditioning. Numerical examples are presented in Section \ref{Sec7}, followed by conclusions in Section \ref{Sec8}.

\section{The compressible Navier--Stokes equations}\label{Sec2}

In this section, we briefly review the governing equations for compressible flow together with the associated entropy framework that forms the basis for the stability analysis of the proposed method. We begin by introducing the conservation form of the equations, followed by a discussion of entropy functions and entropy solutions.

\subsection{Governing equations}\label{Sec21}

The time-dependent compressible Navier--Stokes equations are expressed in non-dimensional conservative form as
\begin{subequations}\label{C4}
 \begin{align}
 \mat q - \nabla \mat u &= 0,\\
 \frac{\partial \mat {u}}{\partial t} + \nabla \cdot (\mat {F} + \mat {G}) &= 0,\label{eq: con law}\\
 \mat u &= \mat{u}_{0}.
 \end{align}
\end{subequations}

The vector of conservative variables $\mat u \in \mathbb{R}^{n_s}$ is defined as $\mat u = (\rho, \rho \mathbf{V}, \rho E)^T$, where $n_s = d + 2$ and $d$ denotes the spatial dimension. Here, $\rho$ is the density, $\mathbf{V}$ the velocity field, and $E$ the total specific energy. The initial condition is given by $\mat{u}_{0} = (\rho_0, \rho_0 \mathbf{V}_0, \rho_0 E_0)^T$.

The inviscid and viscous fluxes are denoted by $\mat{F}(\mat u)$ and $\mat{G}(\mat u, \mat q)$, respectively, and are written as
\begin{subequations}\label{C5}
\begin{align}
\mat {F}(\mat u) &= \left[\rho \mathbf{V}, \ \rho \mathbf{V} \otimes \mathbf{V} + P \mat{I}, \ \rho \mathbf{V} H \right]^T, \\
\mat{G}(\mat{u}, \mat{q}) &= -\left[0, \ \mat{\tau}, \ \mat{\tau}\mathbf{V} - \boldsymbol{\phi} \right]^T.
\end{align}
\end{subequations}

The viscous stress tensor and heat flux are defined as
\begin{subequations}\label{C6}
\begin{align}
\mat {\tau} &= \frac{1}{{ Re}_{c_\infty}} \, \mu \left( \nabla \mathbf{V} + (\nabla \mathbf{V})^T + \lambda (\nabla \cdot \mathbf{V}) \mat{I} \right),\\
\boldsymbol{\phi} &= - \frac{1}{{ Re}_{c_\infty} \, {\rm Pr}} \frac{\kappa}{R} \nabla T.
\end{align}
\end{subequations}

Here, $\mu$ denotes the dynamic viscosity, $\lambda$ the bulk viscosity, $\kappa$ the thermal conductivity, $R$ the specific gas constant, ${\rm Re}_{c_\infty}$ the acoustic Reynolds number, and ${\rm Pr}$ the Prandtl number. The thermodynamic relations are given by
\[
\gamma P = \rho T, \quad e = c_v T, \quad c_v = \frac{R}{\gamma - 1}, \quad E = e + \frac{1}{2}\mathbf{V} \cdot \mathbf{V},
\]
where $P$ is the pressure, $T$ the temperature, $e$ the internal energy, and $\gamma$ the ratio of specific heats. For a detailed derivation, the reader is referred to \cite{badrkhani2025matrix}.

\subsection{Entropy solutions}\label{S22}

For the purpose of entropy analysis, the governing equations can be recast in quasi-linear form as
\begin{align}
\frac{\partial \mat {u}}{\partial t} + \mat{A}_i \nabla_i \mat {u} - \nabla_i \left( \mat{K}_{ij} \nabla_j \mat {u} \right) = 0, \quad i,j = 1,\dots,d,
\end{align}
or, equivalently, in index notation,
\begin{align}
\frac{\partial u_k}{\partial t} + A_{kil} \frac{\partial u_l}{\partial x_i} - \frac{\partial}{\partial x_i} \left( K_{ij} \frac{\partial u_k}{\partial x_j} \right) = 0, 
\quad k = 1,\dots,n_s.
\end{align}
where we used the Einstein summation convention, and have written the viscous flux as $\mat{G}_i=-\mat{K}_{ij}\nabla_j \mat{u}$ (i.e. $G_{ki}=-K_{ij}\partial u_k/\partial x_j$) \cite{hughes1986new}. Performing the arbitrary change of variables $\bm{v}=\bm{v}(\mat{u})$, we arrive at the following system:
\begin{align}\label{eq: transformed system}
    \mat{A}_0 \frac{\partial \bm {v}}{\partial t} + \tilde{\mat{A}}_{i} \nabla_i\bm{v} -\nabla_j (\tilde{\mat{K}}_{ij} \nabla_i \bm{v}) = 0, 
\end{align}
where $\mat{A}_0 = \partial \mat{u}/\partial \bm{v}$, $\mat{A}_i = \partial \mat{F}_{i}/\partial \mat{u}$, $\tilde{\mat{A}}_i=\mat{A}_{i}\mat{A}_0$ and $\tilde{\mat{K}}=\mat{K}\mat{A}_{0}$. The system may be symmetrized if and only if there exists an entropy -- entropy flux pair \cite{mock1980systems,harten1983symmetric}.
A key property of this system is that it can be symmetrized if there exists a convex entropy function $H(\mat u)$ and an associated entropy flux $\boldsymbol{\mathcal{F}}(\mat u)$ satisfying
 \begin{align}\label{s2200}
 \frac{\partial {\boldsymbol{\mathcal{F}}_i}  }{\partial \mat{u}} = \frac{\partial  {\mathbf{F}_i}}{\partial \mathbf{u}} \frac{\partial  H}{\partial \mathbf{u}}= \mat{A}_i\frac{\partial  H}{\partial \mathbf{u}},
 \end{align}
or in index notation:
 \begin{align}\label{s2200}
 \frac{\partial {\mathcal{F}_{i}}  }{\partial u_{l}} = 
 \frac{\partial  {F}_{ki}}{\partial u_{l}} \frac{\partial  H}{\partial u_{k}},
      \quad   k =1, ..., {n}_{s}.
\end{align}
Defining the entropy variables as $\bm{v} = \partial H / \partial \mat{u}$, one obtains the entropy balance relation
\begin{align}\label{eq: transformed system 2}
\frac{\partial H}{\partial t} + \nabla \cdot \boldsymbol{\mathcal{F}} 
- \nabla \cdot \left( \bm{v}^T \tilde{\mathbf{K}} \nabla \bm{v} \right)
+ \nabla \bm{v}^T \tilde{\mathbf{K}} \nabla \bm{v} = 0.
\end{align}

For the compressible Navier--Stokes equations, a suitable entropy--entropy flux pair is given by \cite{harten1983symmetric}
\begin{align}\label{eq:en}
H = -\frac{\rho s}{\gamma - 1}, 
\quad 
\boldsymbol{\mathcal{F}} = - \rho s \mathbf{V},
\end{align}
where $s = \log(P/\rho^\gamma)$ denotes the thermodynamic entropy. The corresponding entropy variables take the form
\begin{equation}\label{s222}
\bm{v}(\mat u) = 
\left[
\frac{\gamma - s}{\gamma - 1} - \beta \frac{\mathbf{V} \cdot \mathbf{V}}{2}, 
\ \beta \mathbf{V}, 
\ -\beta
\right]^T,
\end{equation}
with $\beta = \rho / P$.

The associated entropy potential functions $\Psi_j$ satisfy \cite{harten1983symmetric,fernandez2018entropy}
\begin{align}\label{poen}
\Psi_j = \mat{F}_j^T \bm{v} - \mathcal{F}_j = \rho \mathbf{V},
\end{align}
which provides a useful relation for constructing entropy-consistent numerical fluxes. For viscous flows, the entropy balance reduces to the inequality
\begin{align}\label{eq: entropy evolution diffuse}
\frac{\partial H}{\partial t} + \nabla \cdot \boldsymbol{\mathcal{H}} = -\mathscr{P} \leq 0,
\end{align}
where $\boldsymbol{\mathcal{H}} = \boldsymbol{\mathcal{F}} - \bm{v}^T \tilde{\mathbf{K}} \nabla \bm{v}$ and $\mathscr{P} = \nabla \bm{v}^T \tilde{\mathbf{K}} \nabla \bm{v} \geq 0$.
In the inviscid limit ($\mat{G}=0$), corresponding to the Euler equations, smooth solutions satisfy
\begin{align}\label{eq: entropy evolution Euler smooth}
\frac{\partial H}{\partial t} + \nabla \cdot \boldsymbol{\mathcal{F}} = 0,
\end{align}
while for non-smooth solutions the entropy condition,  the equation \eqref{eq: entropy evolution Euler smooth}, takes the form
\begin{align}\label{eq: entropy evolution Euler non smooth}
\frac{\partial H}{\partial t} + \nabla \cdot \boldsymbol{\mathcal{F}} \leq 0.
\end{align}

Weak solutions that satisfy the entropy inequality \eqref{eq: entropy evolution Euler non smooth} are referred to as entropy solutions. In the remainder of this work, we distinguish between  \textit{entropy-conservative} behavior for smooth solutions and  \textit{entropy-stable} behavior in the presence of discontinuities.

\begin{remark}[Entropy solutions and total variation]
Entropy solutions are closely related to total variation diminishing (TVD) solutions \cite{harten1997high}. In particular, TVD properties can be interpreted within the framework of variation entropy formulations \cite{ten2019variation,ten2020theoretical}.
\end{remark}

The entropy-variable formulation developed in Section \ref{Sec4} is based on these entropy variables and the associated entropy balance.
\section{Baseline macro-element HDG formulation in conservative variables}\label{Sec3}

This section briefly summarizes the conservative-variable macro-element HDG formulation that serves as the computational baseline for the entropy-variable developments presented in Sections~\ref{Sec4} and~\ref{Sec5}. The formulation follows our previous work \cite{badrkhani2023matrix,badrkhani2025matrix} and is included here only to establish the notation, geometric hierarchy, and computational framework required for the subsequent entropy-variable extension.

\subsection{Geometric hierarchy and approximation spaces}\label{S31}

Let $\Omega\subset\mathbb{R}^{d}$ be partitioned into non-overlapping
macro-elements $K$, and denote this partition by $\mathcal{T}_{h}$. Each
$K\in\mathcal{T}_{h}$ is itself resolved by a conforming submesh
$\mathcal{T}_{h}(K)$ of triangles for $d=2$ or tetrahedra for $d=3$. The
internal faces of $\mathcal{T}_{h}(K)$ are not part of the hybrid skeleton:
continuity is imposed across them by the local $C^{0}$ finite element space.
Discontinuities are admitted only across $\partial K$.

For two adjacent macro-elements $K^{+}$ and $K^{-}$, a set
$e=\partial K^{+}\cap\partial K^{-}$ with nonzero $(d-1)$-dimensional measure
is an interior macro-face. A set $e=\partial K\cap\partial\Omega$ of nonzero
measure is a boundary macro-face. Their collections are denoted by
$\varepsilon^{\rm Int}$ and $\varepsilon^{\partial}$, respectively, and the
macro-element skeleton is
\[
    \varepsilon_{h}
    :=\varepsilon^{\rm Int}\cup\varepsilon^{\partial}.
\]
Thus, $\partial\mathcal{T}_{h}$ contains the boundaries of the
macro-elements, rather than the boundaries of all simplices in the local
submeshes.

\begin{figure}
    \centering
    \includegraphics[width=.42\textwidth,trim={0 0 14.2cm 0},clip]
        {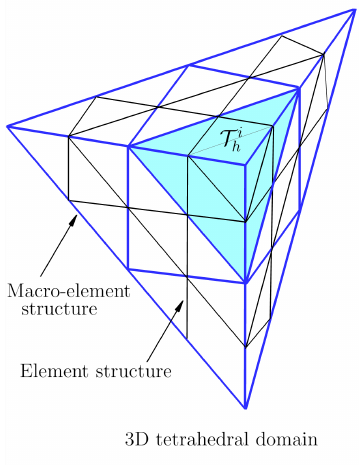}
    \caption{Three-dimensional illustration of eight discontinuous
    macro-elements. Blue lines denote
    macro-element boundaries, whereas black lines denote the boundaries of
    the tetrahedral elements inside $\mathcal{T}_h^i$. The selected
    macro-element $\mathcal{T}_h^i$ contains eight tetrahedral finite
    elements that are continuous within the macro-element.}
    \label{fig:domain}
\end{figure}

Figure~\ref{fig:domain} distinguishes the two continuity levels of the method.
The eight macro-elements are mutually discontinuous and communicate only
through trace unknowns on their blue boundaries. Within the selected
macro-element $\mathcal{T}_h^i$, the eight tetrahedral finite elements are
coupled continuously across the internal black faces. Thus, the black
interfaces belong to the conforming local finite element submesh and are not
part of the hybrid skeleton. If every macro-element contains exactly one
simplex, the construction reduces to standard HDG.

To define the discrete spaces, let $\mathcal{P}_{p}(D)$ be the polynomials of
degree at most $p$ on a simplex or face $D$. For each macro-element, introduce
the local continuous space
\[
\mathcal{W}_{p}(K)
:=
\left\{
w\in C^{0}(K):
w|_{\kappa}\in\mathcal{P}_{p}(\kappa)
\quad\forall\kappa\in\mathcal{T}_{h}(K)
\right\}.
\]
The scalar, vector, gradient, and trace spaces are then
\begin{align}
\mathcal{V}^{k}_{h}
 =&~ \left\{w\in L^{2}(\Omega):
 w|_{K}\in\mathcal{W}_{p}(K)
 \quad\forall K\in\mathcal{T}_{h}\right\}, \nonumber\\[4pt]
\bm{\mathcal{V}}^{k}_{h}
 =&~ \left\{\bm{w}\in[L^{2}(\Omega)]^{n_{s}}:
 \bm{w}|_{K}\in[\mathcal{W}_{p}(K)]^{n_{s}}
 \quad\forall K\in\mathcal{T}_{h}\right\}, \nonumber\\[4pt]
\bm{\mathcal{Q}}^{k}_{h}
 =&~ \left\{\bm{r}\in[L^{2}(\Omega)]^{n_{s}\times d}:
 \bm{r}|_{K}\in[\mathcal{W}_{p}(K)]^{n_{s}\times d}
 \quad\forall K\in\mathcal{T}_{h}\right\},\\[4pt]
\bm{\mathcal{M}}^{k}_{h}
 =&~ \left\{\bm{\mu}\in[L^{2}(\varepsilon_{h})]^{n_{s}}:
 \bm{\mu}|_{e}\in[\mathcal{P}_{p}(e)]^{n_{s}}
 \quad\forall e\in\varepsilon_{h}\right\}, \nonumber\\[4pt]
\mathcal{M}^{k}_{h}
 =&~ \left\{\mu\in L^{2}(\varepsilon_{h}):
 \mu|_{e}\in\mathcal{P}_{p}(e)
 \quad\forall e\in\varepsilon_{h}\right\}. \nonumber
\end{align}
Here, $n_{s}=d+2$ is the number of conservative variables. The notation makes
explicit that functions are continuous on each local submesh but need not be
continuous from one macro-element to the next.

For scalar, vector, and tensor fields in these spaces, the broken
macro-element inner products are
\begin{equation}\label{C31}
 \begin{aligned}
(w,v)_{\mathcal{T}_{h}}
&:=\sum_{K\in\mathcal{T}_{h}}(w,v)_{K}
=\sum_{K\in\mathcal{T}_{h}}\int_{K}wv,\\
(\bm{w},\bm{v})_{\mathcal{T}_{h}}
&:=\sum_{i=1}^{n_{s}}(w_{i},v_{i})_{\mathcal{T}_{h}},\\
(\bm{W},\bm{V})_{\mathcal{T}_{h}}
&:=\sum_{i=1}^{n_{s}}\sum_{j=1}^{d}
(W_{ij},V_{ij})_{\mathcal{T}_{h}}.
 \end{aligned}
\end{equation}
The corresponding pairings on the macro-element boundaries are
\begin{equation}\label{C32}
 \begin{aligned}
\langle\eta,\zeta\rangle_{\partial\mathcal{T}_{h}}
&:=\sum_{K\in\mathcal{T}_{h}}
\langle\eta,\zeta\rangle_{\partial K}
=\sum_{K\in\mathcal{T}_{h}}\int_{\partial K}\eta\zeta,\\
\langle\bm{\eta},\bm{\zeta}\rangle_{\partial\mathcal{T}_{h}}
&:=\sum_{i=1}^{n_{s}}
\langle\eta_{i},\zeta_{i}\rangle_{\partial\mathcal{T}_{h}}.
 \end{aligned}
\end{equation}

\subsection{Semidiscrete conservative formulation}\label{S32}

We now apply the preceding spaces to the first-order form of the compressible
Navier--Stokes equations in \eqref{C4}. The three discrete fields have distinct
roles: $\mathbf{u}_{h}$ approximates the conservative state,
$\mathbf{q}_{h}$ represents its gradient, and
$\widehat{\mathbf{u}}_{h}$ is the independent state trace on the
macro-element skeleton. For $t\in[0,T]$, find
\[
\left(\mathbf{q}_{h}(t),\mathbf{u}_{h}(t),
\widehat{\mathbf{u}}_{h}(t)\right)
\in
\bm{\mathcal{Q}}^{k}_{h}\times
\bm{\mathcal{V}}^{k}_{h}\times
\bm{\mathcal{M}}^{k}_{h}
\]
such that
\begin{subequations}\label{s31}
\allowdisplaybreaks
 \begin{alignat}{2}\label{s31a}
 &\left(\mathbf{q}_{h},\bm{r}\right)_{\mathcal{T}_{h}}
 +\left(\mathbf{u}_{h},\nabla\cdot\bm{r}\right)_{\mathcal{T}_{h}}
 -\left\langle\widehat{\mathbf{u}}_{h},
 \bm{r}\cdot\bm{n}\right\rangle_{\partial\mathcal{T}_{h}}=0,\\
 &\left(\dfrac{\partial\mathbf{u}_{h}}{\partial t},
 \bm{w}\right)_{\mathcal{T}_{h}}
 -\left(\mathbf{F}_{h}(\mathbf{u}_{h})
 +\mathbf{G}_{h}(\mathbf{u}_{h},\mathbf{q}_{h}),
 \nabla\cdot\bm{w}\right)_{\mathcal{T}_{h}}
 +\left\langle\widehat{\mathbf{F}}_{h}
 +\widehat{\mathbf{G}}_{h},\bm{w}\right\rangle_{\partial\mathcal{T}_{h}}
 \nonumber\\
 &\qquad
 +\sum_{e\in K}
 \left(\left(\mathbf{A}\cdot\nabla\right)\bm{w},
 \tau^{\rm SUPG}\mathbf{R}\right)_{\mathcal{T}_{h}}=0,
 \label{s31b}\\
 &\left\langle\widehat{\mathbf{F}}_{h}
 +\widehat{\mathbf{G}}_{h},\bm{\mu}\right\rangle_
 {\partial\mathcal{T}_{h}\setminus\partial\Omega}
 +\left\langle
 \widehat{\mathbf{B}}_{h}
 \left(\widehat{\mathbf{u}}_{h},\mathbf{u}_{h}\right),
 \bm{\mu}\right\rangle_{\partial\Omega}=0.
 \end{alignat}
\end{subequations}
The equations hold for every
$(\bm{r},\bm{w},\bm{\mu})\in
\bm{\mathcal{Q}}^{k}_{h}\times
\bm{\mathcal{V}}^{k}_{h}\times
\bm{\mathcal{M}}^{k}_{h}$. The first relation defines the auxiliary gradient
weakly. The second is the macro-element balance of mass, momentum, and total
energy, including the interior residual stabilization. The third requires the
numerical normal flux to be single-valued across the interior skeleton and
introduces the boundary operator on $\partial\Omega$. It is this last equation
that couples otherwise independent macro-element problems.

The boundary operator
$\widehat{\mathbf{B}}_{h}
(\widehat{\mathbf{u}}_{h},\mathbf{u}_{h})$
weakly imposes the physical boundary data through the hybrid state, following
the standard HDG construction \cite{cockburn2009hybridizable}. On an interior
macro-face, information is exchanged through the inviscid and viscous
numerical fluxes
\begin{subequations}\label{Flux_Con}
 \begin{align}\label{Flux_Con_Inv}
\widehat{\mathbf{F}}_{h}
\left(\widehat{\mathbf{u}}_{h},\mathbf{u}_{h}\right)
&=
\frac{1}{2}
\left(\mathbf{F}_{h}(\widehat{\mathbf{u}}_{h})
+\mathbf{F}_{h}(\mathbf{u}_{h})\right)\cdot\bm{n}
+\frac{1}{2}
\bm{\sigma}
\left(\mathbf{u}_{h},\widehat{\mathbf{u}}_{h}\right)
\cdot[\![\mathbf{u}_{h}]\!],\\
\label{Flux_Con_Vis}
\widehat{\mathbf{G}}_{h}
\left(\widehat{\mathbf{u}}_{h},\bm{q}_{h}\right)
&=
\mathbf{G}_{h}
\left(\widehat{\mathbf{u}}_{h},\bm{q}_{h}\right)\cdot\bm{n}
+\frac{1}{2}
\bm{\sigma}^{\rm vis}
\left(\mathbf{u}_{h},\widehat{\mathbf{u}}_{h}\right)
\cdot[\![\mathbf{u}_{h}]\!].
 \end{align}
\end{subequations}
Here $[\![a]\!]:=\widehat{a}-a$ is the difference between the hybrid and
interior traces, and $\bm{n}$ is the outward normal of the current
macro-element. The two stabilization operators control different mechanisms:
$\bm{\sigma}$ supplies upwind dissipation for the hyperbolic flux, whereas
$\bm{\sigma}^{\rm vis}$ penalizes discrepancies in the diffusive part.

For the inviscid contribution, a scalar Lax--Friedrichs-type stabilization is
used,
\begin{equation}
\bm{\sigma}
\left(\mathbf{u}_{h},\widehat{\mathbf{u}}_{h}\right)
=\lambda_{\max}\left(\widehat{\mathbf{u}}_{h}\right),
\end{equation}
where $\lambda_{\max}$ is the largest characteristic speed of the normal flux
Jacobian
$\mathbf{A}_{n}
=\left(\partial\mathbf{F}/\partial\mathbf{u}\right)\cdot\bm{n}$.
The viscous penalty is
\begin{equation}
\bm{\sigma}^{\rm vis}
\left(\mathbf{u}_{h},\widehat{\mathbf{u}}_{h}\right)
=
\frac{1}{Re}\,
\operatorname{diag}
\left(0,\Upsilon,
\frac{1}{(\gamma-1)M_{\infty}^{2}{\rm Pr}}\right),
\end{equation}
where $\Upsilon$ is the vector of ones associated with the momentum
components, $M_{\infty}$ is the free-stream Mach number, $Re$ is the Reynolds
number, and ${\rm Pr}$ is the Prandtl number.

Because the interior approximation is continuous on
$\mathcal{T}_{h}(K)$, the conservative-variable method supplements the
interface stabilization with the residual-based SUPG contribution in
\eqref{s31b}. In this term, $\mathbf{R}$ denotes the strong local residual,
$\mathbf{A}$ is the inviscid flux Jacobian, and $\tau^{\rm SUPG}$ is the
streamline stabilization parameter; its construction follows
\cite{xu2017compressible,shakib1991new,tezduyar2006stabilization,tezduyar2006computation}.
The term acts only inside a macro-element and therefore does not enlarge the
globally coupled system.

This organization of the unknowns is the principal computational distinction
between the macro-element and standard HDG variants. The state and gradient
fields are eliminated through independent macro-element solves, while the
global problem contains only the hybrid trace degrees of freedom. The same
topology is retained in the entropy-variable formulation introduced next; only
the choice of primary variables and stabilization mechanisms changes.

\section{The macro-element HDG formulation in entropy-variable form}\label{Sec4}

In this section, we introduce the macro-element HDG formulation expressed in entropy variables for the compressible Navier--Stokes system \eqref{C4}. In addition, we analyze the evolution of the discrete total entropy and identify the conditions under which the formulation is entropy-conservative or entropy-stable.

\subsection{Spatial discretization}\label{S32}

We consider the macro-element HDG discretization of the compressible Navier--Stokes equations in entropy-variable form. The discrete problem reads as follows:

Find  
$  \left( \mathbf{q}_{h}\left( t\right) ,\bm{v}_{h}\left( t\right),\widehat{\bm{v}}_{h}\left( t\right)\right)\in \bm{\mathcal{Q}}^{k}_{h} \times \bm{\mathcal{V}}^{k}_{h} \times \bm{\mathcal{M}}^{k}_{h}$ such that
\begin{subequations}\label{s32}
 \begin{align}\label{s32a}
&\left( \mathbf{q}_{h},\bm{r}\right) _{\mathcal{T}_{h}} + 
\left( \bm{v}_{h},\nabla\cdot\bm{r}\right) _{\mathcal{T}_{h}} - 
\left\langle   \widehat{\bm{v}}_{h} ,\bm{r}\cdot\bm{n}\right\rangle  _{\partial \mathcal{T}_{h}}= 0,\\[6pt] \label{s32b}
&\left( \dfrac{\partial \mathbf{u}\left( \bm{v}_{h}\right)}{\partial t},\bm{w}\right) _{\mathcal{T}_{h}} - 
\left( \mathbf{F}_{h}\left( \bm{v}_{h}\right)  + \mathbf{G}_{h}\left( \bm{v}_{h},\mathbf{q}_{h}\right) ,\nabla\cdot \bm{w}\right) _{\mathcal{T}_{h}} + 
\left\langle \widehat{\mathbf{F}}_{h}+\widehat{\mathbf{G}}_{h},\bm{w}\right\rangle _{\partial \mathcal{T}_{h}} = 0,\\[6pt] \label{s32c}
&\left\langle \widehat{\mathbf{F}}_{h}+\widehat{\mathbf{G}}_{h},\bm{\mu}\right\rangle  _{{\partial \mathcal{T}_{h}}\backslash\partial\Omega} 
 + \left\langle   \widehat{\mathbf{B}}_{h}\left( \widehat{\bm{v}}_{h},\bm{v}_{h},\mathbf{q}_{h}\right) ,\bm{\mu}\right\rangle  _{\partial\Omega}= 0  , 
\end{align}   
\end{subequations}
for all $ \left( \bm{r},\bm{w},\bm{\mu}\right) \in \bm{\mathcal{Q}}^{k}_{h} \times \bm{\mathcal{V}}^{k}_{h} \times \bm{\mathcal{M}}^{k}_{h}$ and all $ t\in\left[ 0,T\right] $. Here, $\bm{v}_{h}$ and $\widehat{\bm{v}}_{h}$ denote the discrete approximations of $\bm{v}$ and $\bm{v}\mid_{\varepsilon_{h}}$, respectively. The boundary operator $\widehat{\mathbf{B}}_{h}\left( \widehat{\bm{v}}_{h},\bm{v}_{h},\mathbf{q}_{h}\right)$ imposes the boundary conditions on $\partial\Omega$ within the hybrid entropy-variable framework; see \cite{fernandez2019entropy,fernandez2018entropy}. The viscous numerical flux appearing in \eqref{s32} is given by
 \begin{equation}\label{Flux_En_Vis}
 \widehat{\mathbf{G}}_{h}\left( \widehat{\bm{v}}_{h},\bm{q}_{h}\right)= 
 \mathbf{G}_{h}\left( \widehat{\bm{v}}_{h},\bm{q}_{h}\right) \cdot\bm{n} -\frac{1}{2}
 \bm{\sigma^\text{vis}}\left( \bm{v}_{h},\widehat{\bm{v}}_{h}\right) \cdot [\![ \bm{v}_{h}]\!]  \ .
 \end{equation}
  
Previous studies, including \cite{williams2018entropy,fernandez2018entropy,fernandez2019entropy}, have shown that the standard HDG formulation equipped with the viscous flux \eqref{Flux_En_Vis} is entropy-stable whenever the inviscid numerical flux is chosen in an entropy-stable manner. Since entropy stability is therefore governed primarily by the inviscid contribution, our attention will focus on the construction of appropriate inviscid fluxes. Several such fluxes will be introduced in Section \ref{Sec5}.

\begin{remark}
In contrast to the conservative-variable formulation, no additional stabilization term is introduced inside the $C^0$-continuous macro-elements in \eqref{s32b}. The use of entropy variables already provides the stabilization mechanism required for the macro-element HDG formulation, making supplementary terms such as SUPG unnecessary in this setting.
\end{remark}

\subsection{Entropy-variable properties}

We next derive an identity that governs the evolution of the discrete total entropy. This relation forms the basis for the entropy analysis of the method and clarifies how suitable choices of the inviscid numerical flux lead to entropy conservation or entropy stability.

\begin{proposition}\label{Pro_Time}
The time evolution of total generalized entropy in the macro-element entropy-variable HDG discretization of the compressible Euler equations  is given by
\begin{equation}\label{eq: time evo entropy}
\frac{{\rm d}}{{\rm d}t}\int_{\mathcal{T}_h} H( \mathbf{u}\left( \bm{v}_{h}\right)) 
- \int_{\partial \mathcal{T}_h} ([\![ \bm{v}_{h}]\!] ^T  \cdot \widehat{\mathbf{F}}_{h} -[\![ \bm{\Psi}_{n}]\!])
+ \mathbf{B}_{\partial \Omega} (\widehat{\bm{v}}_h,\bm{v}_{h}; \bm{v}_{h}^{\partial \Omega}) = 0
\end{equation}
where $\bm{\Psi}_{n}  =\bm{\Psi} \cdot \bm{n} $ is defined in \eqref{poen}, and $\mathbf{B}_{\partial \Omega}$ is a boundary term.
\end{proposition}

\begin{proof}
Denote by \( (\bm{v}_{h}(t), \widehat{\bm{v}}_h (t)) \) the numerical solution at time \( t > 0 \). Let \( b_h(\bm{v}_{h}, \widehat{\bm{v}}_h ; \bm{\mu}) \) and
\( a_h(\bm{v}_{h}, \widehat{\bm{v}}_h ; \bm{w}) \) denote shorthand expressions for the left-hand sides of \eqref{s32b} and \eqref{s32c}, respectively, in the inviscid case. For notational simplicity, we suppress the explicit time dependence. After integrating the inviscid flux term in \( a_h \) by parts, we obtain the equivalent expressions \cite{fernandez2019entropy}:
\begin{equation}
\begin{aligned}
    a_h(\bm{v}_{h}, \widehat{\bm{v}}_h; \bm{w}) &= \int_{\mathcal{T}_h} \bm{w}^T  \cdot \frac{\partial \mathbf{u}\left( \bm{v}_{h}\right)}{\partial t} +
     \int_{\mathcal{T}_h} \bm{w}^T  \cdot (\nabla \cdot \mathbf{F}(\bm{v}_h))
     + \int_{\partial  \mathcal{T}_h} \bm{w}^T  \cdot ( \widehat{\mathbf{F}}_{h} -\mathbf{F}_{n})\\
    b_h(\bm{v}_{h}, \widehat{\bm{v}}_h ; \bm{\mu}) &= \int_{\partial \mathcal{T}_h} \bm{\mu}^T \cdot  \widehat{\mathbf{F}}_{h} 
    - \int_{\partial \Omega} \bm{\mu}^T \cdot (\widehat{\mathbf{F}}_{h} -\widehat{\mathbf{B}}_{h} (\widehat{\bm{v}}_h,\bm{v}_{h}; \bm{v}_{h}^{\partial \Omega}) ).
 \end{aligned}
\end{equation}
Since \eqref{s32b} and \eqref{s32c} hold for all \((\bm{w},\bm{\mu}) \in \mathcal{V}_h^k \otimes \mathcal{M}_h^k\), and the trial and test spaces coincide, it follows that
\( a_h(\bm{v}_{h}, \widehat{\bm{v}}_h ; \bm{w}) - b_h(\bm{v}_{h}, \widehat{\bm{v}}_h ; \bm{\mu}) =0 \), that is,
\begin{equation}\label{hprof2}
\begin{aligned}
    & \int_{\mathcal{T}_h} \bm{v}^T  \cdot \frac{\partial \mathbf{u}\left( \bm{v}_{h}\right)}{\partial t} +
     \int_{\mathcal{T}_h} \bm{v}^T  \cdot (\nabla \cdot \mathbf{F}(\bm{v}_h))
     + \int_{\partial  \mathcal{T}_h} \bm{v}^T  \cdot ( \widehat{\mathbf{F}}_{h} -\mathbf{F}_{n})\\
     -& \int_{\partial \mathcal{T}_h} \widehat{\bm{v}}_h^T \cdot  \widehat{\mathbf{F}}_{h} 
    + \int_{\partial \Omega} \widehat{\bm{v}}_h^T \cdot (\widehat{\mathbf{F}}_{h} -\widehat{\mathbf{B}}_{h} (\widehat{\bm{v}}_h,\bm{v}_{h}; \bm{v}_{h}^{\partial \Omega}) )\\
    \overset{\ref{s222}}{=}&\frac{{\rm d}}{{\rm d}t}\int_{\mathcal{T}_h} H( \mathbf{u}\left( \bm{v}_{h}\right)) +  \int_{\mathcal{T}_h} \nabla  \cdot \boldsymbol{\mathcal{F}}\left( \bm{v}_{h}\right)-
    \int_{\partial \mathcal{T}_h} (\widehat{\bm{v}}_h - \bm{v}_{h})^T  \cdot \widehat{\mathbf{F}}_{h}\\
    -&\int_{\partial  \mathcal{T}_h} \bm{v}_h^T  \cdot \mathbf{F}_{n} +\int_{\partial \Omega} \widehat{\bm{v}}_h^T \cdot (\widehat{\mathbf{F}}_{h} -\widehat{\mathbf{B}}_{h} (\widehat{\bm{v}}_h,\bm{v}_{h}; \bm{v}_{h}^{\partial \Omega}) )=0
    \end{aligned}
\end{equation}
Next, we observe that
\begin{equation}\label{eq: doubling faces}
\int_{\partial \mathcal{T}_h \setminus \partial \Omega} \boldsymbol{\mathcal{F}}_n(\widehat{\bm{v}}_h) = 0, \qquad 
\int_{\partial \mathcal{T}_h \setminus \partial \Omega} \widehat{\bm{v}}_h^T \cdot \mathbf{F}_n(\widehat{\bm{v}}_h) = 0,
\end{equation}
which follows from the \(\pm\) duplication of interior faces in \(\partial \mathcal{T}_h \setminus \partial \Omega\).
Substituting \eqref{eq: doubling faces} into \eqref{hprof2} and applying the divergence theorem yields
\begin{equation}\label{hprof3}
\begin{aligned}
 &\frac{{\rm d}}{{\rm d}t}\int_{\mathcal{T}_h} H( \mathbf{u}\left( \bm{v}_{h}\right)) -  \int_{\partial \mathcal{T}_h} \left( \boldsymbol{\mathcal{F}}_n\left( \widehat{\bm{v}}_{h}\right) - \boldsymbol{\mathcal{F}}_n\left( \bm{v}_{h}\right)\right)-
    \int_{\partial \mathcal{T}_h} (\widehat{\bm{v}}_h - \bm{v}_{h})^T  \cdot \widehat{\mathbf{F}}_{h}\\
    +&\int_{\partial  \mathcal{T}_h} (\widehat{\bm{v}}_h - \bm{v}_{h})^T  \cdot \mathbf{F}_{n} 
    + \int_{\partial \Omega}\boldsymbol{\mathcal{F}}_n\left( \widehat{\bm{v}}_{h}\right)
    - \int_{\partial \Omega}\widehat{\bm{v}}_h^T \cdot  \mathbf{F}_{n} \left( \widehat{\bm{v}}_{h}\right)\\
    +&\int_{\partial \Omega} \widehat{\bm{v}}_h^T \cdot (\widehat{\mathbf{F}}_{h} -\widehat{\mathbf{B}}_{h} (\widehat{\bm{v}}_h,\bm{v}_{h}; \bm{v}_{h}^{\partial \Omega}) )
    \overset{\ref{poen}}{=}\frac{{\rm d}}{{\rm d}t}\int_{\mathcal{T}_h} H( \mathbf{u}\left( \bm{v}_{h}\right)) -
 \int_{\partial \mathcal{T}_h} (\widehat{\bm{v}}_h - \bm{v}_{h})^T  \cdot \widehat{\mathbf{F}}_{h}\\
  +&  \int_{\partial \mathcal{T}_h} (\widehat{\bm{\Psi}}_{n }- \bm{\Psi}_{n})
+\mathbf{B}_{\partial \Omega} (\widehat{\bm{v}}_h,\bm{v}_{h}; \bm{v}_{h}^{\partial \Omega}) = 0,
    \end{aligned}
\end{equation}
where 
\begin{equation}\label{BC_Pro}
\mathbf{B}_{\partial \Omega} (\widehat{\bm{v}}_h,\bm{v}_{h}; \bm{v}_{h}^{\partial \Omega}) =  \int_{\partial \Omega}\boldsymbol{\mathcal{F}}_n\left( \widehat{\bm{v}}_{h}\right)
    - \int_{\partial \Omega}\widehat{\bm{v}}_h^T \cdot  \mathbf{F}_{n} \left( \widehat{\bm{v}}_{h}\right)
    +\int_{\partial \Omega} \widehat{\bm{v}}_h^T \cdot (\widehat{\mathbf{F}}_{h} -\widehat{\mathbf{B}}_{h} (\widehat{\bm{v}}_h,\bm{v}_{h}; \bm{v}_{h}^{\partial \Omega}) ).
\end{equation}
This completes the proof.
\end{proof}

\begin{remark}
Proposition \ref{Pro_Time} provides the entropy evolution identity for the macro-element HDG formulation under a general choice of inviscid numerical flux. In the limiting case where each macro-element consists of a single $C^0$-continuous element, the formulation reduces to the standard HDG setting.
\end{remark}

Having established the discrete entropy evolution, we now state the conditions that the inviscid numerical flux must satisfy in order to obtain entropy conservation or entropy stability.

\begin{theorem}\label{EC_ES}
The macro-element entropy-variable HDG discretization of the compressible Euler equations on a periodic domain is entropy-conservative, if the numerical flux function $\widehat{\mathbf{F}}_{h}$ satisfies
\begin{equation} \label{EC_1}
[\![ \bm{v}_{h}]\!] ^T  \cdot \widehat{\mathbf{F}}_{h} -[\![ \bm{\Psi}_{n}]\!]= 0 \,.
\end{equation}
Additionally, it is entropy-stable if the numerical flux function $\widehat{\mathbf{F}}_{h}$ satisfies
\begin{equation} \label{ES_1}
[\![ \bm{v}_{h}]\!] ^T  \cdot \widehat{\mathbf{F}}_{h} -[\![ \bm{\Psi}_{n}]\!] \leq 0 \, .
\end{equation}
\end{theorem}

\begin{proof}
Under periodic boundary conditions, the boundary contribution vanishes, i.e., $\mathbf{B}_{\partial \Omega} = 0$. Consequently, the entropy evolution identity from Proposition~\ref{Pro_Time}, namely equation~\eqref{eq: time evo entropy}, reduces to
\begin{equation}\label{EC_ES_for}
\frac{{\rm d}}{{\rm d}t} \int_{\mathcal{T}_h} H\big( \mathbf{u}( \bm{v}_{h} ) \big) 
= \int_{\partial \mathcal{T}_h} \big( [\![ \bm{v}_{h} ]\!]^T \cdot \widehat{\mathbf{F}}_{h} - [\![ \bm{\Psi}_{n} ]\!] \big) \, .
\end{equation}

If the integrand on the right-hand side vanishes identically on every interface, then the total entropy remains constant in time and the numerical flux $\widehat{\mathbf{F}}_{h}$ is entropy-conservative.

If, on the other hand, the right-hand side is non-positive, that is,
\[
[\![ \bm{v}_{h} ]\!]^T \cdot \widehat{\mathbf{F}}_{h} - [\![ \bm{\Psi}_{n} ]\!] \leq 0 \quad \text{on } \partial \mathcal{T}_h,
\]
then the discrete total entropy is non-increasing, and the scheme is entropy-stable.
\end{proof}

\begin{remark}
The condition stated in Theorem \ref{EC_ES} is analogous to the corresponding criteria used in DG and finite volume formulations for entropy-conservative and entropy-stable fluxes \cite{pazner2019analysis,carpenter2014entropy}. In the present setting, however, the jump in the state may arise either from the traces on opposite sides of a macro-element interface or from variations of the state at different locations within a single macro-element. As in related frameworks, consistency and symmetry of the numerical flux remain fundamental requirements \cite{tadmor1987numerical,fisher2013high}.
\end{remark}
\section{Numerical fluxes and time discretization}\label{Sec5}

In this section, we introduce the inviscid numerical fluxes employed in the proposed formulation and discuss their main stability properties. The constructions considered here are based on modified entropy-conservative fluxes of Tadmor \cite{tadmor2003entropy}, which have been extensively used in finite volume, finite element, and DG discretizations \cite{chen2017entropy,fernandez2014generalized}. We also describe the time integration procedure adopted for the entropy-variable formulation of the compressible Navier--Stokes equations.

\subsection{Numerical fluxes}

We begin by introducing the arithmetic and logarithmic mean operators,
\begin{equation}\label{avg}
\overline{a} =	 \frac{a + \widehat{a}}{2}, \qquad
 {a}^{\ln} = \frac{[\![ a]\!] }{ {\ln}(\widehat{a}/a)} \, .
\end{equation}
In the limiting case $a= \widehat{a}$, the logarithmic mean is evaluated using the numerically stable procedure proposed in \cite{ismail2009affordable}. We also recall the product-jump identity
\begin{equation}\label{eq: jump product}
   [\![ a b]\!] = \overline{a}[\![ b]\!] + \overline{b}[\![ a]\!].
\end{equation}

\subsubsection{Entropy-stable numerical flux}\label{S321}

A first choice for the inviscid flux in \eqref{s32} is the entropy-stable form
 \begin{equation}\label{Flux_En_Inv}
\widehat{\mathbf{F}}_{h}\left( \widehat{\bm{v}}_{h},\bm{v}_{h}\right)= 
\frac{1}{2}\left( \mathbf{F}_{h}\left( \widehat{\bm{v}}_{h}\right) + \mathbf{F}_{h}\left(\bm{v}_{h}\right) \right) \cdot\bm{n} -\frac{1}{2}
 \bm{\sigma}\left( \widehat{\bm{v}}_{h}\right)\mathbf{ A}_{0} \cdot [\![ \bm{v}_{h}]\!] .
 \end{equation}
Here, the matrix $\mathbf{ A}_{0} = {\partial \mathbf{u}}/{\partial  \bm{v}}$ denotes the entropy Jacobian, which is symmetric and positive definite, and provides the mapping between conservative and entropy variables. Although \eqref{Flux_En_Inv} is introduced here in the macro-element HDG setting, the same structure is known to be entropy-stable in the standard HDG framework as well; see, for instance, \cite{williams2018entropy,fernandez2018entropy}.

\begin{remark}
     Proposition \ref{Pro_Time}, together with the numerical inviscid flux \eqref{Flux_En_Inv}, recovers Proposition 1 in \cite{fernandez2019entropy}.
\end{remark}

\subsubsection{Kinetic energy-preserving and entropy-conservative flux}\label{S332}

A considerable body of work has been devoted to the construction of entropy-conservative and entropy-stable fluxes for the Euler equations \cite{chandrashekar2013kinetic,chan2019efficient,hou2007solutions,williams2019analysis}. In the present work, we adapt this class of fluxes to the macro-element HDG framework.

We define the entropy-conservative flux for the macro-element HDG method as
 $\overline{\mathbf{F}} = [\mathbf{f}_{1}^{ec},\  \mathbf{f}_2^{ec},\  \mathbf{f}_3^{ec}]^t$, 
with components
\begin{equation}\label{EC}
\begin{aligned}
\mathbf{f}_{1}^{ec}\left( \widehat{\bm{v}}_{h},\bm{v}_{h}\right) &=  \rho^{\text{ln}}  \ \overline{\mathbf{V}} \cdot \bm{n}, \\[6pt]
\mathbf{f}_{2}^{ec}\left( \widehat{\bm{v}}_{h},\bm{v}_{h}\right) &= \overline{\mathbf{V}} \ \mathbf{f}_{1}^{ec} + \overline{p} \  \bm{n}, \\[6pt]
\mathbf{f}_{3}^{ec}\left( \widehat{\bm{v}}_{h},\bm{v}_{h}\right) &= \left(\frac{1}{\beta ^{\text{ln}} (\gamma - 1)} - \frac{\overline{\mathbf{V} \cdot \mathbf{V}} }{2} \right) \mathbf{f}_{1}^{ec} + \overline{\mathbf{V}} \cdot \mathbf{f}_{2}^{ec}.
\end{aligned}  
\end{equation}
The averaged pressure is defined by \( \overline{P} = {\overline{\rho}}/{\overline{\beta}} \). The flux \( \overline{\mathbf{F}} \) is entropy-conservative for the compressible system \eqref{C4}; a proof is provided in the subsequent subsection.

In addition, the flux \eqref{EC} satisfies the kinetic energy-preserving (KEP) property, which is particularly relevant in under-resolved and turbulent simulations. As shown by Jameson \cite{jameson2008formulation}, the KEP condition for the compressible Euler equations is obtained when the momentum flux \( f_{\rho u} \) takes the form
\begin{equation}\label{KEP}
    f_{\rho u} = p^{ec} + u f_{\rho},
\end{equation}
where \( f_{\rho} \) is the mass flux and \( p^{ec} \) is a consistent average of the pressure. The flux \(\overline{\mathbf{F}}\) satisfies this requirement and therefore preserves kinetic energy by construction \cite{chandrashekar2013kinetic}.

\subsubsection{Kinetic energy-preserving and entropy-stable flux}\label{S333}

A widely used strategy for constructing entropy-stable schemes consists of augmenting an entropy-conservative flux by means of a dissipative correction written in entropy variables \cite{tadmor2003entropy,fjordholm2012arbitrarily,chan2018discretely,winters2017uniquely}. Following this idea, we define a flux \( \mathbf{F}_h \) that combines kinetic energy preservation with entropy stability according to
 \begin{align}\label{KEPES1}
\widehat{\mathbf{F}}_{h}\left( \widehat{\bm{v}}_{h},\bm{v}_{h}\right)= \overline{\mathbf{F}}\left( \widehat{\bm{v}}_{h},\bm{v}_{h}\right) -\frac{1}{2}\mathbf{D}\left( \widehat{\bm{v}}_{h},\bm{v}_{h}\right) \mathbf{ A}_{0} [\![ \bm{v}_{h}]\!] ,
\end{align}
where \( \mathbf{D} \) is a dissipation operator. To enforce entropy stability, the dissipative contribution in \eqref{KEPES1} must be chosen such that the resulting flux satisfies the entropy inequality \eqref{eq: entropy evolution Euler non smooth}.

Different choices of $\mathbf{D}$ give rise to different numerical fluxes. Two classical options are the Lax--Friedrichs (LF) dissipation \cite{badrkhani2023matrix,fleischmann2020low} and Roe-type dissipation \cite{roe1981approximate,stoter2018discontinuous}. The former is more dissipative and typically more robust, whereas the latter is less dissipative and usually more accurate in smooth regions.

Here we combine these two mechanisms through an adaptive hybrid dissipation. The idea is to activate stronger LF-type dissipation near strong discontinuities, while retaining Roe-type dissipation in smooth regions and near contact discontinuities or rarefaction waves. We define
 \begin{align}\label{KEPES3}
\mathbf{D}\left( \widehat{\bm{v}}_{h},\bm{v}_{h}\right) = \overline{\bm{R}} \: \vert  \overline{\bm{ \Lambda}} \vert  \:\overline{\bm{R}}^{-1}.
\end{align}
Here, $\overline{\bm{R}}$ is the matrix of right eigenvectors of the inviscid flux Jacobian for the compressible system. Its averaged form is given by
\begin{equation}
\overline{\bm{R}}\left( \widehat{\bm{v}}_{h},\bm{v}_{h}\right)=
\begin{bmatrix}
1 												&1															&0						&0								&1\\
\overline{\text{v}}_{1} -\overline{c}						&\overline{\text{v}}_{1}												&0						&0								&\overline{\text{v}}_{1} +\overline{c}	\\
\overline{\text{v}}_{2} 								&\overline{\text{v}}_{2}												&1						&0								&\overline{\text{v}}_{2}\\
\overline{\text{v}}_{3} 								&\overline{\text{v}}_{3}												&0						&1								&\overline{\text{v}}_{3}\\
\overline{h}- \overline{\text{v}}_{1}\,\overline{c}			&\frac{1}{2} \overline{{\Vert \text{v}_{tot} \Vert}^2}				&\overline{\text{v}}_{2} 			&\overline{\text{v}}_{3}					&\overline{h} + \overline{\text{v}}_{1}\,\overline{c}	
\end{bmatrix},
\end{equation}
with
\begin{equation}
\overline{c} =	\sqrt \frac{\gamma \overline{P}}{\rho ^{\ln}}, \qquad
\overline{h} = \frac{\gamma}{\beta ^{\ln} (\gamma - 1)} +  \frac{1}{2}  \overline{{\Vert \text{v}_{tot} \Vert}^2}.
\end{equation}
The diagonal matrix of eigenvalues is defined as
\begin{equation}
\vert  \overline{\bm{ \Lambda}} \vert   =	(1-\theta)\,\vert  {\overline{ \bm{ \lambda}}} \vert  + \theta \,\vert  {{ \lambda}_{max}} \vert  \bm{I} , \qquad
\theta = \sqrt{\left| \frac{P - \widehat{P}}{P + \widehat{P}} \right|}
\end{equation}
where the parameter $\theta \in [0,1]$ is determined from a local pressure-based indicator. The matrix $\overline{\bm{\lambda}}$ contains the eigenvalues of the inviscid flux Jacobian,
\begin{equation}
{\overline{  \bm{ \lambda}}\left( \widehat{\bm{v}}_{h},\bm{v}_{h}\right) =  \text{diag}\left(\overline{\text{v}}_n -\overline{c},\overline{\text{v}}_n, \overline{\text{v}}_n,\overline{\text{v}}_n, \overline{\text{v}}_n +\overline{c} \right)},\qquad
\overline{\text{v}}_n= \overline{\text{v}}_1 n_x + \overline{\text{v}}_2 n_y + \overline{\text{v}}_3 n_z .
\end{equation}
To preserve the desired kinetic energy structure, the eigenvalues associated with the mass and energy fluxes must coincide \cite{chandrashekar2013kinetic}. We therefore define the kinetic-energy-preserving and entropy-stable (KEPES) eigenvalues by
\begin{equation}
{\overline{  \bm{ \lambda}}} ={ \bm{ \lambda}^{\textit{KEPES}}} = \text{diag}\left(\overline{\text{v}}_{n} +\overline{c}, \overline{\text{v}}_{n}, \overline{\text{v}}_{n},\overline{\text{v}}_{n}, \overline{\text{v}}_{n} +\overline{c} \right) . 
\end{equation}

To derive an entropy-stable flux with matrix dissipation, one further needs a factorization that connects the entropy Jacobian $\mathbf{A}_0$ with the right eigenvector matrix $\overline{\bm{R}}$. By Barth's eigenvector scaling theorem \cite{barth1999numerical}, there exists a positive diagonal matrix such that
\begin{equation}\label{KEPES4}
\mathbf{A}_0 = \overline{\bm{R}} \:\overline{\bm{T}} \ \overline{\bm{R}}^{t}
\end{equation}
where
\begin{equation}
 \overline{\bm{T}}\left( \widehat{\bm{v}}_{h},\bm{v}_{h}\right) =  \text{diag}\left( \frac{ \rho ^{\ln}}{2 \gamma}, \frac{ \rho ^{\ln} (\gamma - 1) }{\gamma}, \overline{P}, \overline{P}, \frac{ \rho ^{\ln}}{2 \gamma}\right) .
\end{equation}
Using \eqref{KEPES3} together with \eqref{KEPES4}, the dissipation term may be rewritten as
\begin{equation}\label{KEPES5}
\mathbf{D} \mathbf{A}_0  = \overline{\bm{R}} \: \vert  \overline{\bm{ \Lambda}} \vert  \:\overline{\bm{R}}^{-1} ( \overline{\bm{R}} \:\overline{\bm{T}} \ \overline{\bm{R}}^{t}) = \overline{\bm{R}} \: \vert  \overline{\bm{ \Lambda}} \vert  \:\overline{\bm{T}} \ \overline{\bm{R}}^{t} .
\end{equation}
This leads to the final form of the entropy-stable inviscid flux for the macro-element HDG discretization,
 \begin{align}\label{ES}
\widehat{\mathbf{F}}_{h}\left( \widehat{\bm{v}}_{h},\bm{v}_{h}\right)= \overline{\mathbf{F}}\left( \widehat{\bm{v}}_{h},\bm{v}_{h}\right) 
-\frac{1}{2}
  \overline{\bm{R}} \: \vert  \overline{\bm{ \Lambda}} \vert  \:\overline{\bm{T}} \ \overline{\bm{R}}^{t} \ [\![ \bm{v}_{h}]\!] .
\end{align}

\begin{remark}
The inviscid numerical flux defined in \eqref{ES} is conservative. Local conservation follows from the explicit evaluation of the fluxes, while global conservation is enforced through the interface condition \eqref{s32c}.
\end{remark}

\subsection{Stability properties}

We now analyze the properties of the numerical fluxes introduced above. The theoretical results derived in this subsection explain the entropy behavior of the proposed formulation and are corroborated later by the numerical experiments.

\begin{theorem}\label{flux_EC_1}
The macro-element entropy-variable HDG discretization of the compressible Euler equations on a periodic domain with the inviscid numerical flux formulation \eqref{EC} is entropy-conservative, that is, the total generalized entropy is zero over time.
\end{theorem}

\begin{proof}
The flux $\overline{\mathbf{F}}$ is symmetric and consistent because these properties are inherited from the arithmetic and logarithmic mean operators. To verify that \eqref{EC} satisfies the entropy-conservation condition \eqref{EC_1}, we first note that
  \begin{equation}\label{po2}
[\![ \bm{\Psi} \cdot \bm{n}]\!] \overset{\ref{poen}}{=} [\![ \rho \mathbf{V}\cdot \bm{n}]\!] \overset{\ref{eq: jump product}}{=} 
\overline{\rho}[\![ \mathbf{V}]\!]\cdot \bm{n} + [\![ \rho]\!] \overline{\mathbf{V}} \cdot \bm{n} \  .
\end{equation}
Introducing the component notation of \(\overline{\mathbf{F}}\) from \eqref{EC}, we obtain
\begin{equation}\label{ECfluxtot1}
\begin{aligned}
   [\![ \bm{v}_{h}]\!] ^t \cdot \overline{\mathbf{F}}\left( \widehat{\bm{v}}_{h},\bm{v}_{h}\right) &\overset{\ref{s222}}{=} 
    [\![\frac{\gamma - s}{\gamma -1} - \beta \frac{\mathbf{V} \cdot \mathbf{V}}{2}]\!] \mathbf{f}_{1 }^{ec}
     + [\![ \beta \mathbf{V} ]\!] \cdot \mathbf{f}_{2 }^{ec}- [\![ \beta ]\!] \mathbf{f}_{3 }^{ec}\\
    &=\left(\frac{1}{\gamma-1} [\![\text{ln} \beta]\!] + [\![\text{ln} \rho]\!]  \right) \mathbf{f}_{1}^{ec} +\left(\overline{\beta}[\![\mathbf{V} ]\!] + \overline{\mathbf{V}}[\![\beta ]\!]\right) \cdot \mathbf{f}_{2 }^{ec}- [\![ \beta ]\!] \mathbf{f}_{3 }^{ec}.
    \end{aligned}
\end{equation}
Combining \eqref{po2} and \eqref{ECfluxtot1}, the quantity \( [\![ \bm{v}_{h}]\!] ^t  \cdot \overline{\mathbf{F}}_{h} -[\![ \bm{\Psi}_{n}]\!] \) simplifies to
\begin{equation}\label{ECfluxtot2}
\begin{aligned}
   [\![ \bm{v}_{h}]\!] ^t  \cdot \overline{\mathbf{F}}_{h} -[\![ \bm{\Psi}_{n}]\!]&=[\![\text{ln} \rho]\!] \left(\mathbf{f}_{1}^{ec} - \ \rho^{\text{ln}} \ \overline{\mathbf{V}} \cdot \bm{n} \right)  
   + \overline{\beta}[\![\mathbf{V} ]\!] \cdot \left( \mathbf{f}_{2 }^{ec} - \overline{\mathbf{V}} \cdot \mathbf{f}_{1 }^{ec} - \frac{\overline{\rho}}{\overline{\beta}} \bm{n}  \right)\\
   &+ [\![\text{ln} \beta]\!] \left(\frac{1}{\gamma -1}\mathbf{f}_{1}^{ec} - \beta^{\text{ln}} 
   \left(\mathbf{f}_{3 }^{es} - \overline{\mathbf{V}} \cdot \mathbf{f}_{2}^{es} + \frac{\overline{\mathbf{V} \cdot \mathbf{V}}}{2}\mathbf{f}_{1 }^{es}\right) \right)
   \overset{\ref{EC}}{=}0.
   \end{aligned}
\end{equation}
The conclusion follows directly from \eqref{ECfluxtot2}.
\end{proof}

We next show that the flux \eqref{ES} satisfies the entropy-stability condition.

\begin{theorem}\label{flux_ES_1}
The entropy-variable macro-element HDG discretization of the compressible Euler equations on a periodic domain with the inviscid numerical flux formulation \eqref{ES} is entropy-stable, that is, the total generalized entropy is non-increasing over time.
\end{theorem}

\begin{proof}
Starting from the definition of $\widehat{\mathbf{F}}_{h}$ in \eqref{ES}, invoking the entropy-stability condition \eqref{ES_1}, and using Theorem \ref{flux_EC_1}, which establishes the entropy-conservative character of \(\overline{\mathbf{F}}_{h}\), we obtain
    \begin{equation}\label{ESfluxtot1}
\begin{aligned}
   [\![ \bm{v}_{h}]\!] ^t  \cdot \widehat{\mathbf{F}}_{h} -[\![ \bm{\Psi}_{n}]\!]&\overset{\ref{ES}}{=}
   [\![ \bm{v}_{h}]\!] ^t \cdot \overline{\mathbf{F}}_{h} -[\![ \bm{\Psi}_{n}]\!] - \frac{1}{2} [\![ \bm{v}_{h}]\!] ^t \
  \overline{\bm{R}} \: \vert  \overline{\bm{ \Lambda}} \vert  \:\overline{\bm{T}} \ \overline{\bm{R}}^{t} [\![ \bm{v}_{h}]\!]  \\
&\overset{\ref{ECfluxtot2}}{=}- \frac{1}{2}[\![ \bm{v}_{h}]\!] ^t \ 
  \overline{\bm{R}} \: \vert  \overline{\bm{ \Lambda}} \vert  \:\overline{\bm{T}}  \ \overline{\bm{R}}^{t} [\![ \bm{v}_{h}]\!]  \leq 0.
   \end{aligned}
\end{equation}
The term on the right-hand side is a quadratic form multiplied by a negative coefficient, which implies that the discrete entropy inequality holds.
\end{proof}

\begin{remark}
In \eqref{ESfluxtot1}, \( \overline{\bm{T}} \) and \( |\overline{\bm{\Lambda}}| \) are positive diagonal matrices \cite{winters2017uniquely,barth1999numerical}, while \( \overline{\bm{R}} \) is positive definite \cite{roe1981approximate}.
\end{remark}

Theorems \ref{flux_EC_1} and \ref{flux_ES_1} show that the proposed scheme is entropy-conservative or entropy-stable, respectively, for any polynomial degree \( p \geq 0 \) and for arbitrary macro-element partitions \( \mathcal{T}_h \). In the present work, we restrict the analysis to periodic boundary conditions. The construction of general entropy-conservative boundary conditions is left for future investigation.

\subsection{Temporal discretization}\label{S33}

For the temporal discretization, we employ an $s$-stage diagonally implicit Runge--Kutta (DIRK) method \cite{alexander1977diagonally}. Such schemes are widely used for stiff systems due to their high-order accuracy and favorable stability properties. Additional details on high-order time integration in the context of HDG methods can be found in \cite{nguyen2012hybridizable,nguyen2011high}.
The DIRK method is characterized by the Butcher coefficients
\begin{equation}
\begin{array}{c|c}
\bm{c} & A \\
\hline 
        & \bm{b}
\end{array},
\end{equation}
where $A = (a_{ij}) \in \mathbb{R}^{s \times s}$ is a lower triangular matrix with nonzero diagonal entries, and $\bm{b} = (b_1,\dots,b_s)$ and $\bm{c} = (c_1,\dots,c_s)^T$ denote the weights and nodes, respectively. We further define $D = A^{-1}$ with entries $d_{ij}$.

Let $\text{n}$ denote the time level. For each time step, we introduce the stage solutions
\[
\left( \mathbf{q}_h^{\text{n},i}, \bm{v}_h^{\text{n},i}, \widehat{\bm{v}}_h^{\text{n},i} \right), 
\qquad i = 1,\dots,s,
\]
which approximate the solution at the intermediate times $t^{\text{n},i} = t^\text{n} + c_i \Delta t^\text{n}$. Using the transformation between entropy and conservative variables, we obtain the corresponding conservative states $\left( \mathbf{u}_h^{n,i}, \widehat{\mathbf{u}}_h^{n,i} \right)$.
The solution at the next time level is computed as
{\color{black}\begin{equation}\label{C241}
 \mathbf{u}_{h}^{\text{n}+1}=\left(1-\sum_{i=1}^{s} e_j \right)  \mathbf{u}_{h}^{\text{n}} + \sum_{j=1}^{s} \mathbf{u}_{h}^{\text{n},j},
\end{equation}}
where the coefficients $e_j$ are defined by$
e_j = \sum_{i=1}^s b_i d_{ij}.
$

The intermediate stage solutions are obtained by solving, for each stage $i=1,\dots,s$, the HDG system: find
$
 \left( \mathbf{q}_{h}^{\text{n},i},\bm{v}_{h}^{\text{n},i},\widehat{\bm{v}}_{h}^{\text{n},i}\right)\in \bm{\mathcal{Q}}^{k}_{h} \times \bm{\mathcal{V}}^{k}_{h} \times \bm{\mathcal{M}}^{k}_{h} 
$
such that
\begin{equation}\label{C242}
 \begin{aligned}
&\left( \mathbf{q}_{h}^{\text{n},i},\bm{r}\right) _{\mathcal{T}_{h}} + 
\left( \bm{v}_{h}^{\text{n},i},\nabla\cdot\bm{r}\right) _{\mathcal{T}_{h}} - 
\left\langle   \widehat{\bm{v}}_{h}^{\text{n},i} ,\bm{r} \cdot \bm{n}\right\rangle  _{\partial \mathcal{T}_{h}}= 0,\\[6pt]
&\left( \dfrac{\sum_{j=1}^{s}d_{ij}\left(\mathbf{u}_{h}^{\text{n},j}-\mathbf{u}_{h}^{\text{n}} \right) }{\Delta t^{\text{n}}},\bm{w}\right)_{\mathcal{T}_{h}} -
\left( \mathbf{F}_{h}^{\text{n},i}  + \mathbf{G}_{h}^{\text{n},i} ,\nabla \cdot \bm{w}\right) _{\mathcal{T}_{h}} +
\left\langle  \widehat{\mathbf{F}}_{h}^{\text{n},i}+\widehat{\mathbf{G}}_{h}^{\text{n},i},\bm{w}\right\rangle _{\partial \mathcal{T}_{h}} = 0,\\[6pt]
&\left\langle \widehat{\mathbf{F}}_{h}^{\text{n},i}+\widehat{\mathbf{G}}_{h}^{\text{n},i},\bm{\mu}\right\rangle  _{{\partial \mathcal{T}_{h}}\backslash\partial\Omega} 
 + \left\langle   \widehat{\mathbf{B}}_{h}^{\text{n},i}\left( \widehat{\bm{v}}_{h}^{\text{n},i},\bm{v}_{h}^{\text{n},i},\mathbf{q}_{h}^{\text{n},i}\right) ,\bm{\mu} \right\rangle  _{\partial\Omega}= 0,
\end{aligned}   
\end{equation}
for all $ \left( \bm{r},\bm{w},\bm{\mu}\right) \in \bm{\mathcal{Q}}^{k}_{h} \times \bm{\mathcal{V}}^{k}_{h} \times \bm{\mathcal{M}}^{k}_{h}$. 
\begin{remark}
The time discretization of the macro-element HDG method in conservative-variable form differs from that in \eqref{C242}. In this formulation, the intermediate solutions $\left(\mathbf{u}_{h}^{\text{n},i}, \widehat{\mathbf{u}}_{h}^{\text{n},i}\right)$ are computed directly, without requiring a transformation from entropy variables. For further details, see \cite{badrkhani2025matrix}.
\end{remark}
\section{Parallel solution strategy and implementation details }\label{Sec6}

This section describes the computational framework employed to solve the nonlinear system arising from the entropy-variable formulation \eqref{C242}. The overall solution procedure consists of two nested stages. First, the nonlinear system is treated using a Newton-type method with relaxed accuracy requirements. Second, each linearized system is reduced through a static condensation technique that exploits the hybrid structure of the discretization.

A key aspect of the implementation is the use of matrix-free operator evaluations for high-order DG discretizations. This approach avoids the explicit assembly of global matrices and instead computes operator actions on-the-fly. The methodology follows modern matrix-free strategies developed in high-performance finite element frameworks such as deal.II \cite{bangerth2015dealii}, ExaDG \cite{arndt2020exadg}, and Exasim \cite{vila2022exasim}, while extending our previous developments \cite{badrkhani2023matrix,badrkhani2025matrix} to the entropy-variable macro-element HDG formulation.

\subsection{Inexact Newton iteration in entropy-variable form}\label{S51}
At each (sub-)time level \( \text{n} \), the discrete problem defined by \eqref{C242} leads to a nonlinear system for the unknown vector
$
\mathfrak{V}^{\text{n}} =
\left[
\mathbf{q}_{h}^{\text{n}},
\bm{v}_{h}^{\text{n}},
\widehat{\bm{v}}_{h}^{\text{n}}
\right]^T.
$
The system is written in compact residual form as
\begin{equation}\label{C31221}
\mathcal{R}\big(\mathfrak{V}^{\text{n}}\big) = \mathbf{0},
\end{equation}
where the global residual vector is defined as
$
\mathcal{R}(\mathfrak{V}) =
\left[
R_{Q},
R_{V},
R_{\widehat{V}}
\right]^T.
$

To solve \eqref{C31221}, an inexact Newton method combined with a pseudo-transient continuation strategy is employed
\cite{bijl2002implicit,jameson1991time,vakilipour2019developing}. The nonlinear iterations are stabilized by introducing a pseudo-time stepping procedure with adaptive step size control. In particular, the pseudo-time step is dynamically adjusted using a residual-based relaxation strategy \cite{mulder1985experiments}, allowing small steps when the solution is far from convergence and progressively larger steps as the residual decreases. Unless otherwise specified, the initial pseudo-time step is set to \( \tau_{\text{init}} = 1.0 \), while a sufficiently large upper bound  \( \tau_{max} = 10^{8} \) is imposed.

Starting from an initial guess \( \mathfrak{V}^{0,\text{n}} \), the Newton method generates a sequence \( \mathfrak{V}^{m,\text{n}}  \). At each iteration, the residual is linearized about the current state, leading to the incremental problem
\begin{equation}\label{C312}
\mathbf{J}^{m,\text{n}} \, \Delta \mathfrak{V}^{m,\text{n}}
=
- \mathcal{R}^{m,\text{n}},
\end{equation}
where \( \mathbf{J}^{m,\text{n}} \) denotes the Jacobian matrix of \( \mathcal{R}\big(\mathfrak{V}^{\text{m,n}}\big)\), and
\[
\Delta \mathfrak{V}^{m,\text{n}} =
\begin{bmatrix}
\Delta Q^{m,\text{n}} \\[6pt]
\Delta V^{m,\text{n}} \\[6pt]
\Delta \widehat{V}^{m,\text{n}}
\end{bmatrix}, 
\qquad
\mathcal{R}^{m,\text{n}} =
\begin{bmatrix}
R_{Q}^{m,\text{n}} \\[6pt]
R_{V}^{m,\text{n}} \\[6pt]
R_{\widehat{V}}^{m,\text{n}}
\end{bmatrix}.
\]

The solution is updated according to
\begin{equation}\label{C31223}
\mathfrak{V}^{m+1,\text{n}} = \mathfrak{V}^{m,\text{n}} + \Delta \mathfrak{V}^{m,\text{n}}.
\end{equation}

The iterations are repeated until a prescribed nonlinear convergence criterion is satisfied. The resulting linear systems are solved inexactly using a Krylov subspace method, with a tolerance chosen in relation to the nonlinear residual norm.

\begin{remark}
In the entropy-variable formulation, the Newton iterations are performed in entropy variables, while the conservative variables are recovered through a transformation at each iteration. This differs from classical formulations where conservative variables are directly updated \cite{badrkhani2025matrix}.
\end{remark}

\subsection{Second-layer static condensation in entropy-variable form}\label{S52}

The algebraic structure of the linearized system \eqref{C312} admits a natural separation between interior and interface degrees of freedom. Exploiting this property, the macro-element unknowns can be partitioned into interior variables \( (Q,V) \) and trace variables \( \widehat{V} \). Since only the latter are globally coupled in the HDG framework, we eliminate the interior variables locally and construct a reduced system posed exclusively on the interfaces.

To this end, we rewrite the macro-element problem on \( \mathcal{T}_i \) in block form as
\begin{equation}\label{R1}
\begin{bmatrix}
\mathbf{A}_{qq} & \mathbf{A}_{qv} \\
\mathbf{A}_{vq} & \mathbf{A}_{vv}
\end{bmatrix}^{\mathcal{T}_i}
\begin{bmatrix}
\Delta Q \\
\Delta V
\end{bmatrix}^{\mathcal{T}_i}
=
-
\begin{bmatrix}
R_Q \\
R_V
\end{bmatrix}^{\mathcal{T}_i}
-
\begin{bmatrix}
\mathbf{A}_{q\widehat{v}} \\
\mathbf{A}_{v\widehat{v}}
\end{bmatrix}^{\mathcal{T}_i}
\Delta \widehat{V}^{\Gamma_i}.
\end{equation}

The block \( \mathbf{A}_{qq}^{\mathcal{T}_i} \) is diagonal with respect to the flux components, which allows for an efficient inversion at the element level \cite{badrkhani2025matrix}. Eliminating \( \Delta Q^{\mathcal{T}_i} \) from \eqref{R1} yields a reduced relation for \( \Delta V^{\mathcal{T}_i} \),
\begin{equation}\label{R2}
\mathbf{S}^{\mathcal{T}_i} \, \Delta V^{\mathcal{T}_i}
=
- R_V^{\mathcal{T}_i}
+ \mathbf{A}_{vq}^{\mathcal{T}_i} \left(\mathbf{A}_{qq}^{\mathcal{T}_i}\right)^{-1} R_Q^{\mathcal{T}_i}
-
\Big(
\mathbf{A}_{v\widehat{v}}^{\mathcal{T}_i}
-
\mathbf{A}_{vq}^{\mathcal{T}_i}
\left(\mathbf{A}_{qq}^{\mathcal{T}_i}\right)^{-1}
\mathbf{A}_{q\widehat{v}}^{\mathcal{T}_i}
\Big)
\Delta \widehat{V}^{\Gamma_i},
\end{equation}
where the macro-element Schur complement is defined by
\begin{equation}\label{R3}
\mathbf{S}^{\mathcal{T}_i}
=
\mathbf{A}_{vv}^{\mathcal{T}_i}
-
\mathbf{A}_{vq}^{\mathcal{T}_i}
\left(\mathbf{A}_{qq}^{\mathcal{T}_i}\right)^{-1}
\mathbf{A}_{qv}^{\mathcal{T}_i}.
\end{equation}

Substituting \eqref{R2} into the interface equation leads to a condensed problem involving only the trace correction. Summing the macro-element contributions results in the global system
\begin{equation}\label{R4}
\widehat{\mathbf{A}} \, \Delta \widehat{V}
=
\widehat{\mathbf{b}},
\end{equation}
which is assembled from local operators of the form
\begin{subequations}\label{R5}
\begin{align}\label{R5a}
\widehat{\mathbf{A}}^{\mathcal{T}_i}
&=
\mathbf{A}_{\widehat{v}\widehat{v}}^{\mathcal{T}_i}
-
\mathbf{A}_{\widehat{v}q}^{\mathcal{T}_i}
\left(\mathbf{A}_{qq}^{\mathcal{T}_i}\right)^{-1}
\mathbf{A}_{q\widehat{v}}^{\mathcal{T}_i}
\nonumber\\
&\quad
+
\left(
\mathbf{A}_{\widehat{v}v}^{\mathcal{T}_i}
-
\mathbf{A}_{\widehat{v}q}^{\mathcal{T}_i}
\left(\mathbf{A}_{qq}^{\mathcal{T}_i}\right)^{-1}
\mathbf{A}_{qv}^{\mathcal{T}_i}
\right)
\left(\mathbf{S}^{\mathcal{T}_i}\right)^{-1}
\left(
-\mathbf{A}_{v\widehat{v}}^{\mathcal{T}_i}
+
\mathbf{A}_{vq}^{\mathcal{T}_i}
\left(\mathbf{A}_{qq}^{\mathcal{T}_i}\right)^{-1}
\mathbf{A}_{q\widehat{v}}^{\mathcal{T}_i}
\right), \\
\widehat{\mathbf{b}}^{\mathcal{T}_i}
&=
-
R_{\widehat{V}}^{\mathcal{T}_i}
+
\mathbf{A}_{\widehat{v}q}^{\mathcal{T}_i}
\left(\mathbf{A}_{qq}^{\mathcal{T}_i}\right)^{-1}
R_Q^{\mathcal{T}_i}
\nonumber\\
&\quad
+
\left(
-\mathbf{A}_{\widehat{v}v}^{\mathcal{T}_i}
+
\mathbf{A}_{\widehat{v}q}^{\mathcal{T}_i}
\left(\mathbf{A}_{qq}^{\mathcal{T}_i}\right)^{-1}
\mathbf{A}_{qv}^{\mathcal{T}_i}
\right)
\left(\mathbf{S}^{\mathcal{T}_i}\right)^{-1}
\left(
- R_V^{\mathcal{T}_i}
+
\mathbf{A}_{vq}^{\mathcal{T}_i}
\left(\mathbf{A}_{qq}^{\mathcal{T}_i}\right)^{-1}
R_Q^{\mathcal{T}_i}
\right).
\end{align}
\end{subequations}

After solving \eqref{R4}, the interior corrections are reconstructed locally. First, \( \Delta V^{\mathcal{T}_i} \) is obtained from \eqref{R2}, followed by the recovery of \( \Delta Q^{\mathcal{T}_i} \),
\begin{equation}\label{R6}
\Delta Q^{\mathcal{T}_i}
=
\left(\mathbf{A}_{qq}^{\mathcal{T}_i}\right)^{-1}
\left(
- R_Q^{\mathcal{T}_i}
- \mathbf{A}_{qv}^{\mathcal{T}_i} \Delta V^{\mathcal{T}_i}
- \mathbf{A}_{q\widehat{v}}^{\mathcal{T}_i} \Delta \widehat{V}^{\Gamma_i}
\right).
\end{equation}

This formulation highlights that the global coupling is entirely confined to the trace space. Consequently, the dimension of the linear system is significantly reduced compared to classical DG discretizations with element-wise discontinuous variables \cite{cockburn2009unified, badrkhani2025matrix, nguyen2009implicit}. Moreover, all operations involving interior variables are performed locally at the macro-element level, which makes the method well suited for parallel implementations.
The inverse \( \left(\mathbf{S}^{\mathcal{T}_i}\right)^{-1} \) is not formed explicitly; instead, a factorization of \( \left(\mathbf{S}^{\mathcal{T}_i}\right) \) is reused, and its action is obtained by solving \( \mathbf{S}^{\mathcal{T}_i} \mathbf{x} = \mathbf{a} \) through back-substitution.

\subsection{Matrix-free algorithmic framework }\label{S53}

The reduced interface problem and the accompanying local solves are implemented in a parallel matrix-free fashion. The entropy-variable version is summarized in \textbf{Algorithm 1}; the conservative-variable counterpart follows the same pattern with only minor modifications. The purpose of the matrix-free strategy is to avoid explicit assembly of the global reduced operator while still preserving the block structure required by the HDG formulation.

The first stage of the algorithm, namely \textbf{Lines 2--6}, evaluates the macro-element contributions to the reduced right-hand side and, conceptually, to the reduced interface operator. These operations are performed independently over the set of macro-elements assigned to the current process \( n \), denoted by \( \mathcal{T}_{h}^{n} \). At this stage, no communication between processes is required, because the computations depend only on local macro-element data.

The global interface correction is computed in \textbf{Line 7} with a Krylov solver, typically FGMRES, applied to the reduced system \eqref{R4}. Since the number of interface unknowns is still large in realistic three-dimensional simulations, assembling and storing the reduced matrix explicitly would be unnecessarily expensive. Instead, the iterative solve accesses the operator only through matrix-vector products. Once the interface update \( \Delta \widehat{V} \) has been obtained, the macro-element-local corrections \( \Delta Q^{\mathcal{T}_i} \) and \( \Delta V^{\mathcal{T}_i} \) are recovered in \textbf{Lines 8--16}.

The matrix-vector product required in the global solve is carried out in \textbf{Lines 10--14}, following equation~\eqref{R5}. Here, \( \Delta \widehat{V} \) holds the degrees of freedom on the boundaries of macro-elements, denoted as \( \varepsilon \). Each interior interface \( e \in \varepsilon \) sees contributions from both adjacent macro-elements \( \mathcal{T}_+ \) and \( \mathcal{T}_- \), referring to the elements on either side of the interface.

For a given macro-element \( \mathcal{T}_i \), the reduced matrix-vector operation can be decomposed into four stages:
\begin{align*} 
	&1. \ 
	\mathbf{y}_{1}=\left(-\mathbf{A}_{\bm{v}\widehat{\bm{v}}}^{\mathcal{T}_{i}}\: \Delta \widehat{V}^{\Gamma_{i}}+\mathbf{A}_{\bm{v}q}^{\mathcal{T}_{i}}(\mathbf{A}_{qq}^{\mathcal{T}_{i}})^{-1}\mathbf{A}_{q\widehat{\bm{v}}}^{\mathcal{T}_{i}}\: \Delta \widehat{V}^{\Gamma_{i}}\right)
	&&\text{Global elimination }\\
	&2. \ 
	\mathbf{y}_{2}= (\mathbf{S}^{\mathcal{T}_{i}})^{-1} \mathbf{y}_{1}
	&&\text{Local Schur solve}\\
	&3. \
	\mathbf{y}_{3}= \left(\mathbf{A}_{\widehat{\bm{v}}\bm{v}}^{\mathcal{T}_{i}} - \mathbf{A}_{\widehat{\bm{v}}q}^{\mathcal{T}_{i}} (\mathbf{A}_{qq}^{\mathcal{T}_{i}})^{-1} \mathbf{A}_{q\bm{v}}^{\mathcal{T}_{i}}\right) \mathbf{y}_{2}
	&&\text{Global interaction}\\
	&4.\
	\mathbf{y}_{4}= \mathbf{y}_{3} + \left(\mathbf{A}_{\widehat{\bm{v}}\widehat{\bm{v}}}^{\mathcal{T}_{i}}\: \Delta \widehat{V}^{\Gamma_{i}} - \mathbf{A}_{\widehat{\bm{v}}q}^{\mathcal{T}_{i}}(\mathbf{A}_{qq}^{\mathcal{T}_{i}})^{-1}\mathbf{A}_{q\widehat{\bm{v}}}^{\mathcal{T}_{i}}\: \Delta \widehat{V}^{\Gamma_{i}}\right)
	&& \text{Global assembly}	
\end{align*}

These four operations separate naturally into local and interface-related tasks. Step 2 is entirely local to the macro-element and therefore communication-free. By contrast, Steps 1, 3, and 4 contribute to the reduced global operator, and their results must be accumulated consistently across the shared interface degrees of freedom. In particular, the last step requires communication because several macro-elements may contribute to the same global trace entries.
\begin{algorithm}
\caption{Matrix-free macro-element hybridized DG method with iterative solver}
\begin{algorithmic}[1]

\State $n \gets$ Current process

\Comment{\textbf{--- Computing Local Contributions \eqref{R5} ---}}

\State Initialize global residual vector $\widehat{\mathbf{b}} \gets 0$
\ForAll{$\mathcal{T}_{i} \in \mathcal{T}_{h}^{n}$}
    \State $\widehat{\mathbf{b}}^{\mathcal{T}_{i}},\widehat{\mathbf{A}}^{\mathcal{T}_{i}} \gets$ Local contribution from ${Q}^{\mathcal{T}_{i}}$, ${V}^{\mathcal{T}_{i}}$, $\widehat{{V}}^{\Gamma_{i}}$
\State $\widehat{\mathbf{b}}, \ \widehat{\mathbf{A}}  \gets$ 
Assemble $\widehat{\mathbf{b}}^{\mathcal{T}_{i}}$ and $\widehat{\mathbf{A}}^{\mathcal{T}_{i}}$ for all $\mathcal{T}_{i} \in \mathcal{T}_{h}^{n}$
\EndFor

\Comment{\textbf{--- Solving Global System \eqref{R4} ---}}
\State \textbf{Solve} $\widehat{\mathbf{A}}\, \Delta \widehat{V} = \widehat{\mathbf{b}}$ via matrix-free iterative method:
\While{not converged}
    \State $\mathbf{y} \gets 0$
    \ForAll{$\mathcal{T}_{i} \in \mathcal{T}_{h}^{n}$}
        \State $\varepsilon \gets$ Global DOFs on $\partial \mathcal{T}_{i}$
        \State $\Delta \widehat{V}^{\Gamma_{i}} \gets \Delta \widehat{V}[\varepsilon]$
        \State Compute local matrix-vector contribution:
        \[
        \begin{aligned}
        \mathbf{y}[\varepsilon] \mathrel{+}= 
        &\left(\mathbf{A}_{\widehat{\bm{v}}\bm{v}}^{\mathcal{T}_{i}} - \mathbf{A}_{\widehat{\bm{v}}q}^{\mathcal{T}_{i}} (\mathbf{A}_{qq}^{\mathcal{T}_{i}})^{-1} \mathbf{A}_{q\bm{v}}^{\mathcal{T}_{i}} \right)
        (\mathbf{S}^{\mathcal{T}_{i}})^{-1}
        \left( -\mathbf{A}_{\bm{v}\widehat{\bm{v}}}^{\mathcal{T}_{i}} + \mathbf{A}_{\bm{v}q}^{\mathcal{T}_{i}}(\mathbf{A}_{qq}^{\mathcal{T}_{i}})^{-1} \mathbf{A}_{q\widehat{\bm{v}}}^{\mathcal{T}_{i}} \right) \Delta \widehat{V}^{\Gamma_{i}} \\
        &+ \left( \mathbf{A}_{\widehat{\bm{v}}\widehat{\bm{v}}}^{\mathcal{T}_{i}} - \mathbf{A}_{\widehat{\bm{v}}q}^{\mathcal{T}_{i}}(\mathbf{A}_{qq}^{\mathcal{T}_{i}})^{-1} \mathbf{A}_{q\widehat{\bm{v}}}^{\mathcal{T}_{i}} \right) \Delta \widehat{V}^{\Gamma_{i}}
        \end{aligned}
        \]
    \EndFor
    \State Update $\Delta \widehat{V}$ using iterative solver step with $\mathbf{y}$
\EndWhile

\Comment{\textbf{--- Solving Local Systems \eqref{R6} ---}}
\ForAll{$\mathcal{T}_{i} \in \mathcal{T}_{h}^{n}$}
    \State $\Delta Q^{\mathcal{T}_{i}} \gets (\mathbf{A}_{qq}^{\mathcal{T}_{i}})^{-1} \left( -R_{Q}^{\mathcal{T}_{i}} - \mathbf{A}_{q\bm{v}}^{\mathcal{T}_{i}} \Delta V^{\mathcal{T}_{i}} - \mathbf{A}_{q\widehat{\bm{v}}}^{\mathcal{T}_{i}} \Delta \widehat{V}^{\Gamma_{i}} \right)$
    \State $\Delta V^{\mathcal{T}_{i}} \gets (\mathbf{S}^{\mathcal{T}_{i}})^{-1} \left[ 
        \mathbf{A}_{uq}^{\mathcal{T}_{i}} (\mathbf{A}_{qq}^{\mathcal{T}_{i}})^{-1} R_Q^{\mathcal{T}_{i}} - R_V^{\mathcal{T}_{i}} 
        - \left( \mathbf{A}_{\bm{v}\widehat{\bm{v}}}^{\mathcal{T}_{i}} - \mathbf{A}_{\bm{v}q}^{\mathcal{T}_{i}}(\mathbf{A}_{qq}^{\mathcal{T}_{i}})^{-1} \mathbf{A}_{q\widehat{\bm{v}}}^{\mathcal{T}_{i}} \right) \Delta \widehat{V}^{\Gamma_{i}} 
    \right]$
\EndFor

\end{algorithmic}
\end{algorithm}

Algorithm 1 highlights the central advantage of the matrix-free macro-element strategy: only a limited set of objects must be stored explicitly, namely the mappings from reference to physical elements, the local factorizations associated with the Schur complements \( \mathbf{S}^{\mathcal{T}_i} \), and the reduced right-hand side vector \( \widehat{\mathbf{b}} \) \cite{badrkhani2023matrix,badrkhani2025matrix}. The expensive reduced operator itself is never assembled, but accessed only through its action on vectors. This significantly reduces memory consumption and improves scalability in large three-dimensional simulations.

\subsection{Computational configuration}\label{S55}

For the global interface problem, we employ a flexible GMRES (FGMRES) solver \cite{saad1993flexible} tailored to the present discretization. In contrast to standard GMRES \cite{saad1986gmres}, FGMRES permits variable preconditioning and provides greater flexibility in memory management during the iterative process. In matrix-free flow solvers, multilevel preconditioning has repeatedly been shown to improve convergence for both compressible and incompressible problems \cite{diosady2009preconditioning, fidkowski2005p,badrkhani2017development}. Following this line of work, we use FGMRES with GMRES as an inner preconditioner. In addition, the inverse of the global block \( (\hat{A}_{\hat{\bm{v}}\hat{\bm{v}}})^{-1} \) is employed as an auxiliary preconditioning component.

This inverse is never formed explicitly. Instead, we exploit the fact that the local blocks \( \mathbf{A}_{\widehat{\bm{v}}\widehat{\bm{v}}}^{\mathcal{T}_{i}} \) are symmetric positive definite and store their Cholesky factorizations for later use in the preconditioner application. For a discretization containing \( N_{\text{Mcr}} \) macro-elements, the preconditioner acts separately on the \( N_{\text{Mcr}} \) local blocks associated with the corresponding macro-elements. Thus, although the Krylov iteration is global, the action of this preconditioner remains localized at the macro-element level. Additional details may be found in Section 3.3.2 of \cite{badrkhani2023matrix}.

The implementation is written in Julia\footnote{The Julia Programming Language, \href{https://julialang.org/}{https://julialang.org/}}, with custom element-level routines handling the local algebra. These tasks include the assembly of local matrices, the computation of LU factorizations of the local Schur complements \( \mathbf{S}^{\mathcal{T}_i} \), and the extraction of local data from the global interface vector. The global linear systems are solved using FGMRES and GMRES as provided by PETSc\footnote{Portable, Extensible Toolkit for Scientific Computation, \href{https://petsc.org/}{https://petsc.org/}}. In the standard HDG setting, where matrix blocks remain relatively small, dense LAPACK routines\footnote{Linear Algebra PACKage, \href{https://netlib.org/lapack/}{https://netlib.org/lapack/}} are used to accelerate local linear algebra. In the macro-element setting, however, the larger block sizes make sparse direct solvers more suitable; we therefore use UMFPACK \cite{davis2004algorithm} to handle these systems efficiently.

All computations were performed on the Lichtenberg II cluster at the High-Performance Computing Center of the Technical University of Darmstadt. The code was compiled with GCC version 9.2.0 and linked against Portable Hardware Locality version 2.7.1 and OpenMPI version 4.1.2. Each compute node of the cluster contains two Intel Xeon Platinum 9242 processors, yielding 48 cores per node at a base clock frequency of 2.3~GHz, together with 384~GB of main memory\footnote{High Performance Computing Center at TU Darmstadt, \href{https://www.hrz.tu-darmstadt.de}{https://www.hrz.tu-darmstadt.de}}.
\section{Numerical experiments} \label{Sec7}

In this section, we evaluate the performance of the proposed macro-element HDG method through a set of representative benchmark problems. The focus is placed on three main aspects: accuracy of the numerical solution, robustness of the formulation under challenging flow conditions, and computational efficiency in a parallel environment.

The test cases include inviscid and viscous flow configurations, covering both laminar and turbulent regimes. In addition, comparisons are carried out between the standard HDG formulation and the macro-element variant, as well as between conservative-variable and entropy-variable approaches.

Time integration of the semi-discrete system \eqref{C242} is performed using a third-order diagonally implicit Runge–Kutta scheme (DIRK(3,3)) \cite{alexander1977diagonally}. The different numerical fluxes considered in this work are denoted as follows: the conservative-variable formulation is referred to as \textit{Con}, while entropy-variable formulations are distinguished by the choice of flux, namely the entropy-stable (ES) and kinetic-energy-preserving entropy-stable (KEPES) variants.

The parameter $m$ denotes the number of continuous elements per macro-element edge, with $m=1$ corresponding to the standard HDG discretization.

\begin{figure}[h!]
    \centering
    \subfloat[Geometry and density field (\(\epsilon = 2.5\), \(M_\infty = 0.50\)).\label{fig:Test_Vortexa}]{\includegraphics[width=0.6\textwidth]{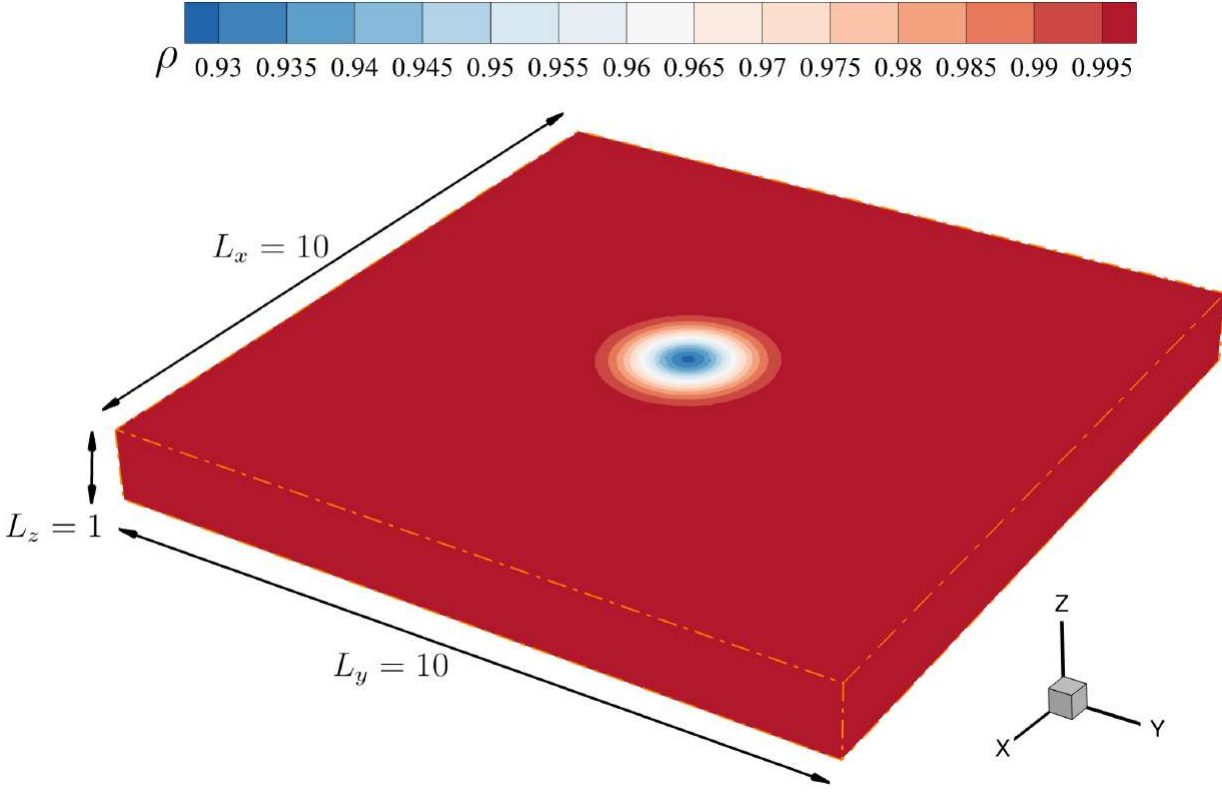}}\\
    \subfloat[Structured macro-element HDG mesh for $m=4$.\label{fig:Test_Vortexb}]{\includegraphics[width=0.6\textwidth]{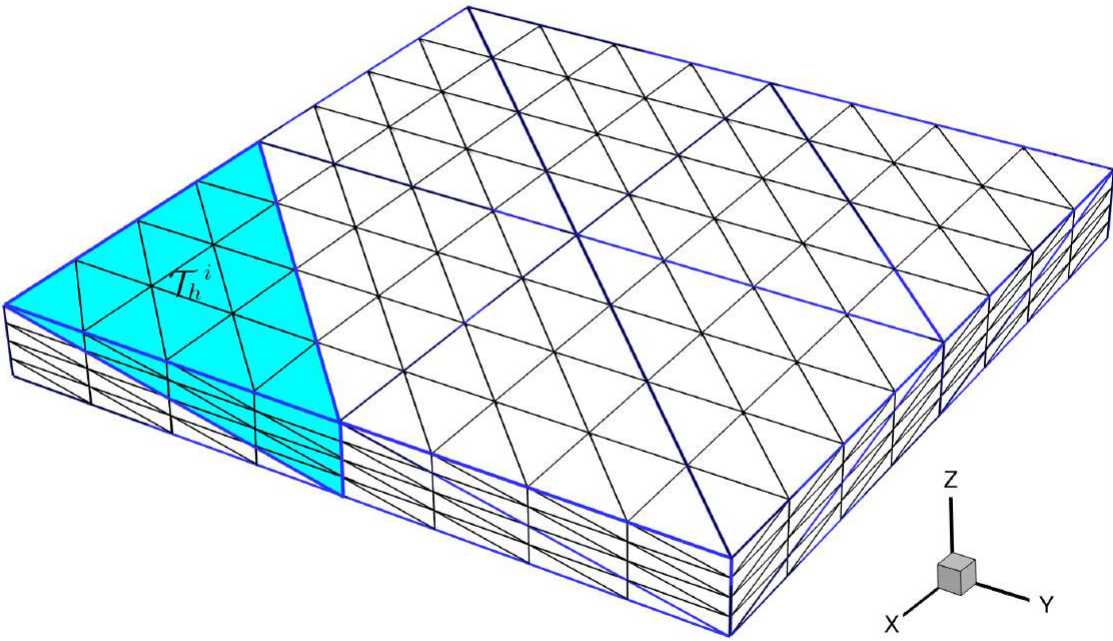}}
    \caption{3D isentropic vortex problem. HDG macro-elements are plotted in blue, $C^0$-continuous elements within each macro-element are plotted in black.}
    \label{fig:Test_Vortex}
\end{figure}

\subsection{3D isentropic vortex problem}\label{S60}

As a first validation case, we consider the three-dimensional isentropic vortex governed by the compressible Euler equations. This benchmark is widely used to assess the accuracy and long-time stability of high-order numerical methods.

The computational domain consists of a periodic cube, and the vortex is initialized at the center of the domain. It is subsequently transported by a uniform background flow with velocity $\mathbf{V}_\infty$ at an angle $\alpha$. The analytical solution is known and allows for a precise evaluation of discretization errors.

The exact solution is given by

\begin{equation}\label{S601}
 \begin{split}
    r (x,y,z,t) &= \left( \left(x - x_0 - \text{V}_\infty \cos\left( \alpha\right) t\right)^2 + \left(y - y_0 - \text{V}_\infty \sin\left( \alpha\right)t\right)^2 \right) ^{1/2}\\
    \rho (x,y,z,t)  &=\rho_{\infty} \left( 1 - \frac{\epsilon^2 (\gamma - 1) M_{\infty}^2}{8\pi^2} \exp\left( 1 - r^{2}\right) \right) ^{\frac{1}{\gamma - 1}} ,\\
    \text{V}_1 (x,y,z,t) &= \text{V}_\infty \left( \cos\left( \alpha\right) -  \frac{\epsilon  \left( y - y_0 - \text{V}_\infty \sin\left( \alpha\right)t\right) }{2 \pi} \exp\left(\frac{1 - r^{2}}{2}\right) \right), \\
    \text{V}_2 (x,y,z,t) &= \text{V}_\infty \left( \sin\left( \alpha\right) +  \frac{\epsilon  \left( x - x_0 - \text{V}_\infty \cos\left( \alpha\right)t\right) }{2 \pi} \exp\left(\frac{1 - r^{2}}{2}\right) \right), \\
    \text{V}_3 (x,y,z,t) &= 0,\\
 P (x,y,z,t) &= P_\infty  \ \rho^{\gamma}
\end{split}   
\end{equation}
where \(\epsilon\) represents the vortex strength, and \(\rho_\infty\), \(P_\infty\) and \(M_\infty\) are the freestream density, pressure and Mach number, respectively. 

\begin{figure}
    \centering
    \subfloat[\centering \(\epsilon = 2.5\), \(M_\infty = 0.5\)]{{\includegraphics[width=0.45\textwidth]{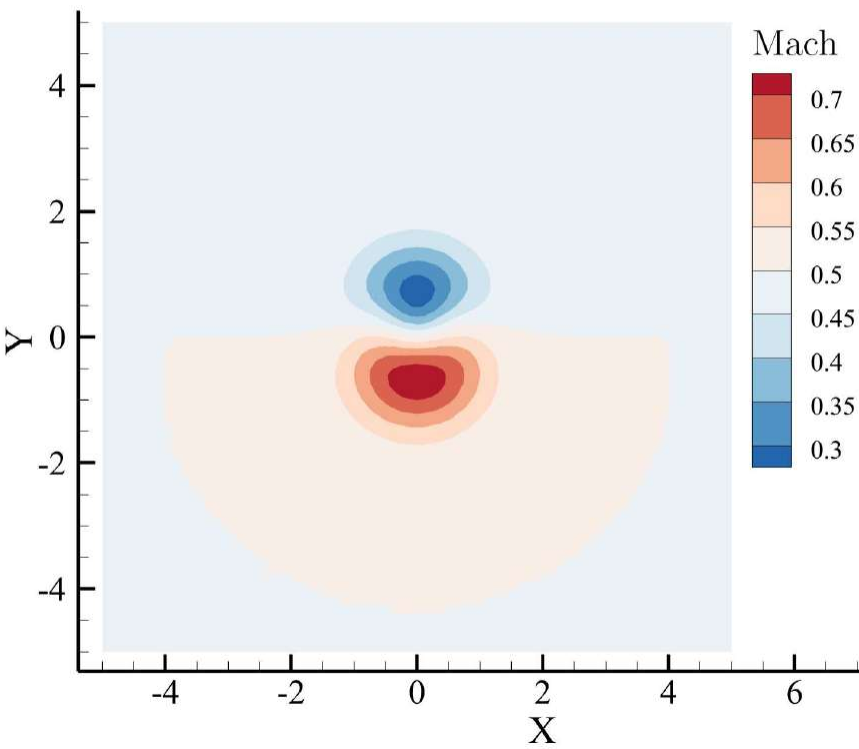} }}\hspace{1.0cm }
    \subfloat[\centering \(\epsilon = 5.0\), \(M_\infty = 0.85\)]{{\includegraphics[width=0.45\textwidth]{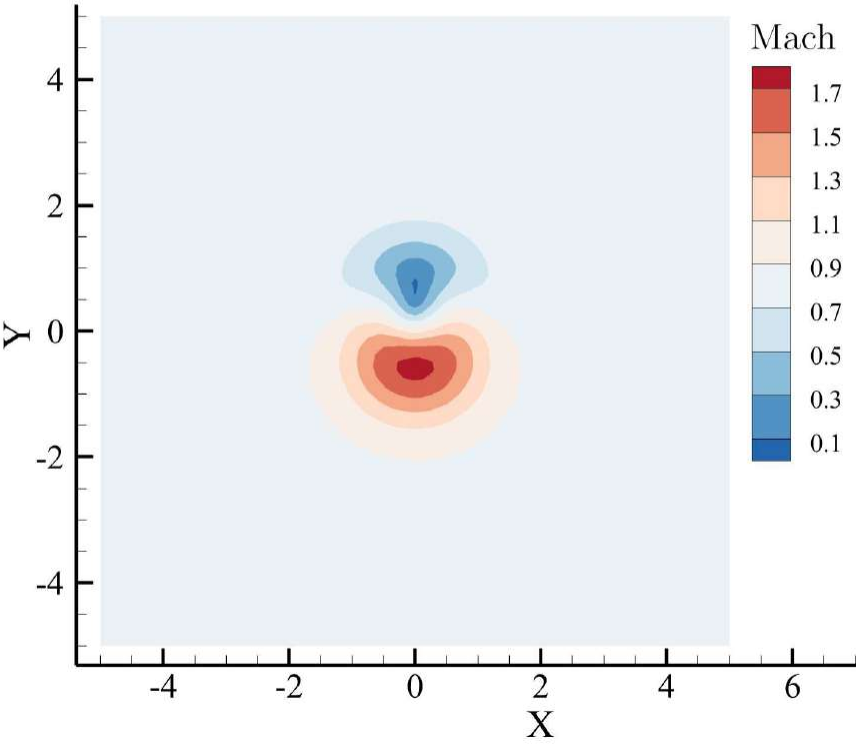} }}\\
    \subfloat[\centering Density, $\rho$]{{\includegraphics[width=0.45\textwidth]{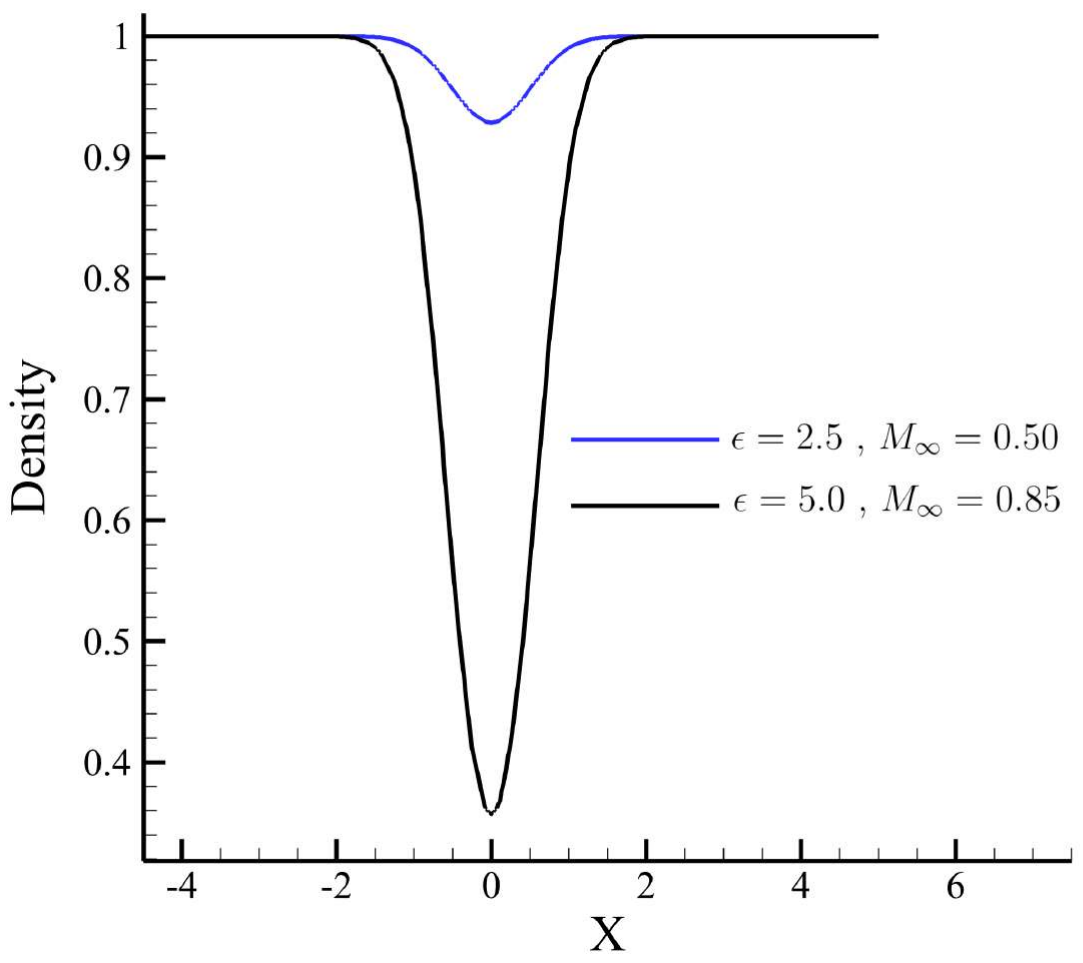} }}
\caption{3D isentropic vortex problem: (a) and (b) depict 2D slices of Mach number contours, while (c) presents the density profile along the line \( y = 0 \).}
    \label{fig:Vortex_M}%
\end{figure}

In the following, we generally choose the parameters $\alpha = 0.0$ and $\text{V}_\infty=1$. To evaluate the convergence rates in space and time, we specifically set \(\epsilon = 2.5\) and \(M_\infty = 0.50\). To analyze the stability of different schemes, we specifically set \(\epsilon = 5.0\) and \(M_\infty = 0.85\).  Figure~\ref{fig:Vortex_M} plots the initial Mach number contours and density profiles for both parameter setups.

\subsubsection{Assessment of the solution accuracy}\label{S611}

We consider macro-element HDG meshes as shown in Figure~\ref{fig:Test_Vortexb}, where we specify the number of $C^2$-continuous elements along a macro-element edge in each direction as $m=4$. For comparison, we also consider the standard HDG case, which we obtain for $m=1$.
The simulations were performed using a very small time step, \(\Delta t = 0.001\) to ensure negligible temporal discretization error. An absolute tolerance of \(10^{-9}\) was set for the nonlinear solver, with a relative tolerance of \(10^{-3}\) for the iterative linear solver.

\begin{figure}
    \centering
    \subfloat[\centering Standard HDG ($m=1$).]{\includegraphics[width=0.47\textwidth]{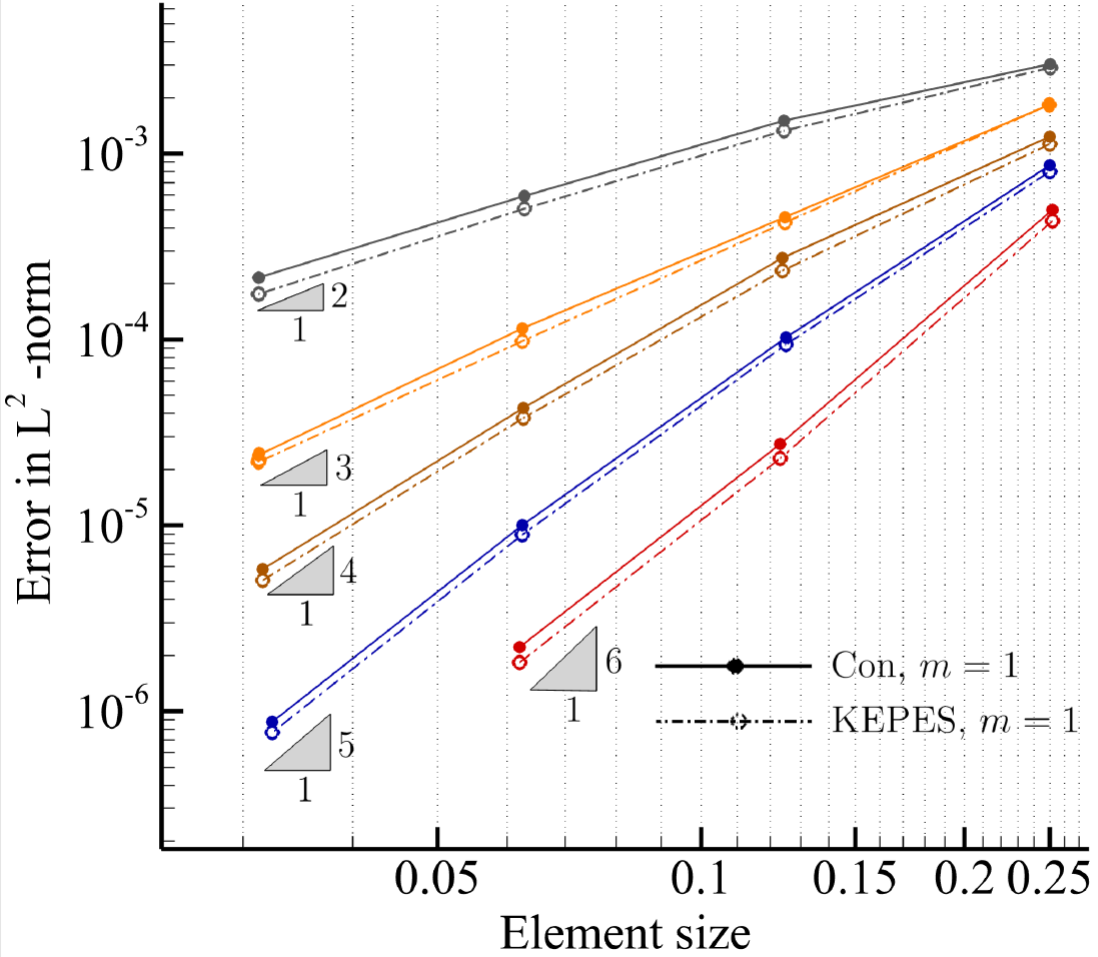} \label{fig:Test_Vortex_Cona} }\hspace{.20cm}
    \subfloat[\centering Macro-element HDG ($m=4$).]{{\includegraphics[width=0.48\textwidth]{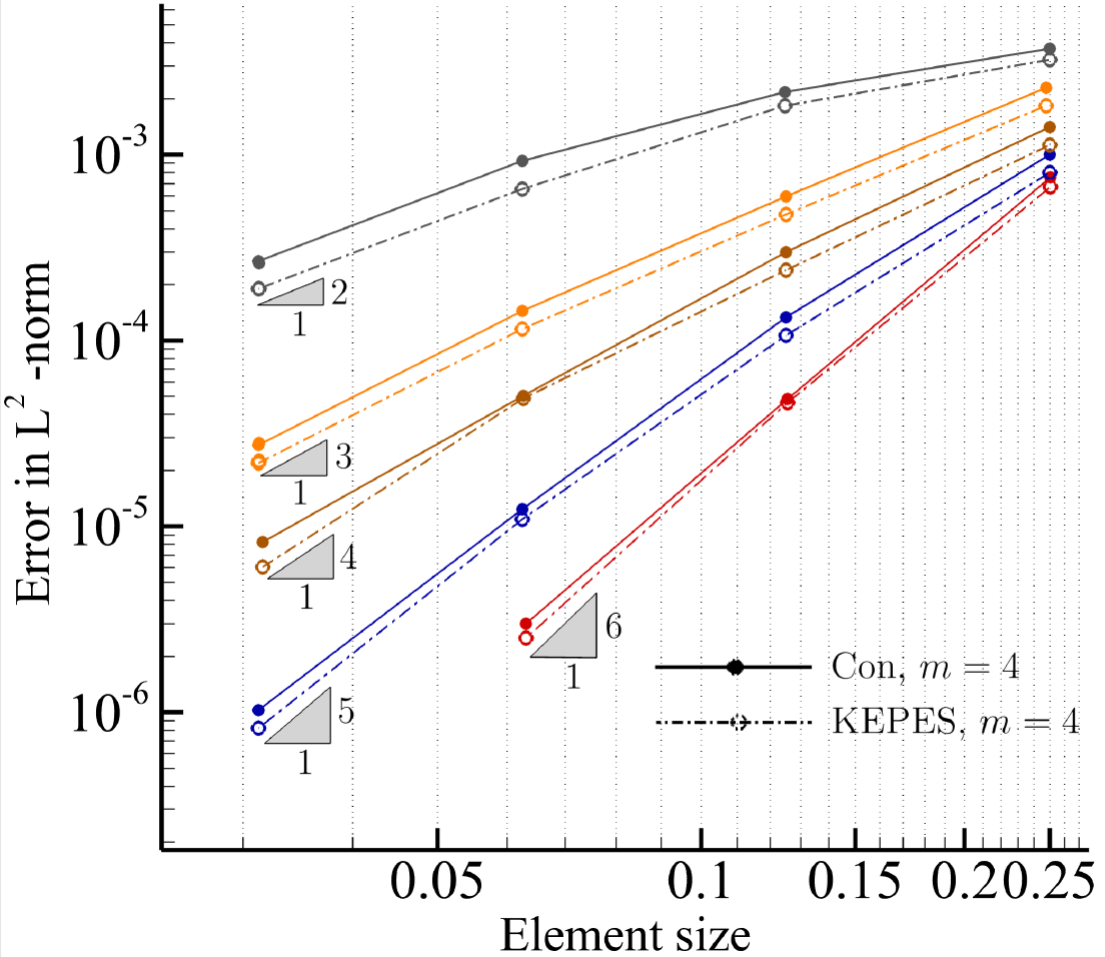} } \label{fig:Test_Vortex_Conb}}
\caption{3D isentropic vortex problem (\(\epsilon = 2.5\), \(M_\infty = 0.50\)): Convergence of the $L^2$ error of the density field in space, obtained the standard conservative-variable formulation (solid lines) and the entropy-variable formulation using the KEPES numerical inviscid flux (dashed lines).}
    \label{fig:Test_Vortex_Con}%
\end{figure}

In Figure~\ref{fig:Test_Vortex_Con}, we plot the resulting \(L^2\)-norm of the error as a function of the characteristic element size under uniform refinement of macro-elements for $m=1$ and $m=4$, evaluated after 10\% of the convective period. We note that the characteristic element size governing the resolution of the solution refers to the $C^0$-continuous elements within each macro-element patch.
The results show that for polynomial degrees $p=1$ to 5, optimal convergence rates of order \(p+1\) are achieved in all cases. The results clearly demonstrate that the macro-element HDG method with $m=4$ achieves practically the same accuracy in terms of error magnitude and convergence rate as the standard HDG method, when plotted against the characteristic element size.

\begin{table}[h!]
\centering
\caption{3D isentropic vortex problem (\(\epsilon = 2.5\), \(M_\infty = 0.50\)): convergence in time for the conservative-variable formulation (Con) and the entropy-variable formulation using the inviscid flux \eqref{ES} (KEPES).}
\begin{tabularx}{\textwidth}{l *{8}{>{\centering\arraybackslash}X}}\toprule
& \multicolumn{2}{c}{KEPES, $m = 1$} & \multicolumn{2}{c}{KEPES, $m = 4$} & \multicolumn{2}{c}{Con, $m = 1$}& \multicolumn{2}{c}{Con, $m = 4$} \\
\cmidrule(lr){2-3}\cmidrule(lr){4-5}\cmidrule(lr){6-7}\cmidrule(lr){8-9}
$T/\Delta t $& Error & rate & Error & rate & Error & rate & Error & rate  \\\midrule           
$4$   &1.33e-1  &      &1.33e-1  &      &1.34e-1  &     &1.34e-1  & \\[4pt]
$8$   &2.89e-2  &2.21  &2.64e-2  &2.34  &2.67e-2  &2.21 &2.65e-3  &2.34 \\[4pt]
$16$  &4.82e-3  &2.58  &3.78e-3  &2.81  &3.78e-3  &2.58 &3.79e-3  &2.81 \\[4pt]
$32$  &6.39e-4  &2.91  &5.74e-4  &2.72  &5.31e-4  &2.91 &5.36e-4  &2.72 \\\bottomrule				
\end{tabularx}
\label{Tab:Con_Time}
\end{table}

For the temporal integration, the convergence behavior of the DIRK(3,3) scheme is presented in  Table~\ref{Tab:Con_Time}. The results were obtained using $24,576$ elements with polynomial degree $p=3$, applied as standard HDG elements ($m=1$) or clustered in macro-element HDG patches with $m=4$. Integration in time was performed until a specific convection time, $T$, specified as a ratio with respect to the time step $\Delta t$. Both conservative-variable and entropy-variable formulations were considered, using the inviscid numerical flux KEPES \eqref{ES}. Across all spatial discretizations, the DIRK(3,3) scheme achieves the expected third-order convergence rate in time, with comparable error magnitudes in all methods.

\begin{figure}[h!]
    \centering
    \subfloat[Time evolution of thermodynamic entropy.]{{\includegraphics[width=0.5\textwidth]{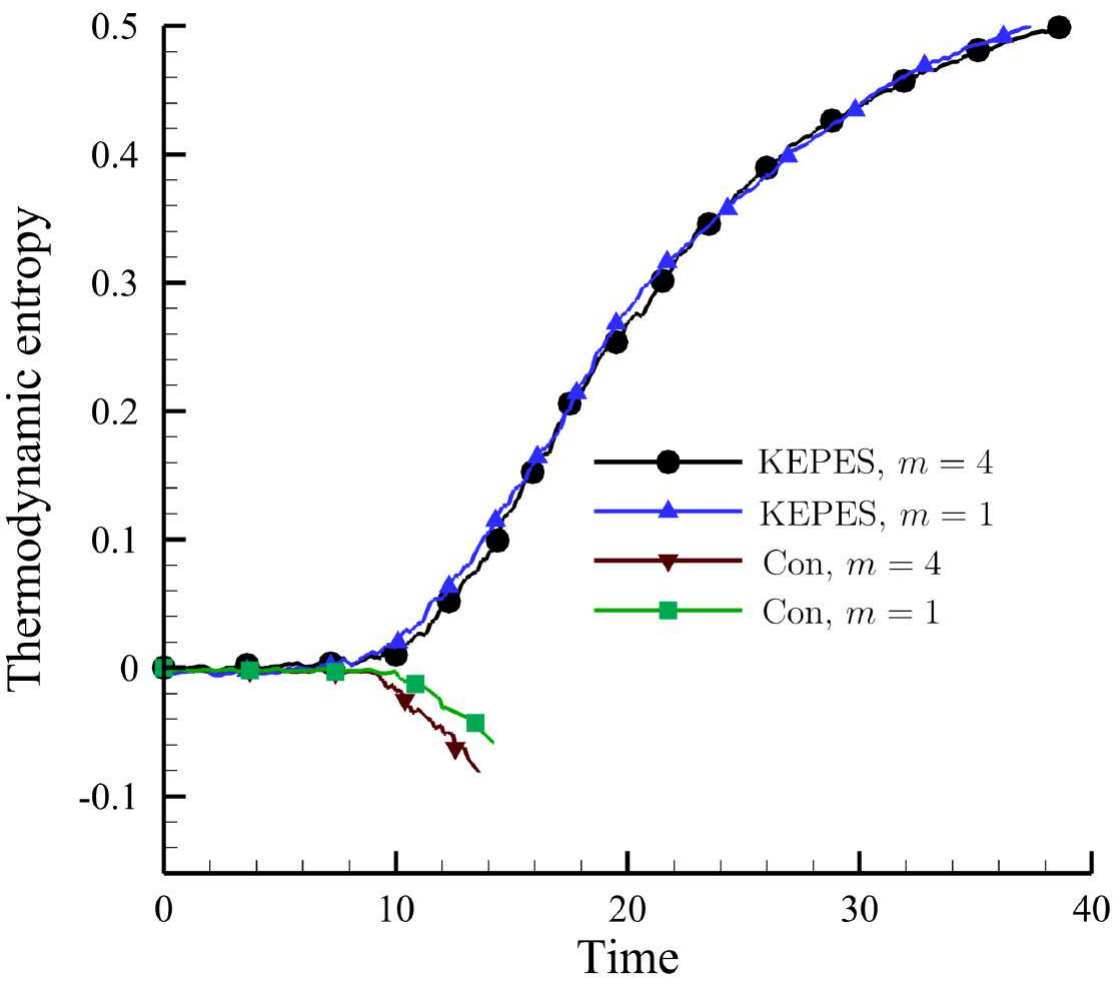} }}%
    \subfloat[Time evolution of density error.]{{\includegraphics[width=0.5\textwidth]{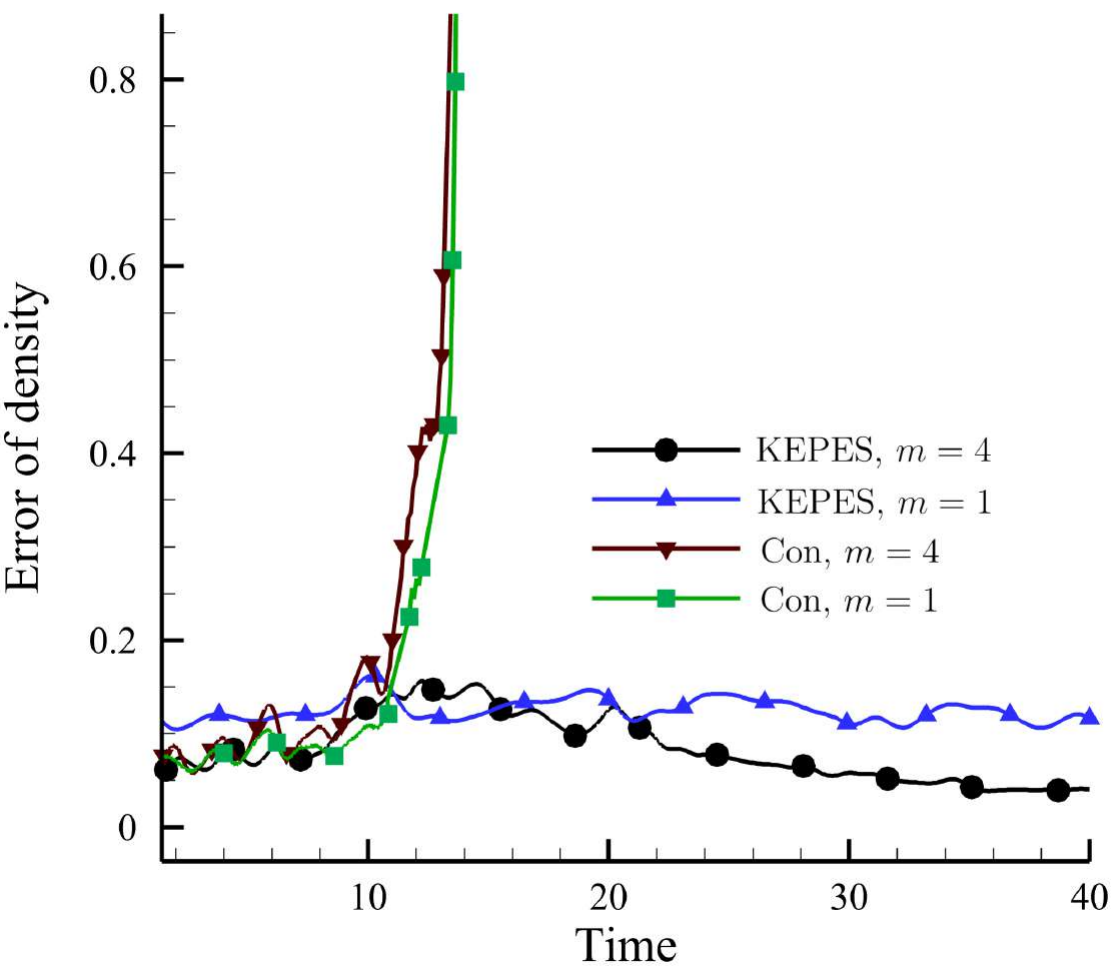} }}%
\caption{3D isentropic vortex problem ($\epsilon = 5.0, M_0 = 0.85$): conservative-variable formulations become unstable shortly after $t = 10.0$, while entropy-variable formulations remain stable.}
    \label{fig:Vortex_ES}%
\end{figure}

\subsubsection{Assessment of stability}\label{S612}

To analyze the behavior of different types of numerical fluxes, we consider HDG meshes with 1,200 elements for $m=1$ and $m=4$ and polynomial degree \( p = 2 \). First, we examine the smooth case ($\epsilon = 2.5, M_{\infty} = 0.5$), where both standard and macro-element HDG variants remain stable. However, when increasing the vortex strength and Mach number to ($\epsilon = 5.0, M_{\infty} = 0.85$), both standard and macro-element HDG variants become unstable. In contrast, entropy-variable formulations combined with numerical fluxes such as ES and KEPES maintain stability for all HDG variants.

This is illustrated in Figure~\ref{fig:Vortex_ES} that plots the time evolution of thermodynamic entropy, computed through \eqref{eq:en}, and the time evolution of the corresponding density error, integrated over the complete domain. We observe that the standard HDG ($m=1$) and the macro-element HDG ($m=4$) variants in KEPES format lead to an increasing entropy over time, thus satisfying the second law of thermodynamics -- a behavior theoretically supported by Theorem \ref{flux_ES_1}. In addition, they lead to a steady density error that does not explode in time. In contrast, the corresponding conservative-variable formulations decrease entropy, violating the second law of thermodynamics. This leads to unstable behavior, illustrated by the exploding density error shortly after $t = 10.0$.

\begin{figure}
    \centering
    \includegraphics[width=0.65\textwidth]{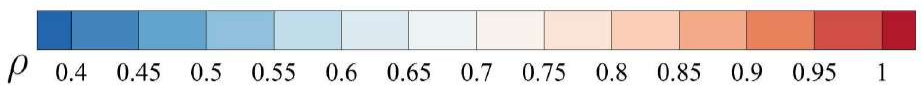}  \\
    \subfloat[\centering Analytical reference for the density field.]{{\includegraphics[width=0.33\textwidth]{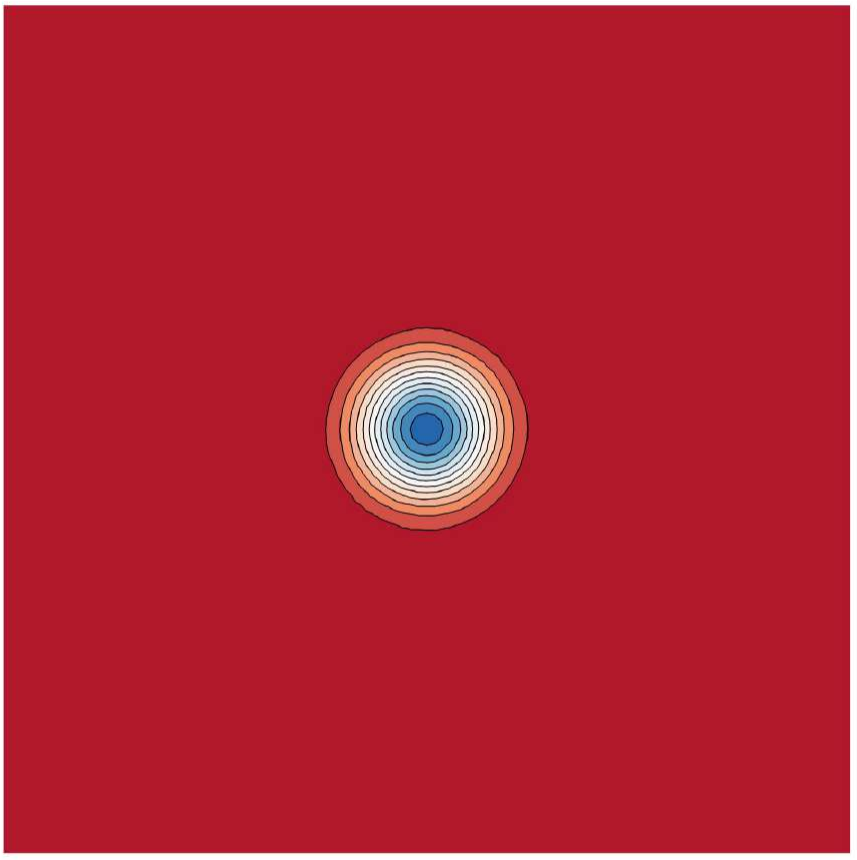} }}%
    \subfloat[\centering Macro-element HDG with ES flux, $t=10$.]{{\includegraphics[width=0.33\textwidth]{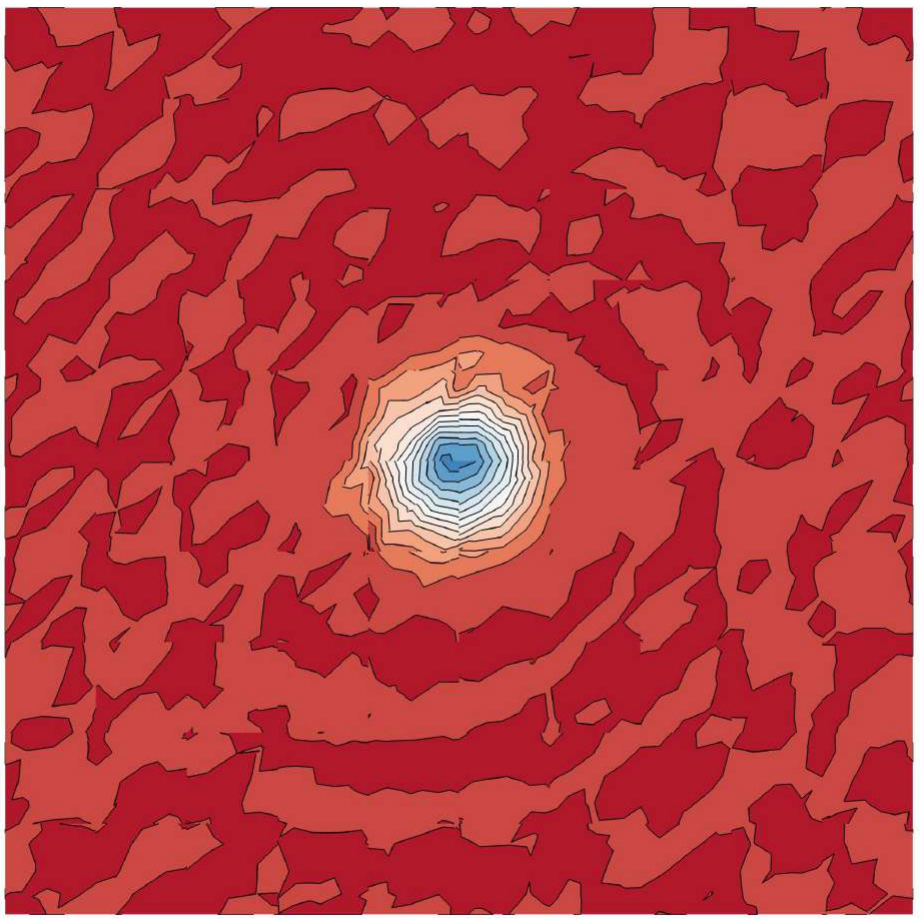} }}
    \subfloat[\centering Macro-element HDG with KEPES flux, $t=10$.]{{\includegraphics[width=0.33\textwidth]{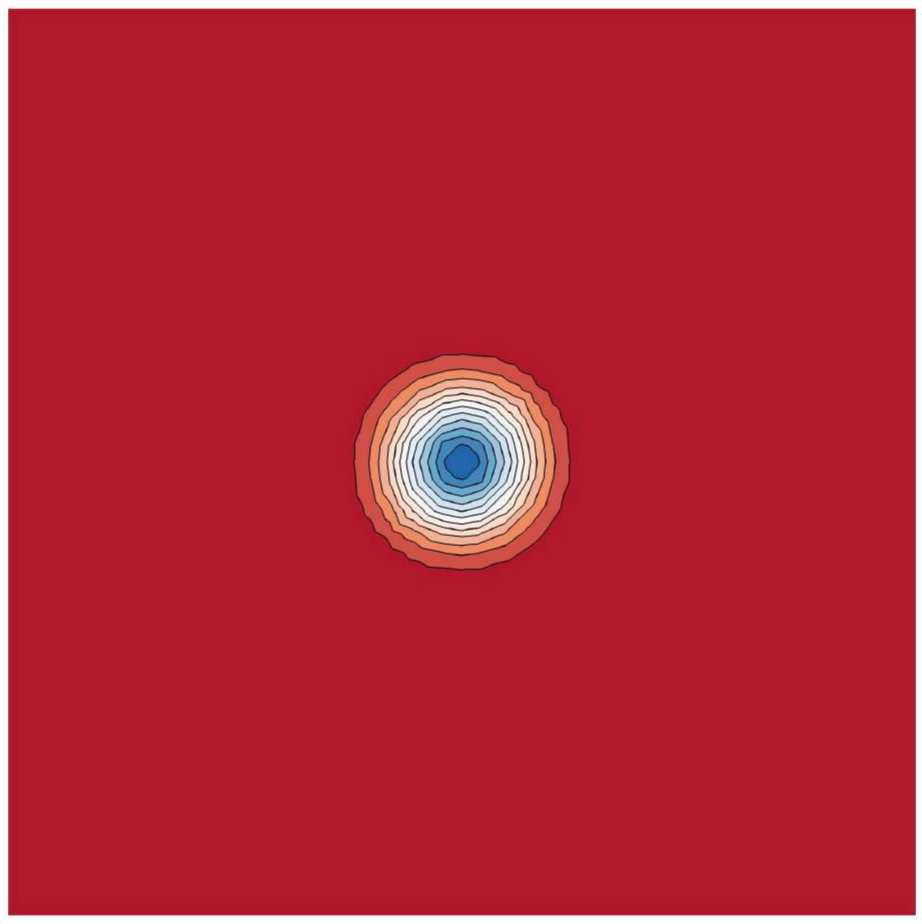} }}\\\vspace{1.0cm }
    \includegraphics[width=0.65\textwidth]{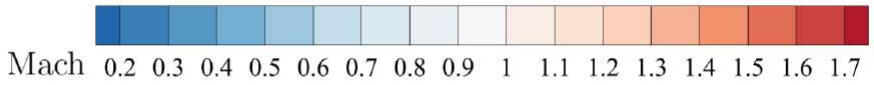} \\
    \subfloat[\centering Analytical reference for the Mach number contours.]{{\includegraphics[width=0.33\textwidth]{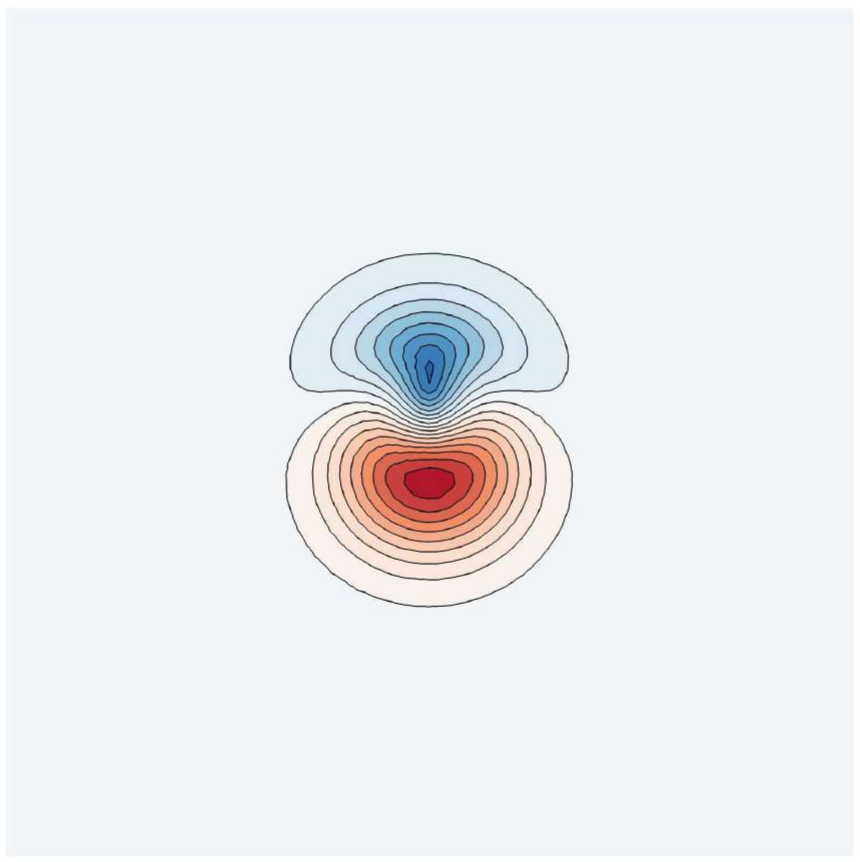} }}%
    \subfloat[\centering Macro-element HDG ($m=4$) with ES flux, $t=10$.]{{\includegraphics[width=0.33\textwidth]{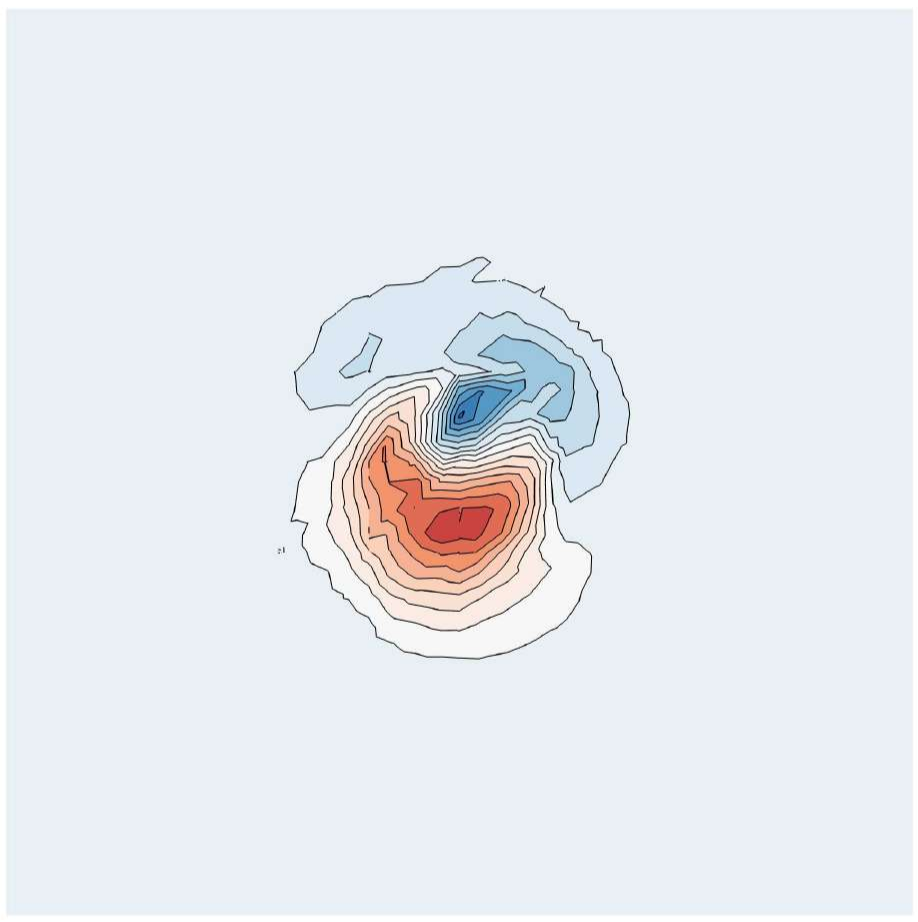} }}
    \subfloat[\centering Macro-element HDG ($m=4$) with KEPES flux, $t=10$.]{{\includegraphics[width=0.33\textwidth]{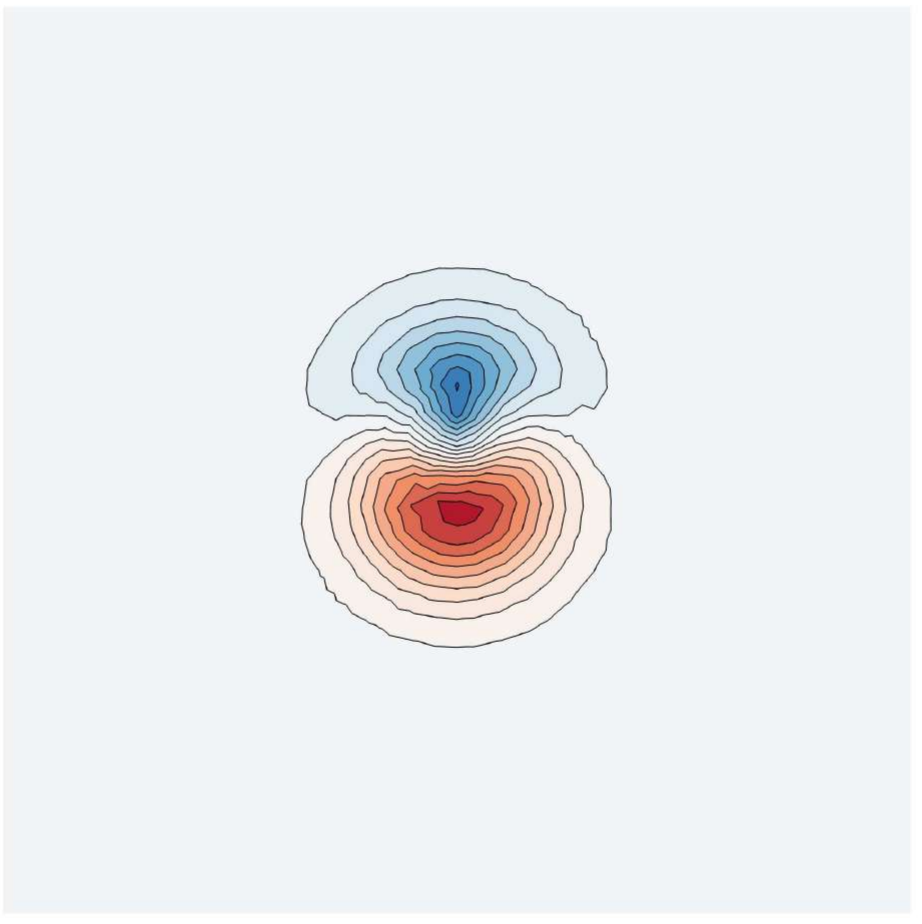} }}
\caption{3D isentropic vortex ($M_\infty = 0.85$, $\epsilon = 5.0$): comparison of the density and Mach number field solutions obtained with the macro-element HDG method in entropy-variable format ($m=4$, $p=2$)  using the inviscid flux \eqref{Flux_En_Inv} (ES) and the inviscid flux \eqref{ES} (KEPES) against the analytical reference. The results show the advantage of the KEPES flux formulation.}
    \label{fig:Vortex_b5}%
\end{figure}

The stability property of all tested variants in the different formulations for the two different test problems are summarized in Table~\ref{Tab:3D_Vortex_Stability}. 
While both entropy-variable formulations using the inviscid flux \eqref{Flux_En_Inv} (ES) and the inviscid flux \eqref{ES} (KEPES) are stable, only the latter produces accurate results. This is further illustrated in Figure~\ref{fig:Vortex_b5} that compares the density field and Mach number contour solutions obtained with the macro-element HDG method in entropy-variable format with the ES formulation and the KEPES formulation against the analytical reference. The results show that the macro-elment HDG variant based on the KEPES formulation preserves kinetic energy and remains free of spurious oscillations, while the ES formulation fails to accurately capture the vortex structure under long time integration.

\begin{table}[h!]
\centering
\caption{3D isentropic vortex problem: stability for standard HDG ($m=1$) and macro-element HDG ($m=4$) at polynomial degree $p=2$, formulated in conservative variable format (Con) and entropy-variable format using the inviscid flux \eqref{Flux_En_Inv} (ES) and the inviscid flux \eqref{ES} (KEPES).}
\begin{tabularx}{\textwidth}{c *{6}{>{\centering\arraybackslash}X}} \toprule
& \multicolumn{3}{c}{$m = 1$} & \multicolumn{3}{c}{$m = 4$} \\
\cmidrule(lr){2-4}\cmidrule(lr){5-7}
& Con & ES & KEPES & Con & ES & KEPES \\\midrule           
$M_\infty = 0.50$, $\epsilon = 2.5$  & \cmarkgreen & \cmarkgreen & \cmarkgreen & \cmarkgreen & \cmarkgreen & \cmarkgreen \\[5pt]
$M_\infty = 0.85$, $\epsilon = 5.0$  & \xmarkred & \cmarkgreen & \cmarkgreen & \xmarkred & \cmarkgreen & \cmarkgreen \\\bottomrule				
\end{tabularx}
\label{Tab:3D_Vortex_Stability}
\end{table}

\subsection{Inviscid Taylor-Green vortex} \label{S62}
\begin{figure}
    \centering
    \subfloat[Macro-element mesh with the effective resolution $N_{\text{eff}} = 70^3$.]{{\includegraphics[width=0.4\textwidth]{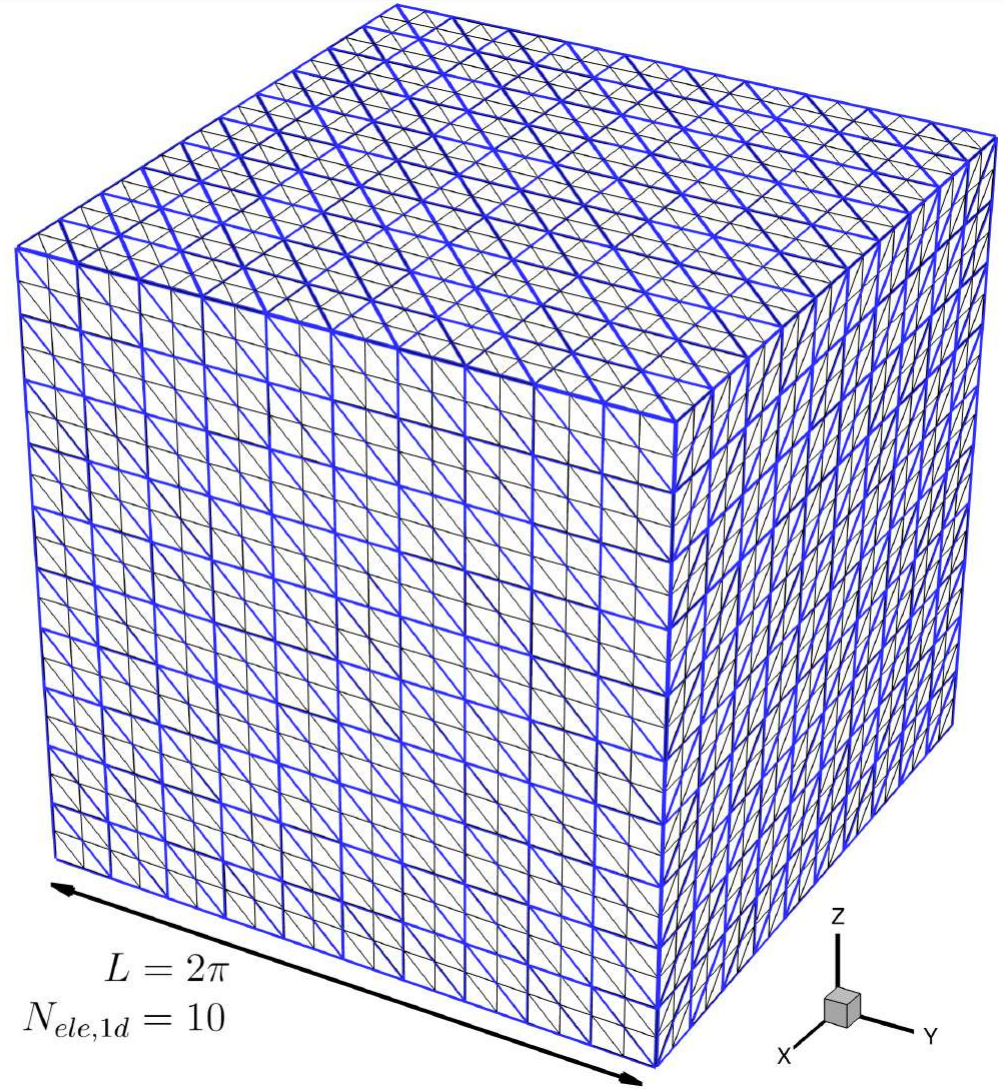} }}\hspace{1.0cm }
    \subfloat[Initial Mach number contour, scaled for the case ${M}_{0}=0.8$.]{{\includegraphics[width=0.5\textwidth]{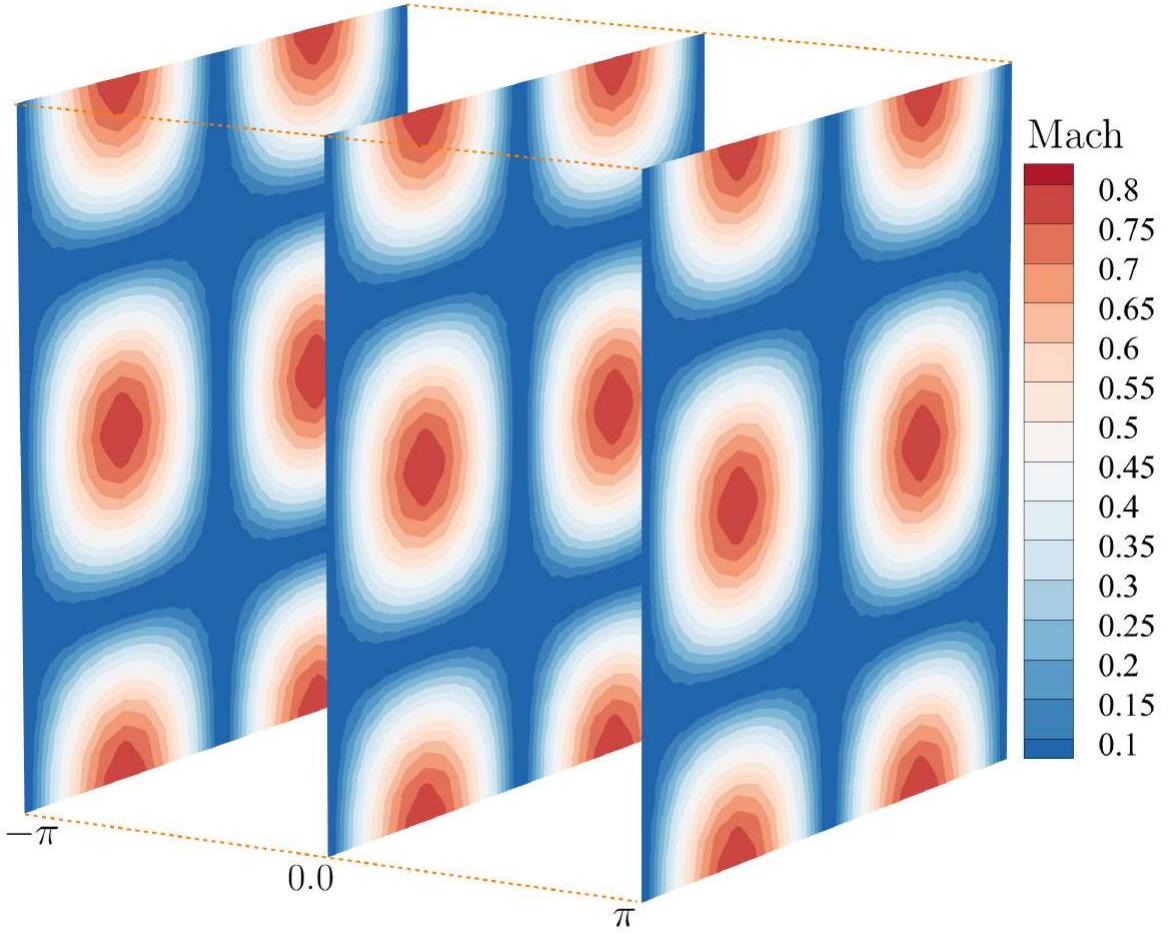} }}
    \caption{Inviscid Taylor–Green vortex: HDG discretization ($m=2$) and intial condition. The discontinuous macro-element edges are plotted in blue, the $C^0$-continuous element edges in black.}%
    \label{fig:Inv_TGV_case}%
\end{figure}

As the next benchmark, we consider the inviscid Taylor-Green vortex flow \cite{taylor1937mechanism,brachet1991direct,fehn2018efficiency} on a cube $\left[ -\pi L,\pi L\right]^3$ for two different Mach numbers ${M}_{0}=0.1$ (nearly incompressible) and ${M}_{0}=0.8$ (significant compressibility). The stability
of DG discretizations for the under-resolved simulation of this benchmark was
studied in \cite{moura2017eddy,winters2018comparative}. We prescribe periodic boundary conditions in all coordinate directions, and use the following initial conditions:
\begin{equation}\label{TGV0}
\allowdisplaybreaks
 \begin{split}
 \text{V}_{1}(x,y,z,t) &= \ \ \text{V}_{0} \sin\left(\dfrac{x_{1}}{L}\right)\cos\left(\dfrac{x_{2}}{L}\right)\cos\left(\dfrac{x_{3}}{L}\right) ,\\
 \text{V}_{2}(x,y,z,t) &=-\text{V}_{0} \cos\left(\dfrac{x_{1}}{L}\right)\sin\left(\dfrac{x_{2}}{L}\right)\cos\left(\dfrac{x_{3}}{L}\right),\\
 \text{V}_{3}(x,y,z,t) &=0,\\
 P(x,y,z,t) &= p_{0}  + \dfrac{\rho_{0} \text{V}_{0}^{2}}{16} \left(\cos\left(\dfrac{2x_{1}}{L}\right)+\cos\left(\dfrac{2x_{2}}{L}\right) \right) \left( \cos\left(\dfrac{2x_{3}}{L}\right)+2\right),
\end{split}   
\end{equation}
where $L = 1$, $\rho_{0} = 1$ and $P_{0} =1/\gamma$. The flow is initialized to be isothermal, that is, $P / \rho=P_{0} / \rho_{0}$. The unsteady simulation is performed for a duration of $10 \, t_{c}$, where $t_{c}=L/ \text{V}_{0}$ is the characteristic convective time, and $\text{V}_{0}=M_{0} c_{0}$, where $c_{0}$ is the speed of sound corresponding to $P_{0} $ and $\rho_{0}$, ${c_{0}}^2 = \gamma P_{0}/ \rho_{0}$.

\begin{figure}
    \centering
    \subfloat[$t/t_{c} = 2.0$.]{{\includegraphics[width=0.33\textwidth]{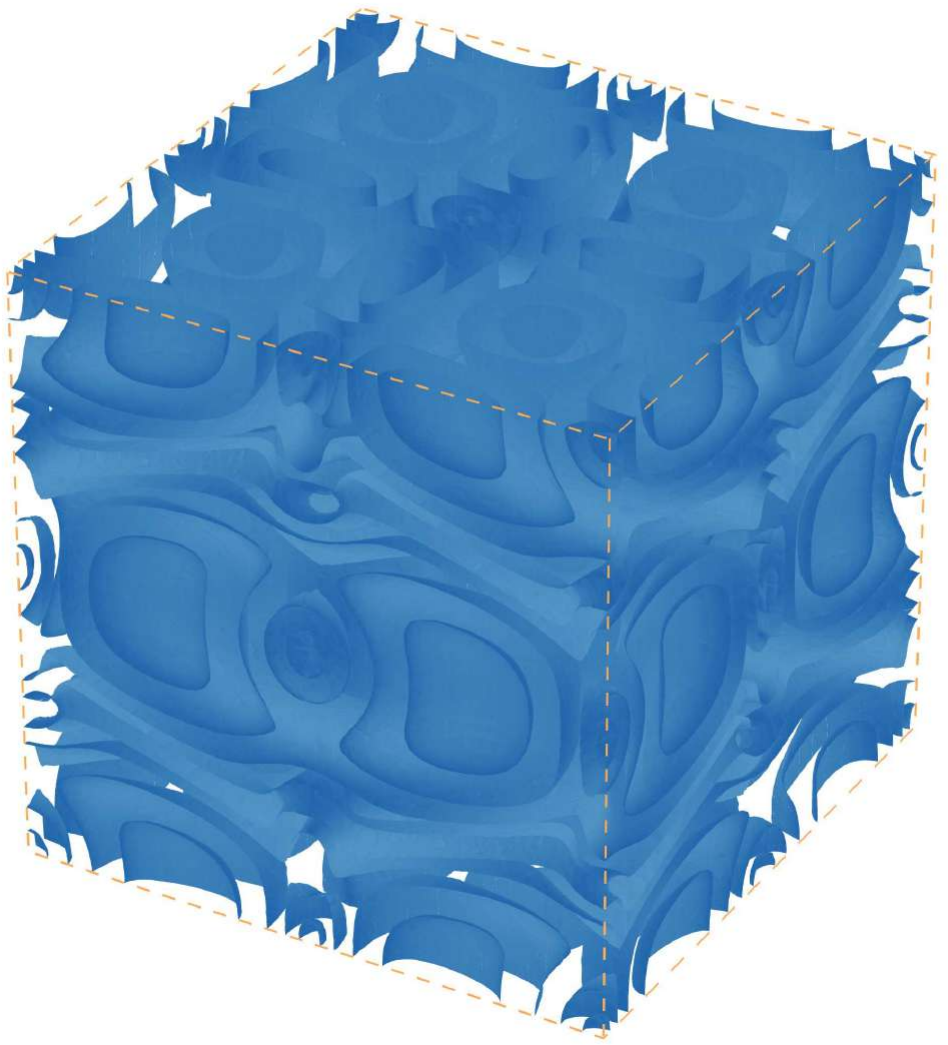} }}   
    \subfloat[$t/t_{c} = 4.0$.]{{\includegraphics[width=0.33\textwidth]{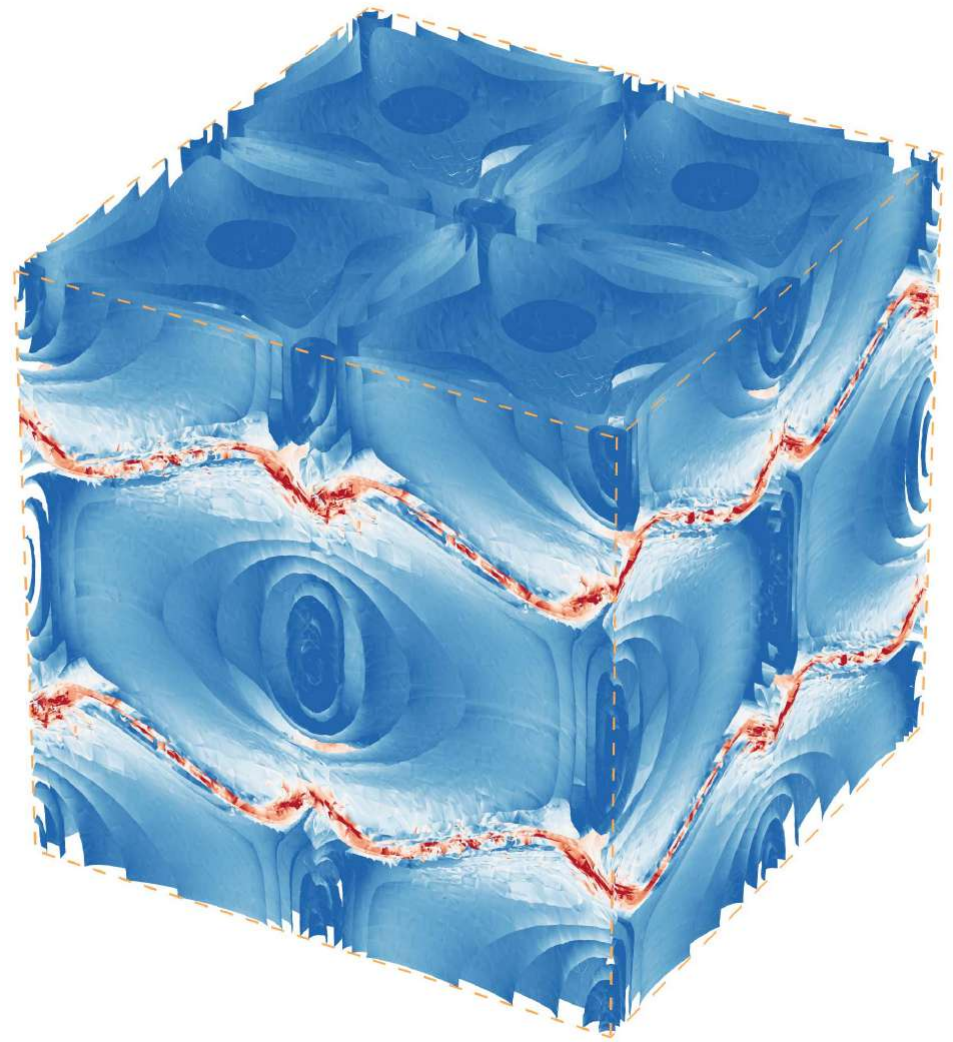} }}  
    \subfloat[$t/t_{c} = 6.0$.]{{\includegraphics[width=0.33\textwidth]{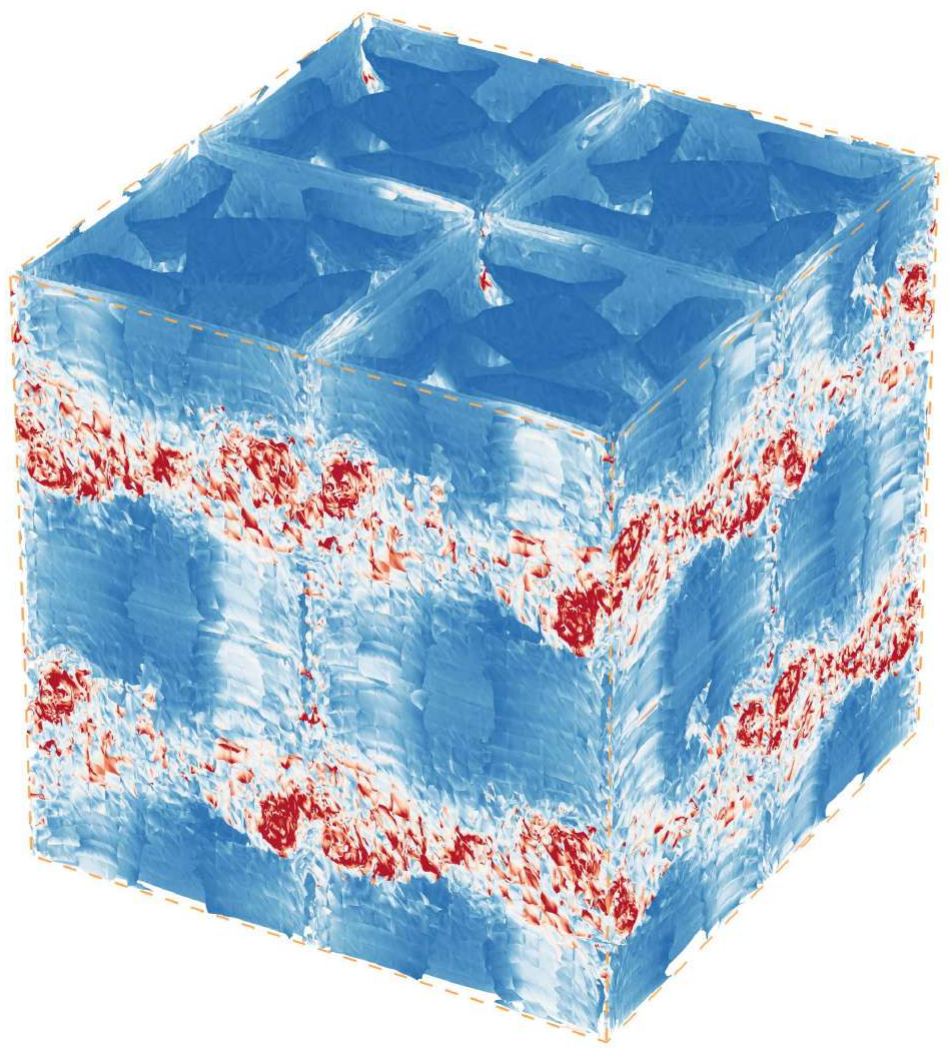} }}\\
    \subfloat{{\includegraphics[width=0.4\textwidth]{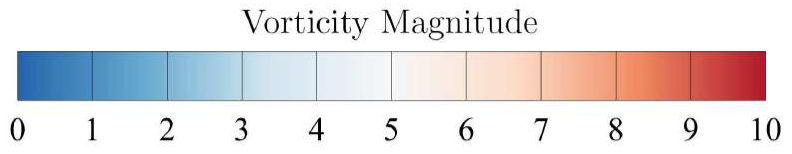} }}
    \caption{Inviscid Taylor–Green vortex (low-Mach number case with $M_0=0.1$): Vorticity magnitude plotted on iso-contours of the velocity field,  obtained with macro-element HDG ($m=2, p=3$) in entropy-variable format and the KEPES flux formulation \eqref{ES} on the mesh resolution $N_{\text{eff}}=98^3$.}%
    \label{fig:Inv_TGV_M01}%
\end{figure} 

We compare the standard HDG method ($m=1$, $p=3$) and the macro-element HDG method ($m=2$, $p=3$), using the numerical inviscid flux formulatios given in \eqref{Flux_En_Inv} (ES) and in \eqref{ES} (KEPES). The effective resolution of the computational domain is defined as $N_{\text{eff}} = \left(N_{ele,1d} \cdot (mp+1)\right)^3$, where $N_{ele,1d}$ represents the number of macro-elements per spatial direction. We choose the time step as $\Delta t = 0.057 \ t_{c}$. 

\begin{figure}
    \centering
    \subfloat[$t/t_{c}= 3.0$.]{{\includegraphics[width=0.32\textwidth]{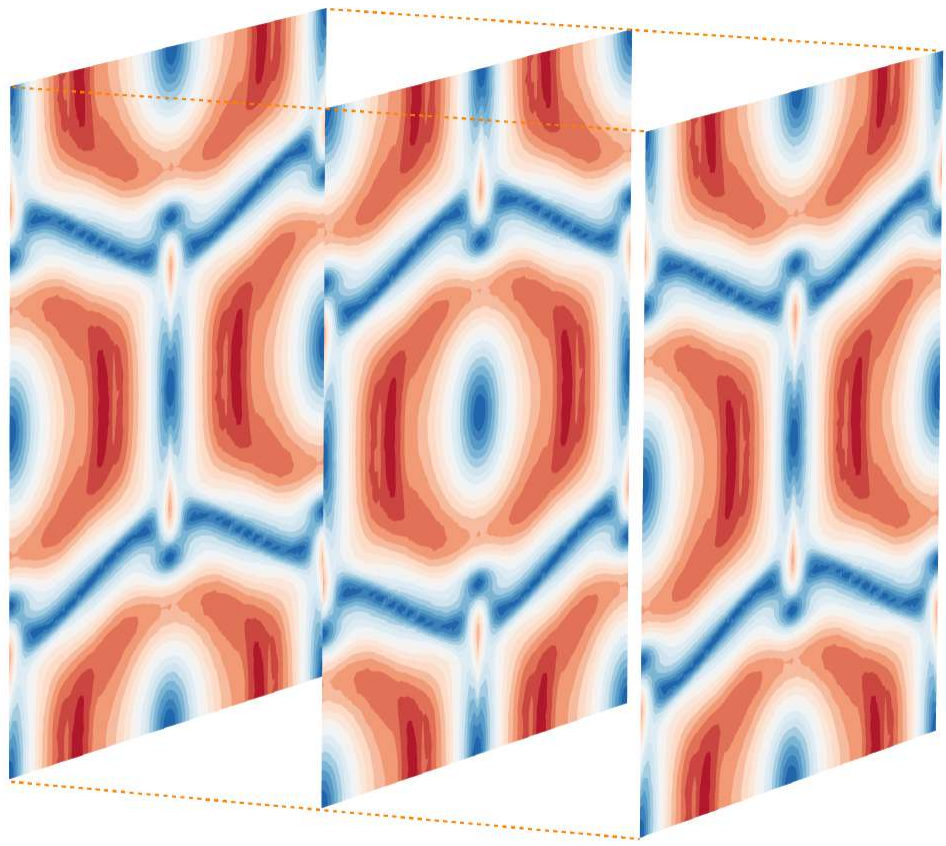} }}\hspace{0.00cm }    
    \subfloat[$t/t_{c}= 6.0$.]{{\includegraphics[width=0.32\textwidth]{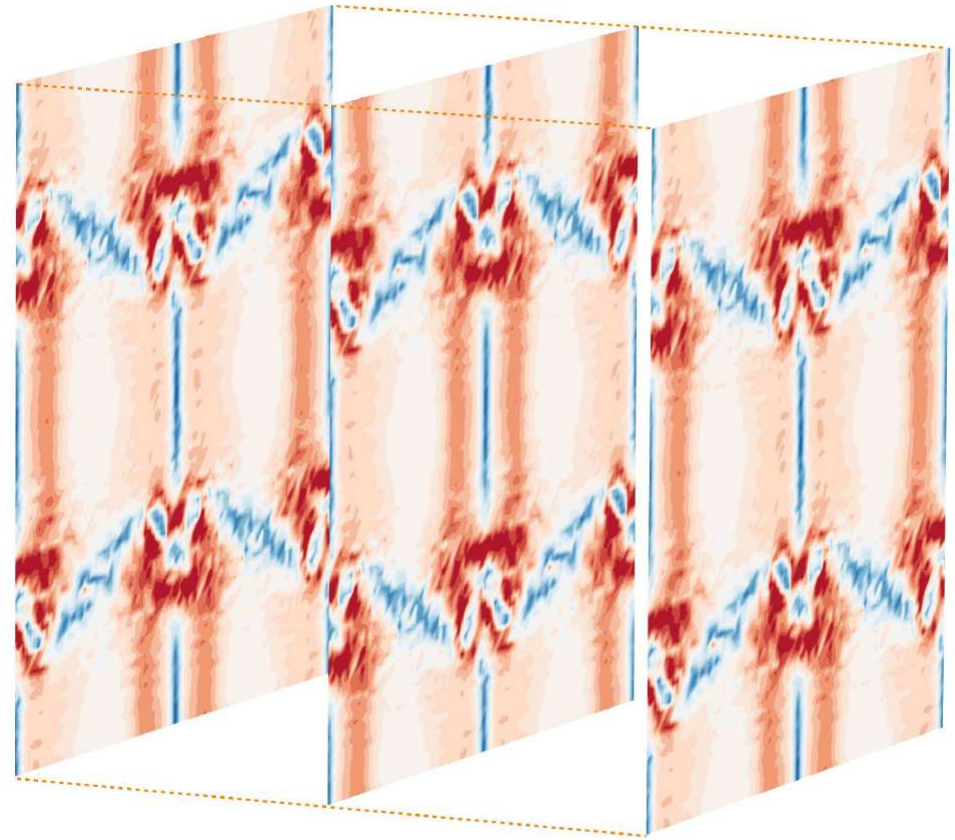} }}\hspace{0.00cm }    
    \subfloat[$t/t_{c}= 9.0$.]{{\includegraphics[width=0.32\textwidth]{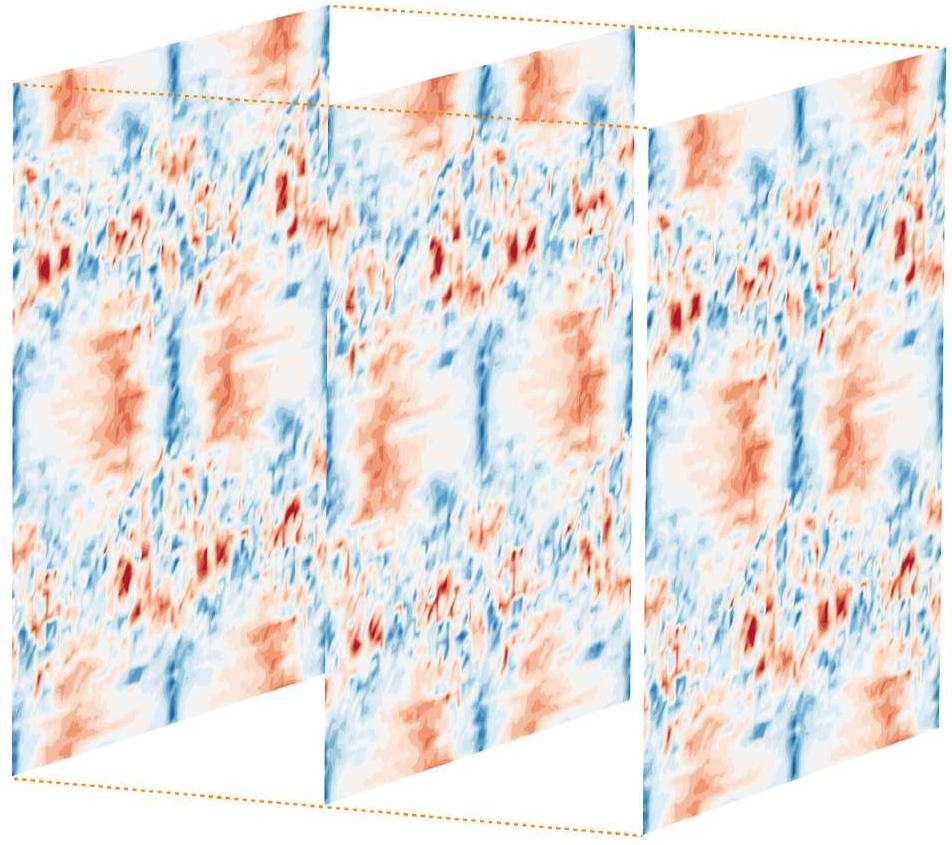} \label{fig:Inv_TGV_Snac}}}\\\vspace{.5cm }
    \includegraphics[width=0.65\textwidth]{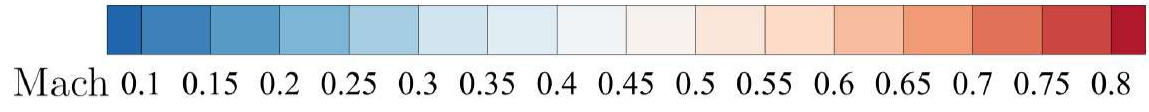} \\
    \caption{Inviscid Taylor–Green vortex (moderate-Mach number case with $M_0=0.8$): Mach number field on 2D slices at $x = -\pi$, $0$, and $\pi$, obtained with macro-element HDG ($m=2, p=3$) in entropy-variable format and the KEPES flux formulation \eqref{ES} on the mesh resolution $N_{\text{eff}}=98^3$.}%
    \label{fig:Inv_TGV_Sna}%
\end{figure}

Figure \ref{fig:Inv_TGV_case} illustrates the mesh structure and the initial condition, whose scaling corresponds to the case ${M}_{0}=0.8$. In the following, we consider two different effective resolutions: $N_{\text{eff}} = 70^3$ and $N_{\text{eff}} = 98^3$. Figures \ref{fig:Inv_TGV_M01} and \ref{fig:Inv_TGV_Sna} plot solutions fields at three different time instances, obtained with the macro-element HDG variant in entropy-variable format and the KEPES flux formulation on the finer mesh with $N_{\text{eff}} = 98^3$. Figure \ref{fig:Inv_TGV_M01} refers to the low-Mach-number case $M_0=0.1$, plotting the vorticity magnitude on iso-contours of the velocity field. Figure \ref{fig:Inv_TGV_Sna} refers to the moderate-Mach-number case $M_0=0.8$, plotting the Mach number field on three different slices of the cube.

We can observe that the flow evolution can be broadly categorized into three phases. Initially, for $t < 3.5 \, t_{c}$, the flow remains laminar and free of subgrid-scale activity. Therefore, the solution is well-resolved by the grid. This is followed by a transitional phase, where small-scale structures begin to emerge. The flow is still predominantly laminar but shows signs of under-resolution, that is, the numerical mesh is too coarse to resolve the fine-scale features. This phase, signaling the onset of turbulence, extends up to approximately $t = 8 \, t_{c}$. Beyond this point, the flow becomes fully turbulent with chaotic and multiscale behavior. It is now strongly under-resolved, with many features occuring at scales smaller than the mesh size. The features shown here, particularly in Figure \ref{fig:Inv_TGV_Snac}, are in agreement with the observations reported in \cite{fernandez2019entropy}.

\subsubsection{Assessment of stability}\label{S611}

\begin{table}
\centering
\caption{Inviscid Taylor–Green vortex: Stability of the macro-element HDG method ($m = 2$, $p=3$) for the low-Mach-number and moderate-Mach-number cases on two mesh resolutions. The entropy-variable format uses either the entropy-stable flux \eqref{Flux_En_Inv} (ES) or the energy-preserving entropy-stable flux \eqref{ES} (KEPES).}
\begin{tabularx}{\textwidth}{>{\centering\arraybackslash}p{0.85cm} >{\centering\arraybackslash}p{0.85cm} *{3}{>{\centering\arraybackslash}X}} \toprule
& & \multicolumn{3}{c}{\small Macro-element HDG ($m = 2$, $p=3$)} \\
\cmidrule(lr){3-5}
$M_0$ & $N_{\text{eff}}$ & \small Conservative-variable format & \small Entropy-variable format (ES flux) & \small Entropy-variable format (KEPES flux) \\\midrule 
0.1 & $70^3$  & \cmarkgreen & \cmarkgreen & \cmarkgreen \\[5pt]
0.1 & $98^3$  & \cmarkgreen & \cmarkgreen & \cmarkgreen \\\midrule  
0.8 & $70^3$  & \xmarkred & \cmarkgreen & \cmarkgreen \\[5pt]
0.8 & $98^3$  & \xmarkred & \cmarkgreen & \cmarkgreen \\\bottomrule				
\end{tabularx}
\label{Tab:Inv_TGV_Stability}
\end{table}

In the scope of this paper, the stability of the numerical scheme is of particular interest. Table \ref{Tab:Inv_TGV_Stability} provides an overview of stability for the different combinations of test cases, conservative-variable vs.\ entropy-variable formats, mesh resolutions and flux formulations. We observe that for the low-Mach-number case ($M_0 = 0.1$), both conservative-variable and entropy-variable formats produce stable results.
For the moderate-Mach-number case ($M_0 = 0.8$), however, 
the macro-element HDG scheme in conservative-variable format becomes unstable shortly after $t = 4.7 \, t_{c}$. In contrast, the macro-element HDG scheme in entropy-variable format maintains stability throughout the entire simulation duration.

For a more detailed analysis of stability, our focus lies on the evolution of the thermodynamic entropy and the kinetic energy dissipation rate. For compressible flow at low and moderate Mach numbers, the kinetic energy dissipation rate, $\epsilon$, can be approximated as 
\begin{equation}\label{C5321}
\epsilon = 2\dfrac{\mu}{\Omega} \int_{\Omega} \rho\: \dfrac{\bm{\omega} \cdot \bm{\omega}}{2} \:\mathrm{d}\Omega,
\end{equation} 
where the vorticity, $\bm{\omega}$, is defined as $\bm{\omega}=\nabla \times \mathbf{V}$.

\begin{figure}
    \centering
    \subfloat[Evolution of thermodynamic entropy.]{{\includegraphics[width=0.45\textwidth]{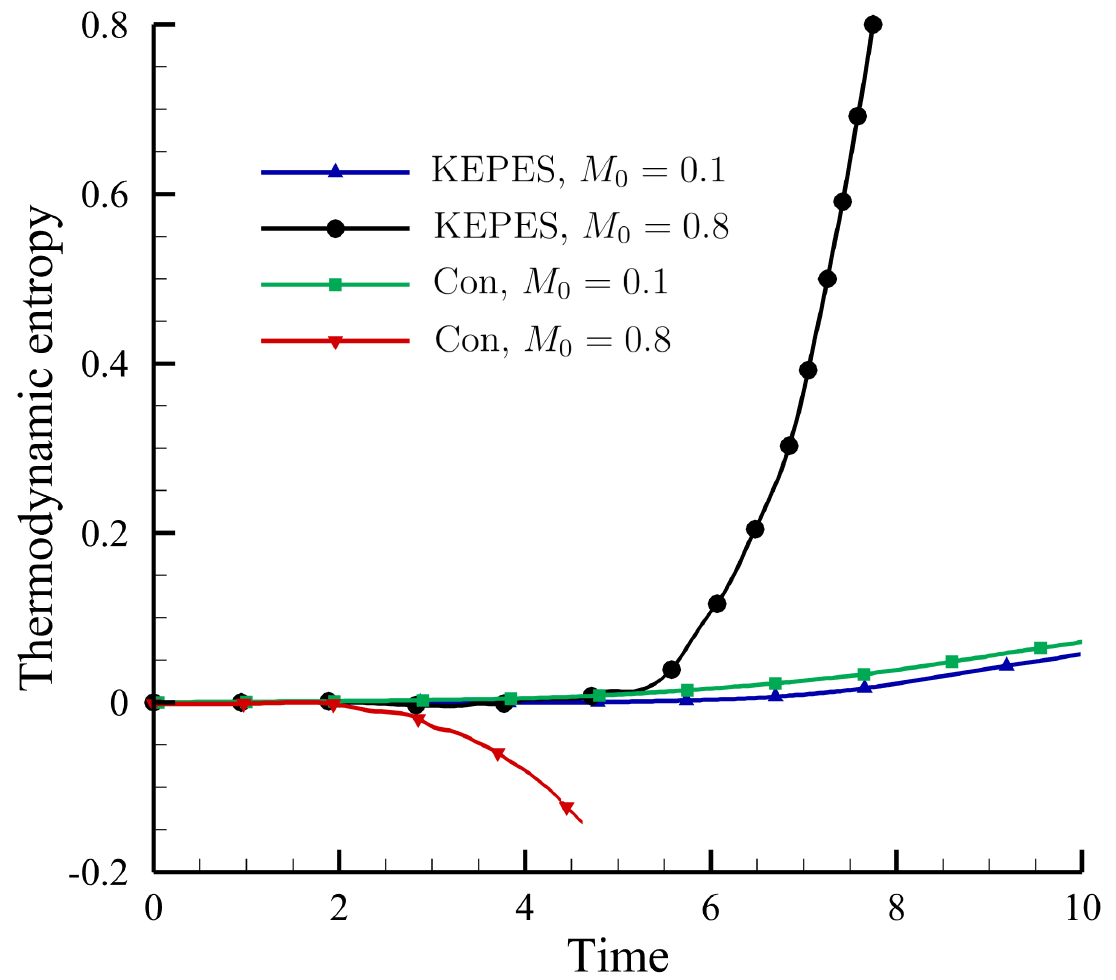} }} \hspace{1.0cm }
    \subfloat[Evolution of kinetic energy dissipation rates.]{{\includegraphics[width=0.45\textwidth]{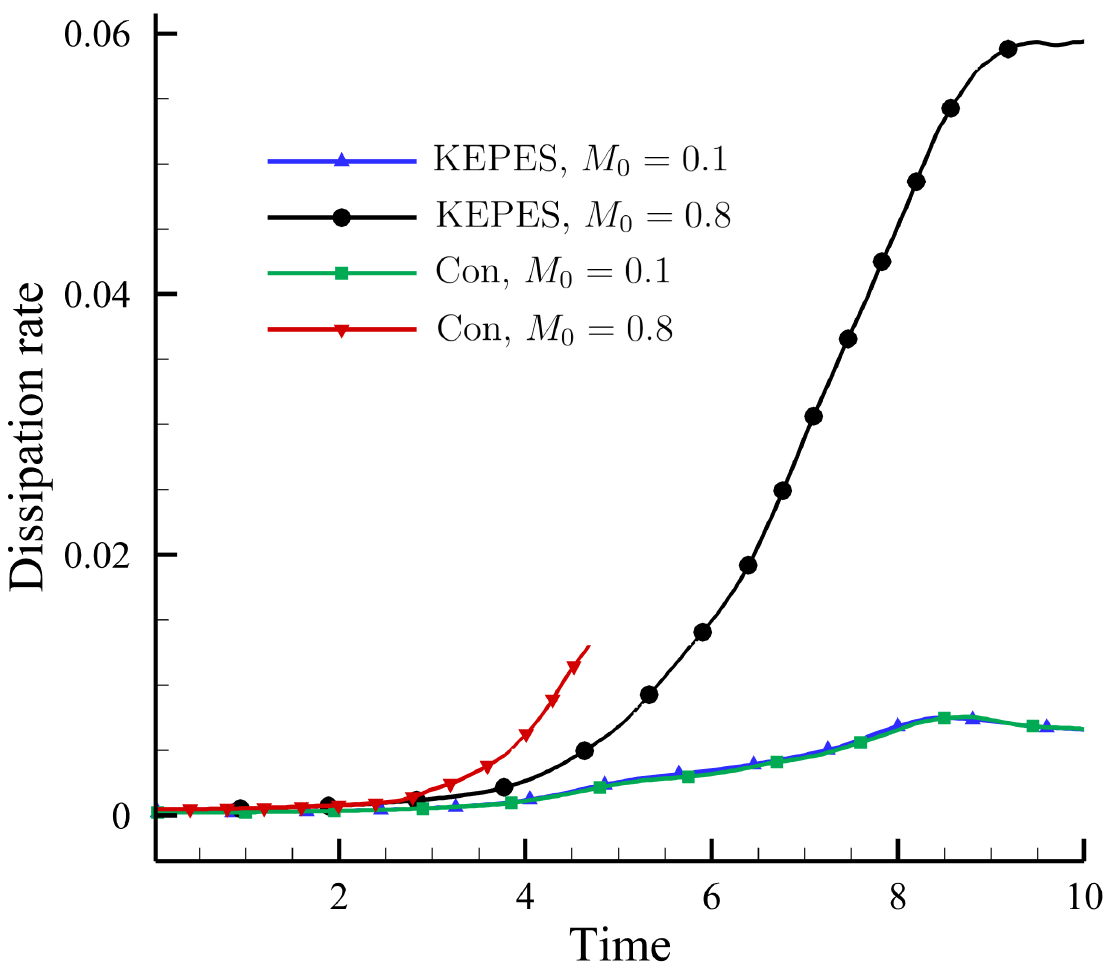} }}
    \caption{Inviscid Taylor–Green vortex: Entropy and energy dissipation behavior over time, as reproduced by macro-element HDG ($m = 2$, $p=3$) in conservative-variable (Con) and entropy-variable (KEPES) formats. For $M_0=0.8$, the former becomes unstable shortly after $t = 4.7 \, t_{c}$. }%
    \label{fig:Inv_TGV_EK_ES}%
\end{figure}

Figure \ref{fig:Inv_TGV_EK_ES} plots the thermodynamic entropy and the kinetic energy dissipation rate over time for the macro-element HDG method ($m=2$, $p=3$) in conservative-variable and entropy-variable formats on the mesh resolution $N_{\text{eff}}=98^3$, where the latter uses the KEPES flux formulation \eqref{ES}. Prior to $t = 3.7 \, t_{c}$, when the flow resolution remains well-resolved, the solutions obtained in conservative variables and entropy variables show good agreement. As subgrid-scale features begin to emerge and the flow solutions begin to be under-resolved, discrepancies between the two formulations become evident.

In the presence of unresolved scales, entropy production becomes significant, see also \cite{fernandez2019entropy,shu2005numerical}. In the one hand, the entropy-variable formulation captures the correct entropy behavior through a consistent increase in total thermodynamic entropy, reflecting the expected physical dissipation. This trend continues over time, aligning with the theoretical behavior of the entropy-variable approach. On the other hand, the conservation-variable formulation does not inherently account for entropy growth when the flow solution is under-resolved. As a result, it fails to capture the correct dissipation behavior and eventually becomes unstable, leading to a breakdown of the simulation. This highlights the critical role of entropy stability, particularly in the context of under-resolved turbulent flows \cite{fernandez2017subgrid}.

\begin{table}[h!]
\caption{Inviscid Taylor–Green vortex: number of local and global degrees of freedom resulting from standard HDG ($m=1$) and macro-element HDG ($m=2$) at different mesh resolutions and polynomial degree $p=3$.}
\centering
\begin{tabularx}{\textwidth}{>{\centering\arraybackslash}p{1.cm} >{\centering\arraybackslash}p{1.cm} >{\centering\arraybackslash}p{2.cm} *{4}{>{\centering\arraybackslash}X}}\toprule
\multicolumn{3}{c}{Mesh resolution} & \multicolumn{2}{c}{$\text{dof}^{local}$ } & \multicolumn{2}{c}{$\text{dof}^{global}$}
\\\cmidrule(r){1-3}\cmidrule(lr){4-5}\cmidrule(lr){6-7}
$N_{ele,1d}$  &$N_\text{eff}$  & $N_{ele}$     & $m = 1$ &$m = 2$   &$m = 1$  &$m = 2$    \\\midrule    
10            & $70^3$	       & 48,000	     & 19,200,000   & 10,080,000         & 4,800,000       & 1,680,000  	      \\[5pt]
14	         & $98^3$	       & 131,712	     & 52,684,800	  & 27,659,520         & 13,171,200      & 4,609,920       \\\bottomrule	 						
\end{tabularx}
\label{Tab:Inv_TGV_DOF}
\end{table}

\subsubsection{Computational performance}\label{S611}

We now briefly address the computational advantages of the macro-element HDG method ($m=2$) in comparison with the standard HDG method ($m=1$). In standard HDG, all elements are fully discontinuous, while macro-element HDG combines groups of eight $C^0$-continuous tetrahedra in onbe discontinuous macro-element. 

We first consider Table \ref{Tab:Inv_TGV_DOF} that summarizes the number of degrees of freedom associated with the local problems and the global problem, as well as the number of parallel processes for standard HDG and macro-element HDG at both mesh resolutions. A key observation is that the macro-element concept leads to a substantial reduction in the total number of degrees of freedom in both the local and global systems.

\begin{table}[h!]
\caption{Inviscid Taylor–Green vortex ($M_{0} = 0.8$): computing times for local operations and global operations, the number of global solver iterations, and the total computing time for standard HDG ($m=1$) and macro-element HDG ($m=2$); $p = 3$, $N_\text{eff}=70^3$.}
\centering
\begin{tabularx}{\textwidth}{
  >{\arraybackslash}p{4.5cm}  
  >{\centering\arraybackslash}p{2.cm}  
  >{\centering\arraybackslash}p{2.cm}  
  >{\centering\arraybackslash}p{2.2cm}  
  >{\centering\arraybackslash}p{1.8cm}  
  >{\centering\arraybackslash}p{1.8cm}  
}
\toprule
& \footnotesize Time local op's [min] 
& \footnotesize Time global op's [min] 
& \footnotesize \# Global solver iter's 
& \footnotesize Total time [min] 
& \footnotesize \# Proc's \\\midrule
\footnotesize Standard HDG ($m = 1$) & 113.6 & 68.3  & 35,162  & 181.9 & 2,304 \\
\footnotesize Macro-element HDG ($m = 2$) & 44.6  & 23.9  & 23,962  & 68.5  & 2,304 \\
\bottomrule
\end{tabularx}
\label{Tab:Inv_TGV_Time}
\end{table}

Table \ref{Tab:Inv_TGV_Time} reports 
the computing times (in minutes) for operations that are carried out locally per HDG element or macro-element HDG patch, versus the remaining components of the matrix-free global solver. The former contain the local solver, the local component of the matrix-free global solver (see \cite{badrkhani2023matrix,badrkhani2025matrix} for details), and the local formation of the Jacobian matrix. Computing times are reported for the standard HDG method ($m=1$) and the macro-element HDG method ($m=2$), both at polynomial degree $p = 3$ and on the mesh resolution $N_\text{eff}=70^3$. We observe that for both methods the computing times for local and global operations are in the same order and hence well balanced. In terms of the overall computing cost, however, we observe a clear advantage for the macro-element HDG method, which reduces the computing times by approximately a factor of three. This performance gain can be attributed to the reduction in the number of degrees of freedom as reported in Table \ref{Tab:Inv_TGV_DOF}, which also involves a significant reduction in the number of global solver iterations. For a more detailed analysis of the computational performance of the macro-element HDG method, we refer the interested reader to \cite{badrkhani2023matrix,badrkhani2025matrix}.



\subsection{Viscous Taylor-Green vortex}\label{S63}

\begin{figure}[t!]
    \centering
    \subfloat[\centering $N_\text{eff} = 84^3$]{{\includegraphics[width=0.45\textwidth]{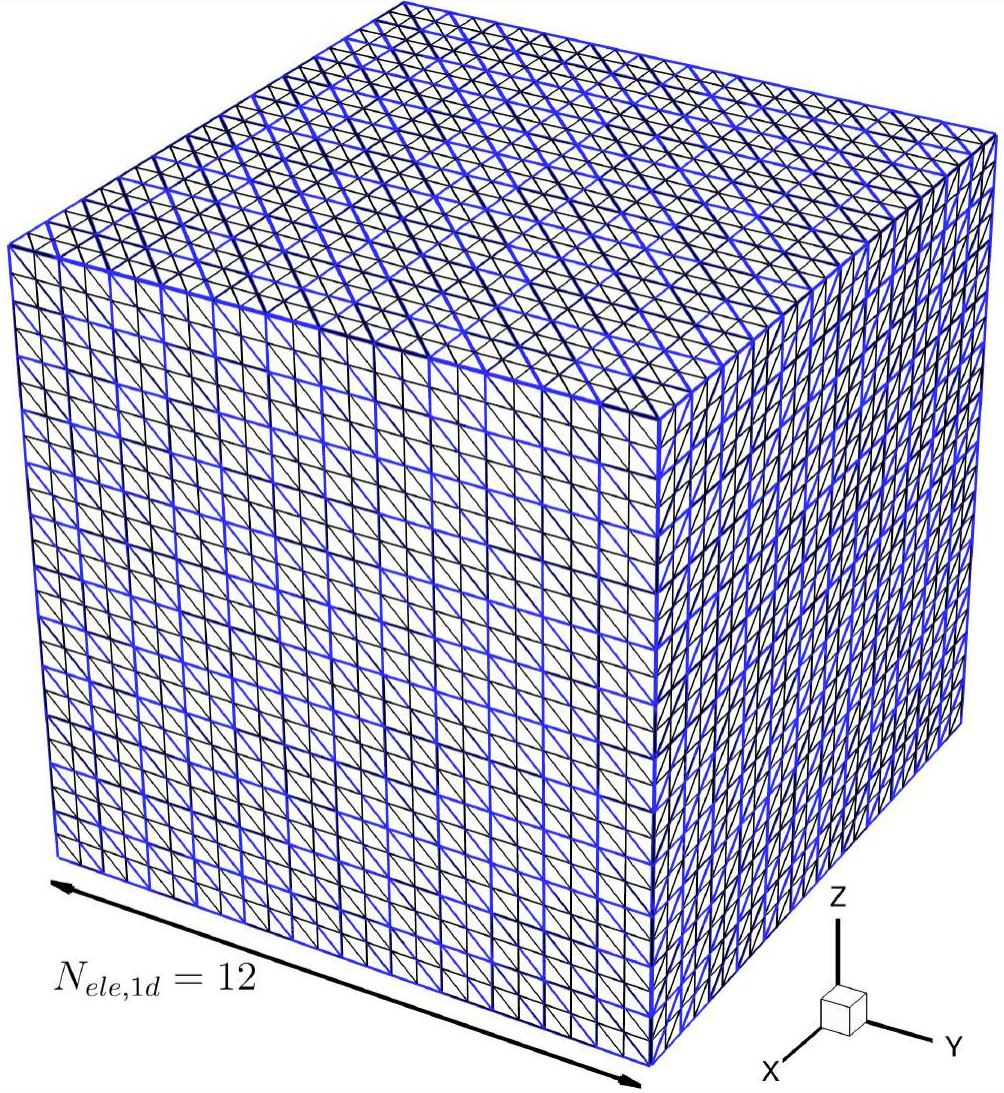} }}\hspace{1.0cm }
    \subfloat[\centering $N_\text{eff} = 126^3$]{{\includegraphics[width=0.45\textwidth]{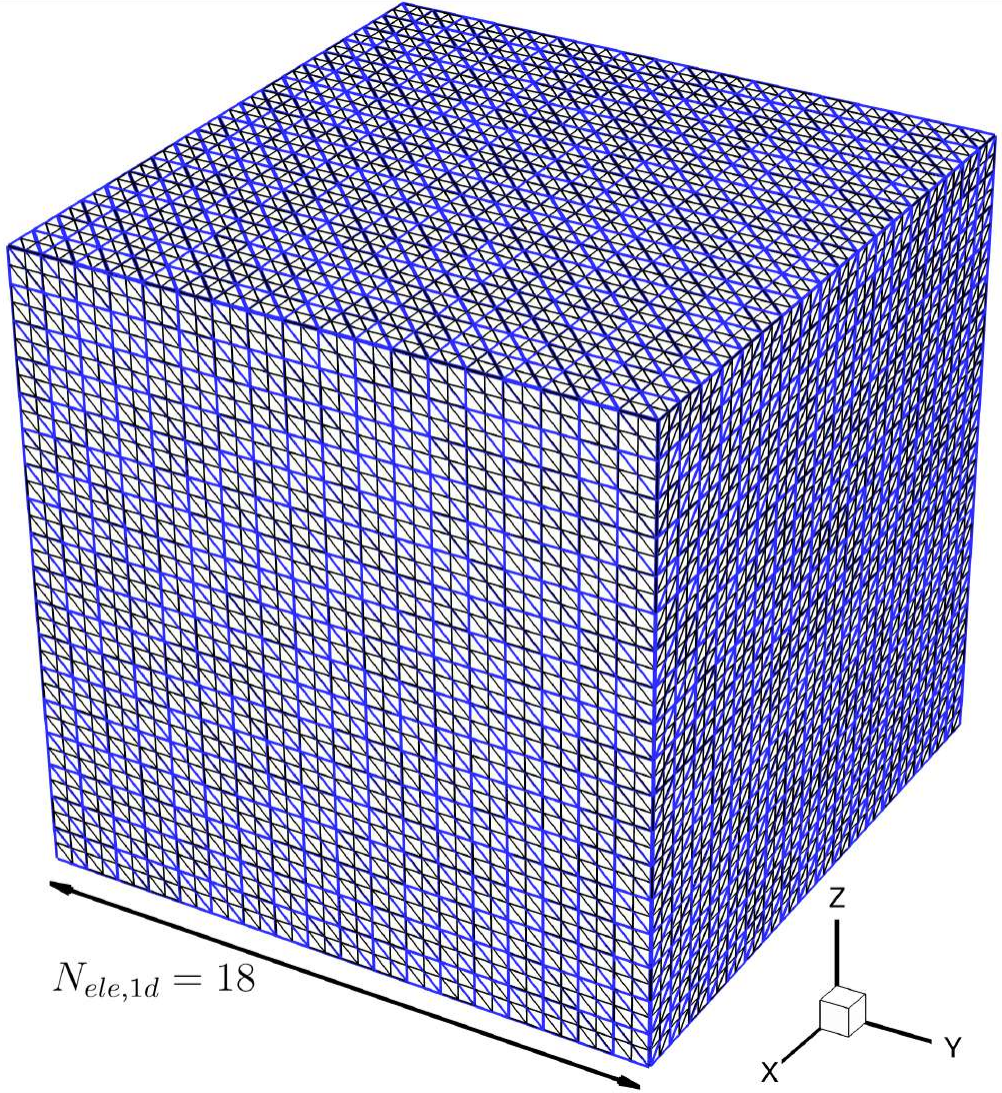} }}
    \caption{Viscous Taylor–Green vortex (${Re}_{\infty}=1,600$): Macro-element HDG discretizations ($m = 2$). The discontinuous macro-element edges are plotted in blue, the C$^0$-continuous element edges in black.}
    \label{fig:TGV_Re}%
\end{figure}

We now consider viscous Taylor Green vortex flow, based on the same setup outlined in detail in Section \ref{S62}. We focus on a low-Mach-number transitional flow regime characterized by ${Re}_{\infty} = 1,600$ and ${M}_{0} = 0.1$. To investigate the performance of our method, we consider two effective resolutions: $N_\text{eff} = 84^3$ and $N_\text{eff} =126^3$. The corresponding meshes are illustrated in Figure \ref{fig:TGV_Re}. We again compare the standard HDG method ($m=1$) and our macro-element HDG method ($m=2$) with polynomial degree $p = 3$.

\begin{figure}
    \centering
    \subfloat[$t/t_{c} = 0.0$.]{{\includegraphics[width=0.33\textwidth]{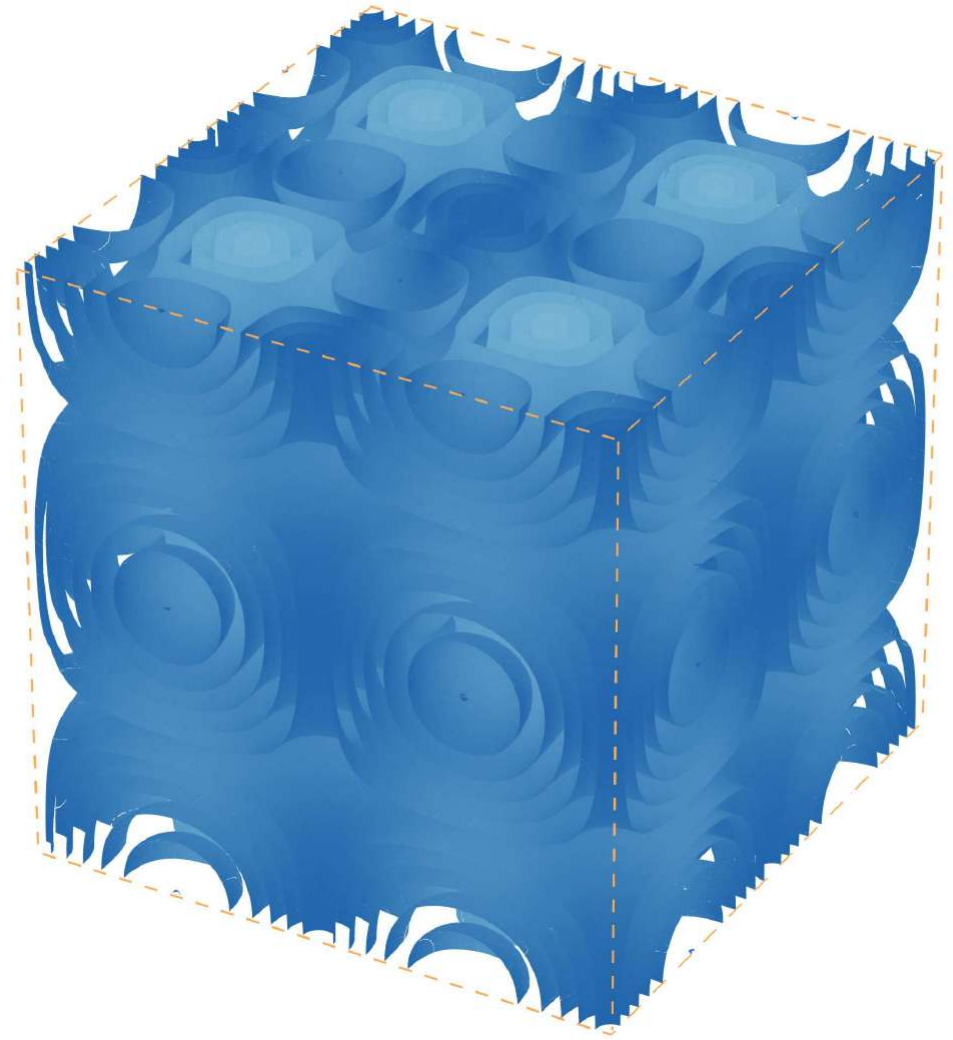} }}   
    \subfloat[$t/t_{c} = 3.0$.]{{\includegraphics[width=0.33\textwidth]{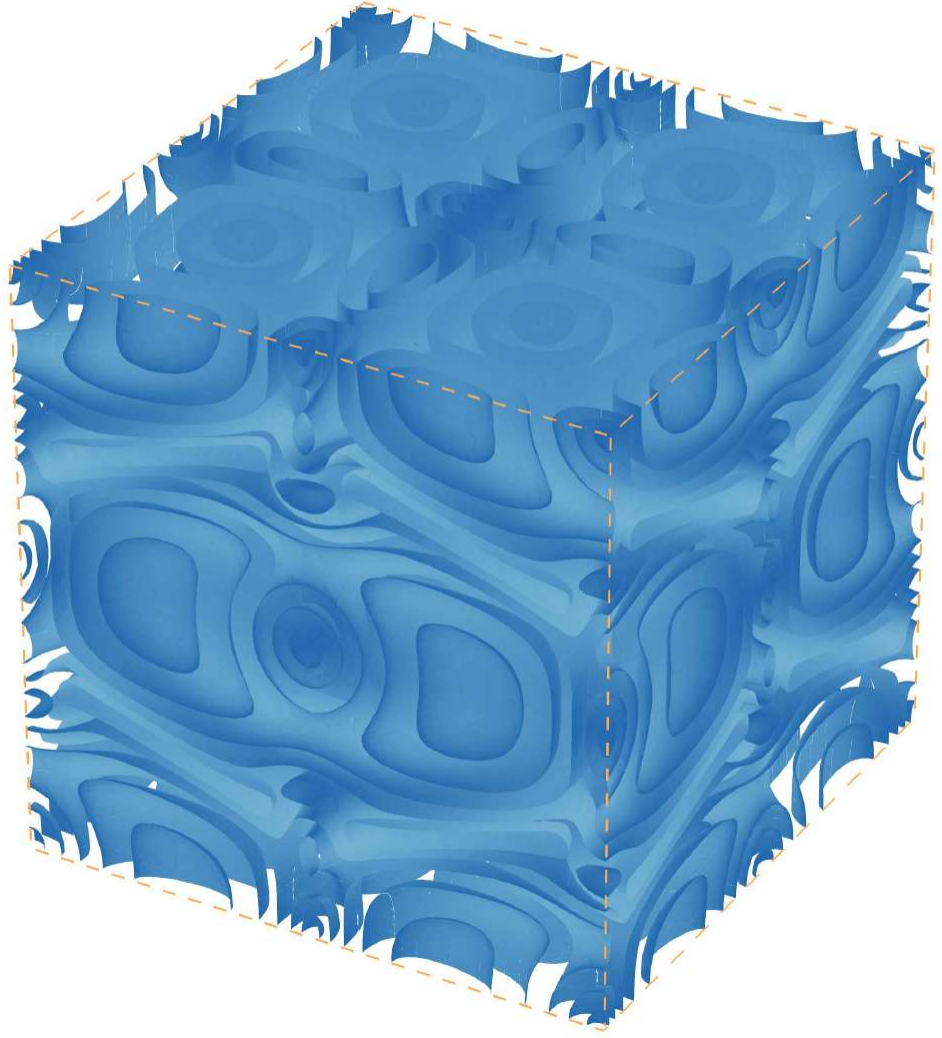} }}    
    \subfloat[$t/t_{c} = 6.0$.]{{\includegraphics[width=0.33\textwidth]{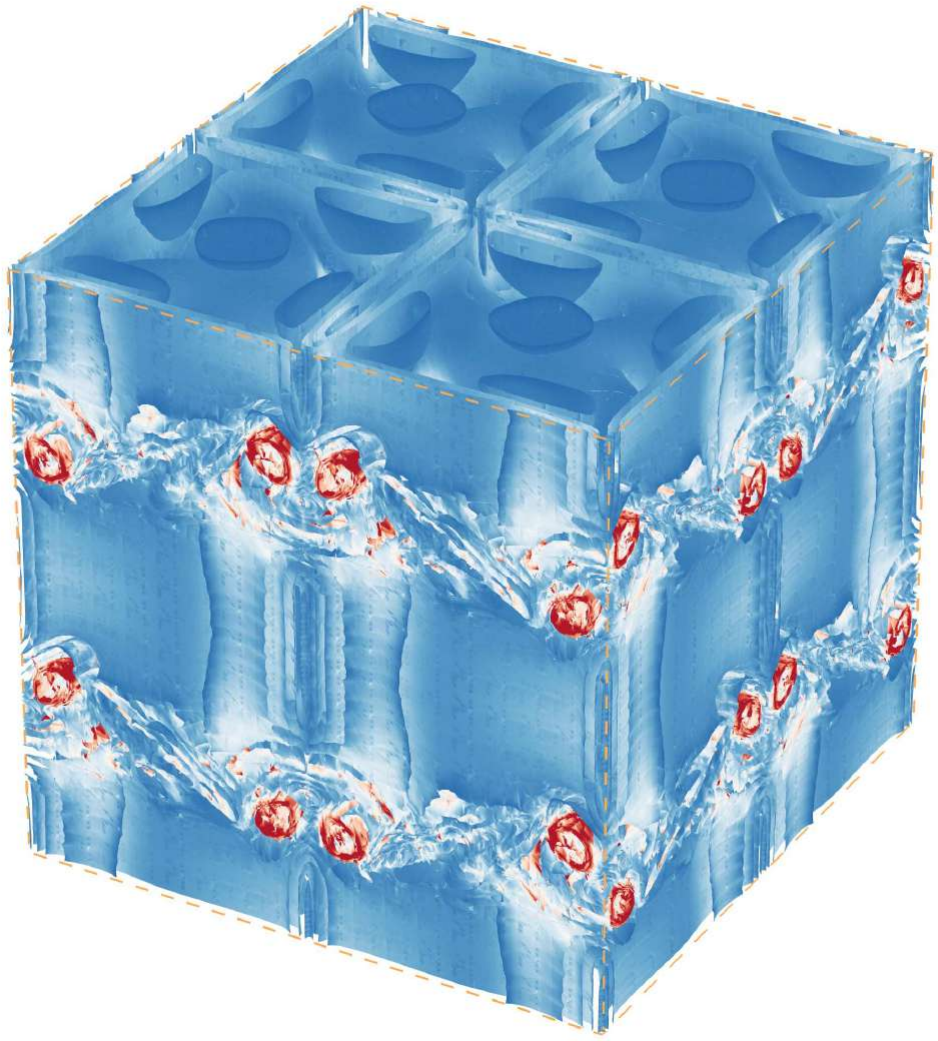} }}\\
    \subfloat[$t/t_{c} = 9.0$.]{{\includegraphics[width=0.33\textwidth]{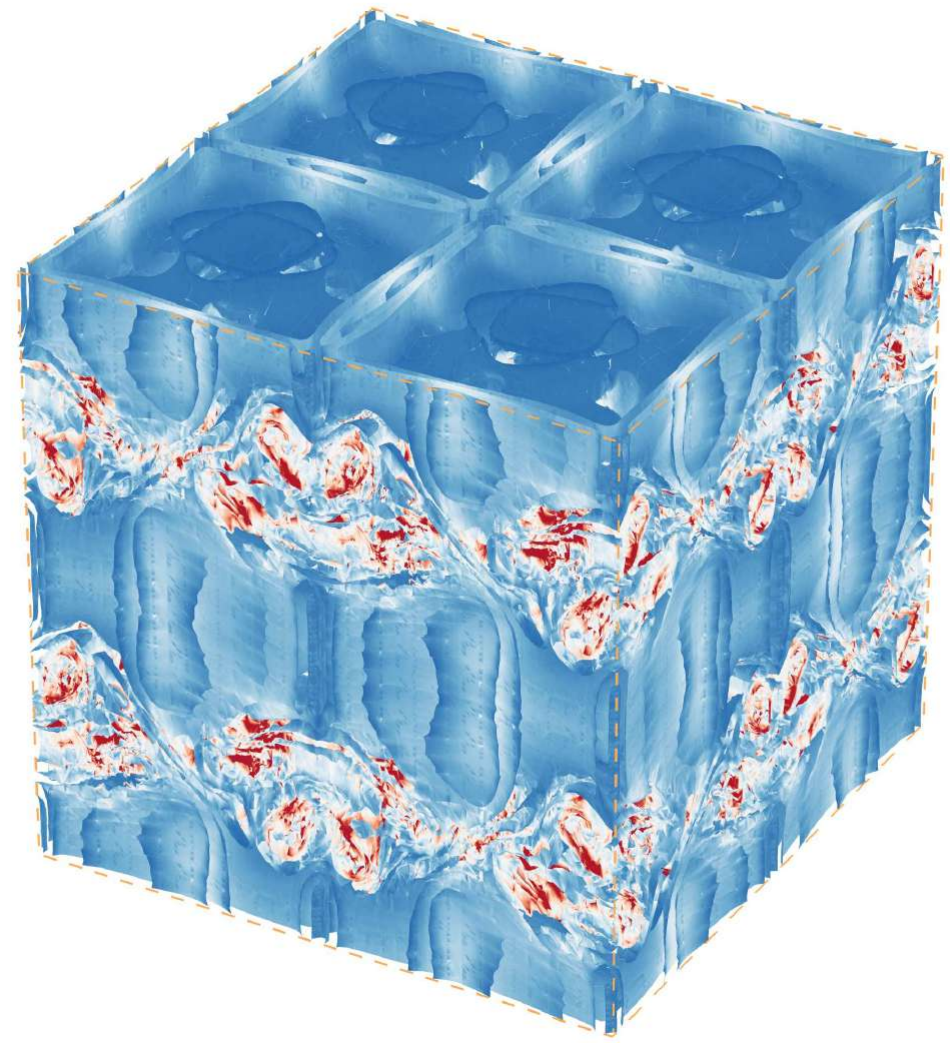} }}    
    \subfloat[$t/t_{c} = 12.0$.]{{\includegraphics[width=0.33\textwidth]{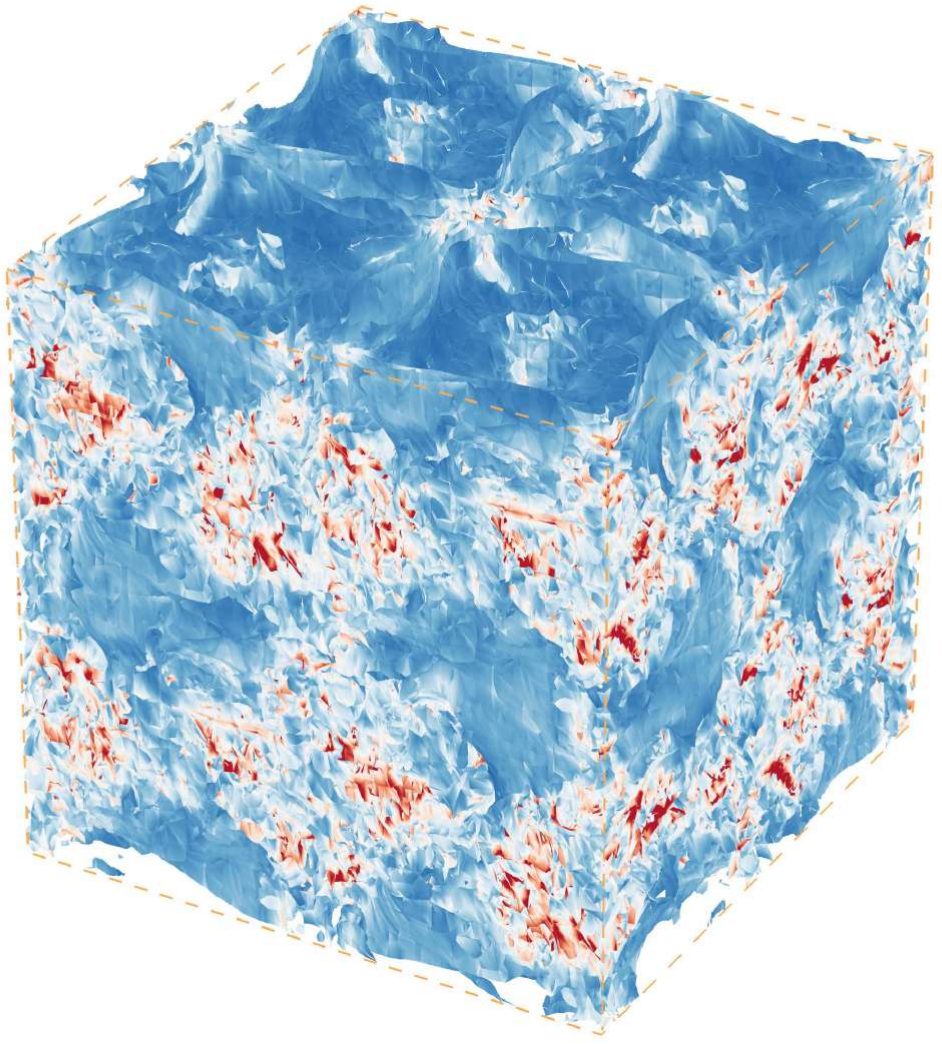} }}   
    \subfloat[$t/t_{c} = 15.0$.]{{\includegraphics[width=0.33\textwidth]{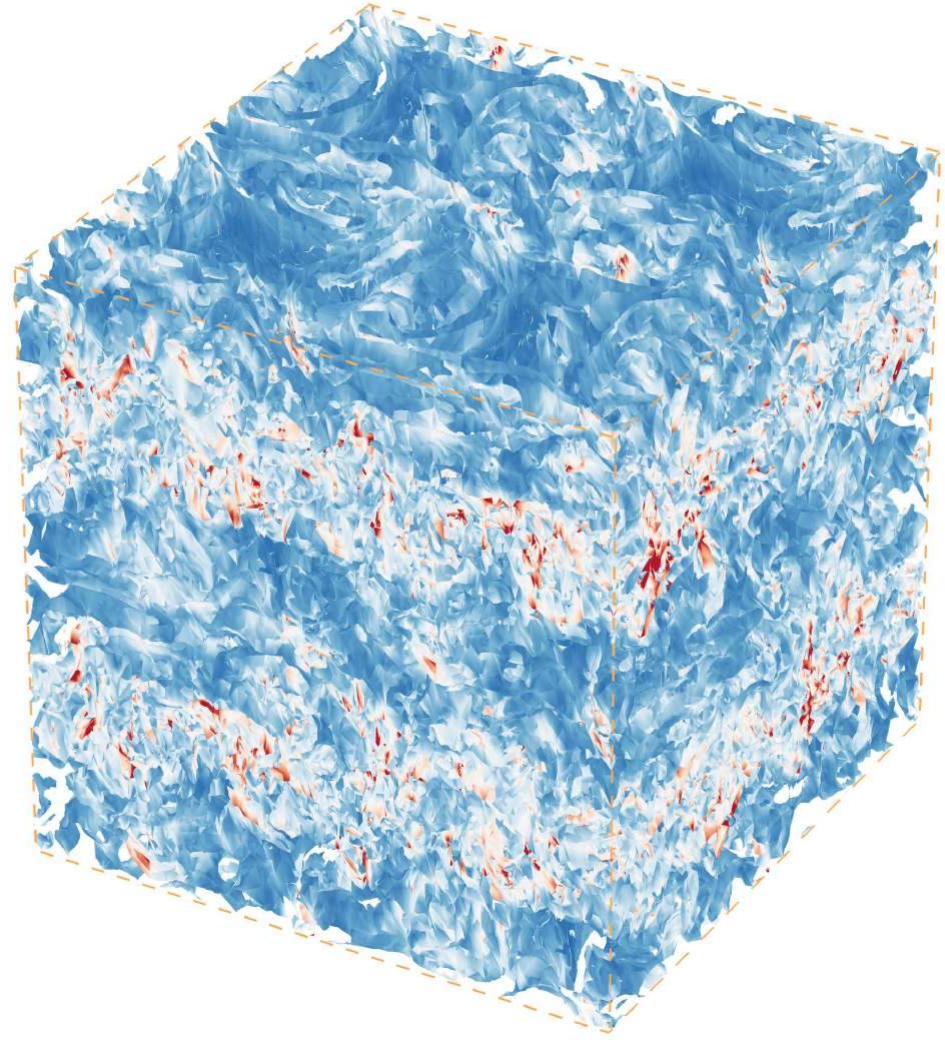} }} \\
    \subfloat{{\includegraphics[width=0.5\textwidth]{p3m2K1D18_legend.pdf} }}    
    \caption{Viscous Taylor–Green vortex (${Re}_{\infty}=1,600$): Vorticity magnitude plotted on iso-contours of the velocity field, obtained with macro-element HDG ($m=2, p=3$) in entropy-variable format and the KEPES flux formulation \eqref{ES} on the mesh resolution $N_{\text{eff}}=126^3$.}%
    \label{fig:Inv_TGV_Re_Sna}%
\end{figure}

The simulations are performed over the time interval $t_0 = 0.0$ to $t_f = 15 \, t_c$. For the viscous terms, we apply the numerical fluxes \eqref{Flux_Con_Vis} for the conservative-variable formulation and the numerical fluxes \eqref{Flux_En_Vis} for the entropy-variable formulation. We will examine the behavior of both formulations in the under-resolved flow setting, using our macro-element HDG framework. Figure \ref{fig:Inv_TGV_Re_Sna} illustrates the temporal evolution of the flow field, plotting the vorticity magnitude on iso-contours of the velocity field. These results were computed with the macro-element HDG method ($m=2, p=3$) in entropy-variable format and the KEPES flux formulation \eqref{ES} on the mesh resolution $N_{\text{eff}}=126^3$.

\subsubsection{Assessment of accuracy and stability}
\begin{figure}
    \centering  \subfloat[Evolution of thermodynamic entropy.]{{\includegraphics[width=0.43\textwidth]{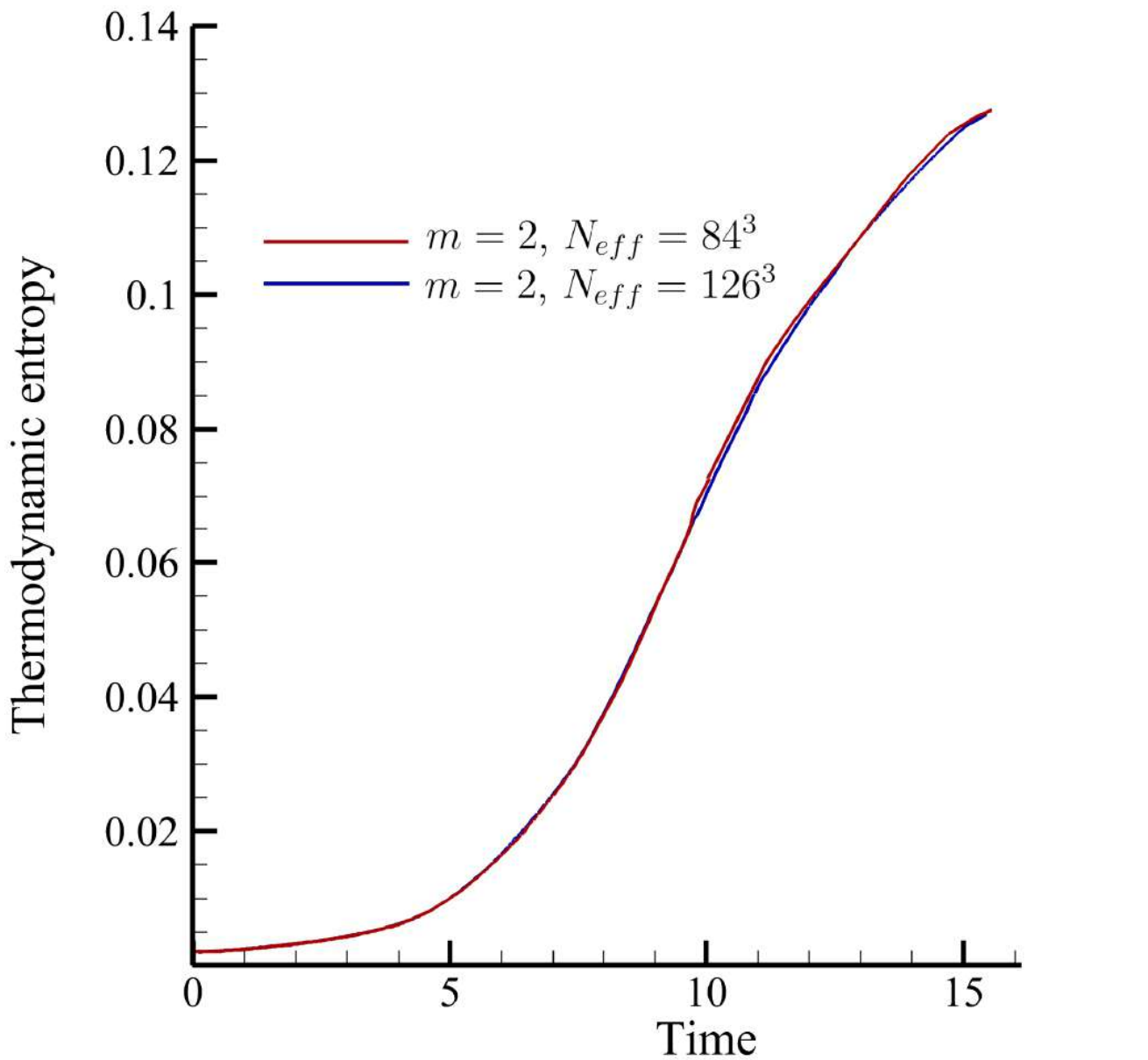} \label{fig:TGV_EK_Ka} }}   \hspace{0.5cm} \subfloat[Evolution of kinetic energy dissipation rate.]{{\includegraphics[width=0.5\textwidth]{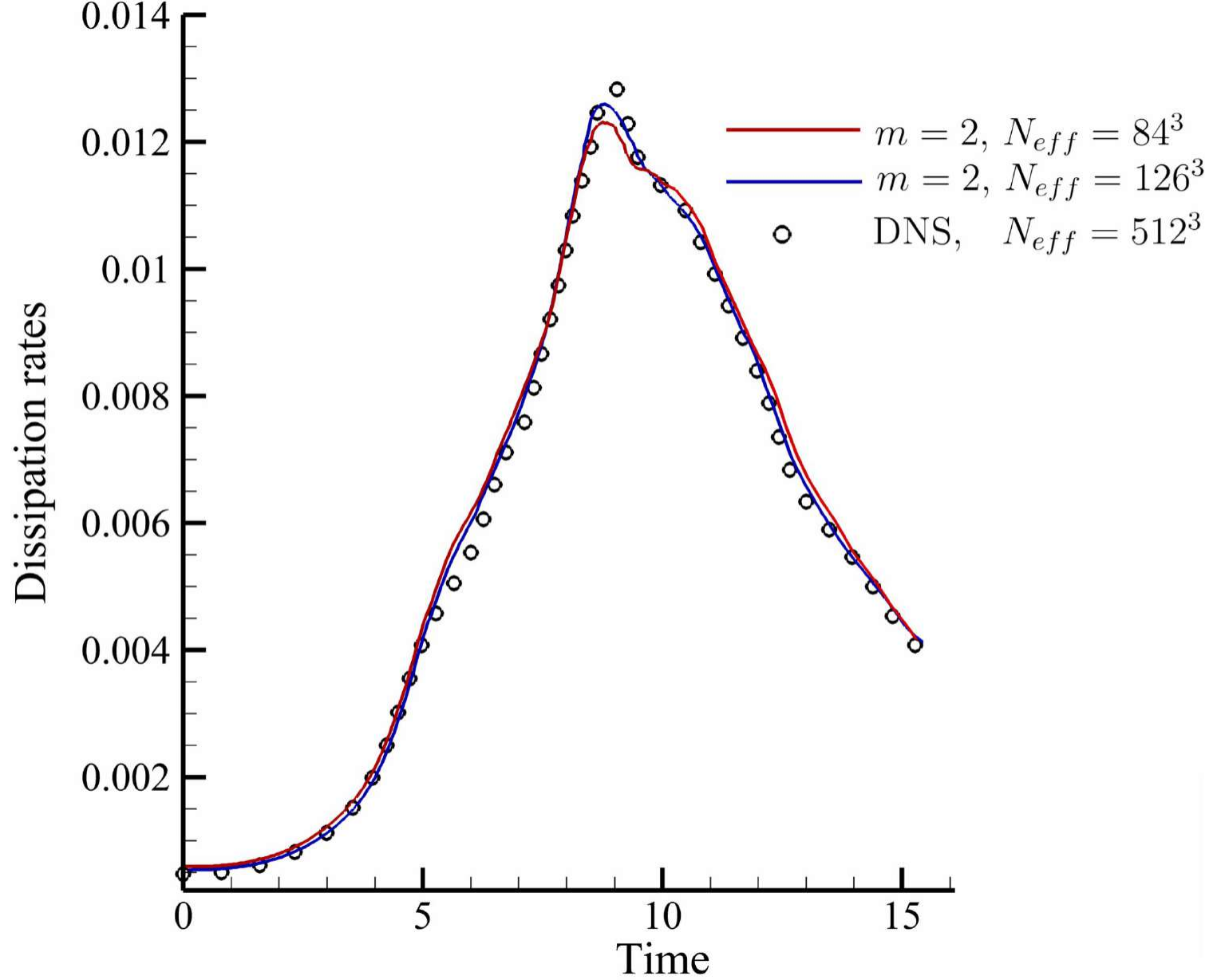} }}
    \caption{Viscous Taylor–Green vortex ($Re_{\infty}=1,600$): Entropy and energy dissipation behavior over time, reproduced by macro-element HDG ($m = 2$, $p=3$) in entropy-variable format with KEPES flux \eqref{ES}.}%
    \label{fig:TGV_EK_K}%
\end{figure}

To assess the accuracy of our macro-element HDG method, we place particular focus on the evolution of the thermodynamic entropy and the kinetic energy dissipation rate, as illustrated in Figure \ref{fig:TGV_EK_K}. We compare our results against a high-fidelity DNS reference solution reported in \cite{van2011comparison}, which was computed via a spectral method at an effective resolution of $N_{\text{eff}} = 512^3$.
The results demonstrate that the macro-element HDG method in entropy-variable format, combined with the inviscid flux \eqref{ES} and the viscous flux \eqref{Flux_En_Vis}, delivers high accuracy across the both mesh resolutions considered here, for which turbulent flow features are not fully resolved. This confirms the method’s ability to capture essential flow dynamics in an under-resolved setting, while maintaining computational efficiency.

\begin{table}[b!]
\centering
\caption{Viscous Taylor–Green vortex (${Re}_{\infty}=1,600$): Stability of the macro-element HDG method ($m = 2$, $p=3$) on two mesh resolutions. The entropy-variable format uses either the entropy-stable flux \eqref{Flux_En_Inv} (ES) or the energy-preserving entropy-stable flux \eqref{ES} (KEPES).}
\begin{tabularx}{\textwidth}{>{\centering\arraybackslash}p{1.2cm} *{3}{>{\centering\arraybackslash}X}} \toprule
& \multicolumn{3}{c}{\small Macro-element HDG ($m = 2$, $p=3$)} \\
\cmidrule(lr){2-4}
$N_{\text{eff}}$ & \small Conservative-variable format & \small Entropy-variable format (ES flux) & \small Entropy-variable format (KEPES flux) \\\midrule 
$84^3$  & \xmarkred & \cmarkgreen & \cmarkgreen \\[5pt]
$126^3$  & \cmarkgreen & \cmarkgreen & \cmarkgreen  \\\bottomrule				
\end{tabularx}
\label{Tab:TGV_Stability}
\end{table}

In Table \ref{Tab:TGV_Stability}, we see that the macro-element HDG method in conservative-variable format is unstable for the coarser mesh. On the finer mesh resolution, the method becomes stable. In contrast, the entropy-variable formulations remain stable across both mesh resolutions. This highlights the robustness of the entropy-variable formulations within our macro-element HDG framework, irrespective of the mesh resolution and in the under-resolved setting.
Figure \ref{fig:TGV_EK_Ka} confirms that the macro-element HDG method in entropy-variable format consistently satisfies the second law of thermodynamics for both mesh resolutions, evidenced by an increase in total thermodynamic entropy over time. This behavior aligns with the theoretical guarantees established in Theorem \ref{flux_ES_1}, which applies not only to the inviscid Euler equations but also extends to the compressible Navier–Stokes equations. 

\begin{table}[h!]
\caption{Viscous Taylor–Green vortex (${Re}_{\infty}=1,600$): number of local and global degrees of freedom resulting from standard HDG ($m=1$) and macro-element HDG ($m=2$) at different mesh resolutions and polynomial degree $p=3$.}
\centering
\begin{tabularx}{\textwidth}{>{\centering\arraybackslash}p{1.cm} >{\centering\arraybackslash}p{1.cm} >{\centering\arraybackslash}p{2.cm} *{4}{>{\centering\arraybackslash}X}}\toprule
\multicolumn{3}{c}{Mesh resolution} & \multicolumn{2}{c}{$\text{dof}^{local}$ } & \multicolumn{2}{c}{$\text{dof}^{global}$}
\\\cmidrule(r){1-3}\cmidrule(lr){4-5}\cmidrule(lr){6-7}
$N_{ele,1d}$  &$N_{\text{eff}}$  & $N_{ele}$     & $m = 1$ &$m = 2$   &$m = 1$  &$m = 2$ \\\midrule    
12            &$84^3$	 &  82,944	  &  33,177,600   &  17,418,240            &  8,294,400       &  2,903,040  	      \\[5pt]
18	         &$126^3$	 & 279,936	  &  111,974,400  &  58,786,560	            &  27,993,600     &  9,797,760      \\\bottomrule	 						
\end{tabularx}
\label{Tab:TGV_DOF}
\end{table}

\begin{table}[h!]
\caption{Viscous Taylor–Green vortex (${Re}_{\infty}=1,600$): computing times for local operations and global operations, the number of global solver iterations, and the total computing time for standard HDG ($m=1$) and macro-element HDG ($m=2$); $p = 3$, $N_\text{eff}=70^3$.}
\centering
\begin{tabularx}{\textwidth}{
  >{\arraybackslash}p{4.5cm}  
  >{\centering\arraybackslash}p{2.cm}  
  >{\centering\arraybackslash}p{2.cm}  
  >{\centering\arraybackslash}p{2.2cm}  
  >{\centering\arraybackslash}p{1.8cm}  
  >{\centering\arraybackslash}p{1.8cm}  
}
\toprule
& \footnotesize Time local op's [min] 
& \footnotesize Time global op's [min] 
& \footnotesize \# Global solver iter's 
& \footnotesize Total time [min] 
& \footnotesize \# Proc's \\\midrule
\footnotesize Standard HDG ($m = 1$) & 421.5 & 286.4  & 63,620  & 707.9 & 1,728  \\
\footnotesize Macro-element HDG ($m = 2$) & 167.1  & 83.8  & 33,210  & 250.9  & 1,728  \\
\bottomrule
\end{tabularx}
\label{Tab:TGV_Time}
\end{table}

\subsubsection{Computational performance}

Table \ref{Tab:TGV_DOF} summarizes the degrees of freedom associated with the two mesh resolutions. Table \ref{Tab:TGV_Time} reports the computing times (in minutes) for operations that are carried out locally per HDG element or macro-element HDG patch, versus the remaining components of the matrix-free global solver. The former contain the local solver, the local component of the matrix-free global solver (see \cite{badrkhani2023matrix,badrkhani2025matrix} for details), and the local formation of the Jacobian matrix. Computing times are reported for the standard HDG method ($m=1$) and the macro-element HDG method ($m=2$), both at polynomial degree $p = 3$ and on the mesh resolution $N_\text{eff}=84^3$. 

We observe again the same advantage for the macro-element HDG method, which reduces the computing times by approximately a factor of three compared to the standard HDG method on the same mesh resolution. This performance gain can be attributed to the reduction in the number of degrees of freedom as reported in Table \ref{Tab:TGV_DOF}, which also involves a significant reduction in the number of global solver iterations. 
These findings suggest that for large-scale simulations, adopting discontinuous macro-elements composed of $C^0$-continuous elements instead of individual discontinuous elements offers an effective strategy to reducing the number of degrees of freedom and to balance local and global computations, ultimately leading to improved overall efficiency.
\section{Conclusions and future work\label{Sec8}}

This work establishes an entropy-variable formulation for macro-element hybridized discontinuous Galerkin (HDG) discretizations of the compressible Navier--Stokes equations. By extending the previously developed macro-element HDG framework with entropy variables together with entropy-conservative, entropy-stable, and kinetic-energy-preserving numerical fluxes, the proposed formulation combines the computational efficiency of macro-element discretizations with the robustness of entropy-stable methods.

The derived discrete entropy evolution equation provides conditions for entropy conservation and entropy stability for arbitrary polynomial degrees and general macro-element discretizations. Within this framework, two inviscid numerical fluxes have been developed. The first achieves entropy stability through a dissipation mechanism based on the symmetric entropy Jacobian, while the second extends this construction by additionally preserving kinetic energy at the semidiscrete level, resulting in a kinetic-energy-preserving and entropy-stable (KEPES) flux. In contrast to the corresponding conservative-variable formulation, the entropy-variable formulation provides intrinsic stabilization within the $C^0$-continuous macro-elements, eliminating the need for additional SUPG stabilization while preserving the computational structure of the macro-element HDG framework.

The numerical investigations corroborate the theoretical developments. The proposed formulation achieves optimal convergence for smooth solutions and demonstrates substantially improved robustness in under-resolved, transonic, and turbulent compressible flow simulations while retaining the order-of-magnitude computational advantage of the macro-element HDG framework over conventional HDG discretizations. In particular, the entropy-stable and KEPES fluxes provide robust discretizations in challenging flow regimes where unresolved flow scales can otherwise compromise numerical stability. Collectively, these results demonstrate that entropy-variable formulations can be successfully integrated into the macro-element HDG framework while preserving its defining computational advantages. The resulting formulation combines entropy stability, kinetic-energy preservation, improved robustness, and computational efficiency within a unified high-order discretization framework, thereby extending the applicability of macro-element HDG methods to challenging compressible flow simulations.

Future work will focus on extending the proposed formulation to more demanding applications, including high-Reynolds-number turbulent flows, implicit large-eddy simulations, multi-component and reacting compressible flows \cite{badrkhani2025entropy,ten2025compressible}, and coupled multi-physics problems. In addition, the entropy-variable framework developed here provides a natural foundation for adaptive refinement strategies and next-generation matrix-free implementations for large-scale high-performance computing.

\section*{CRediT authorship contribution statement}
\textbf{Vahid Badrkhani:} Writing – review \& editing, Writing – original draft, Visualization, Validation, Software, Methodology, Investigation, Formal analysis, Data curation, Conceptualization. \textbf{Marco F.P. ten Eikelder:} Writing – review \& editing, Formal analysis. \textbf{Christian Hasse:} Writing – review \& editing, Formal analysis, Conceptualization. \textbf{Dominik Schillinger:} Writing – review \& editing, Project administration, Funding acquisition, Conceptualization.

\section*{Declaration of competing interest}
The authors declare that they have no known competing financial interests or personal relationships that could have appeared to influence the work reported in this paper.

\section*{Acknowledgments}

The authors gratefully acknowledge financial support from the German Research Foundation (Deutsche Forschungsgemeinschaft) through the DFG Emmy Noether Grant SCH 1249/2-1. The authors also gratefully acknowledge the computing time provided to them on the high-performance computer Lichtenberg at the NHR Centers NHR4CES at TU Darmstadt. This is funded by the Federal Ministry of Education and Research and the State of Hesse. We also acknowledge open access funding enabled and organized by Projekt DEAL.

\section*{Data availability}
The data that support the findings of this study are available from the corresponding author upon reasonable request.


\bibliographystyle{elsarticle-num}
\bibliography{references_names}

\end{document}